\documentclass[conference]{IEEEtran}
\IEEEoverridecommandlockouts

\usepackage{microtype}
\usepackage{graphicx}
\usepackage{subfig}
\usepackage{booktabs} 
\usepackage{amsmath}
\usepackage{comment}

\usepackage{amsmath,amssymb,amsfonts}
\usepackage{algorithmic}
\usepackage{microtype}
\usepackage{comment}
\usepackage{graphicx}
\usepackage{textcomp}
\usepackage{xcolor}
\usepackage{url}

\usepackage{breakurl}
\usepackage[breaklinks]{hyperref}
\usepackage{biblatex}
\newcommand{\MLP}{\text{MLP}}
\newcommand{\Ln}{\text{Norm}}
\newcommand{\Attn}{\text{Attn}}

\makeatletter
\newcommand{\linebreakand}{%
  \end{@IEEEauthorhalign}
  \hfill\mbox{}\par
  \mbox{}\hfill\begin{@IEEEauthorhalign}
}
\makeatother

\begin{document}

\title{The Case for Co-Designing Model Architectures with Hardware}

\author{
    \IEEEauthorblockN{Quentin Anthony}
    \IEEEauthorblockA{EleutherAI\\ Ohio State University}\and
    \IEEEauthorblockN{Jacob Hatef}
    \IEEEauthorblockA{EleutherAI\\ Ohio State University}\and
    \IEEEauthorblockN{Deepak Narayanan}
    \IEEEauthorblockA{NVIDIA}\and
    \IEEEauthorblockN{Stella Biderman}
    \IEEEauthorblockA{EleutherAI}\and
    \IEEEauthorblockN{Stas Bekman}
    \IEEEauthorblockA{Contextual AI}\linebreakand
    \IEEEauthorblockN{Junqi Yin}
    \IEEEauthorblockA{Oak Ridge National Lab}\and
    \IEEEauthorblockN{Aamir Shafi}
    \IEEEauthorblockA{Ohio State University}\and
    \IEEEauthorblockN{Hari Subramoni}
    \IEEEauthorblockA{Ohio State University}\and
    \IEEEauthorblockN{Dhabaleswar K. Panda}
    \IEEEauthorblockA{Ohio State University}
}

\maketitle


\vskip 0.3in

\begin{abstract}
While GPUs are responsible for training the vast majority of state-of-the-art deep learning models, the implications of their architecture are often overlooked when designing new deep learning (DL) models. As a consequence, modifying a DL model to be more amenable to the target hardware can significantly improve the runtime performance of DL training and inference. In this paper, we provide a set of guidelines for users to maximize the runtime performance of their \emph{transformer} models. These guidelines have been created by carefully considering the impact of various model hyperparameters controlling model shape on the efficiency of the underlying computation kernels executed on the GPU. We find the throughput of models with ``efficient'' model shapes is up to 39\% higher while preserving accuracy compared to models with a similar number of parameters but with unoptimized shapes.
\end{abstract}



\section{Introduction}
\label{sec:intro}

Transformer-based \parencite{vaswani2017attention} language models have become widely popular for language and sequence modeling tasks. Consequently, it is extremely important to train and serve large transformer models such as GPT-3 \parencite{brown2020language} and Codex as efficiently as possible given their scale and wide use. At the immense scales that are in widespread use today, efficiently using computational resources becomes a complex problem and small drops in hardware utilization can lead to enormous amounts of wasted compute, funding, and time. In this paper, we tackle a frequently ignored aspect of training large transformer models: how the shape of the model can impact runtime performance. We use first principles of GEMM optimization to optimize individual parts of the transformer model (which translates to improved end-to-end runtime performance as well). Throughout the paper, we illustrate our points with extensive computational experiments demonstrating how low-level GPU phenomenon impact throughput throughout the language model architecture.

Many of the phenomena remarked on in this paper have been previously documented, but continue to plague large language model (LLM) designers to this day. We hypothesize that there are three primary causes of this:

\begin{enumerate}
    \item Few resources trace the performance impacts of a transformer implementation all the way to the underlying computation kernels executed on the GPU.
    \item The existing documentation on how transformer hyperparameters map to these kernels is not always in the most accessible formats, including tweets \parencite{karpathy2023tweet,horace2023tweet}, footnotes \parencite{shoeybi2019megatron}, and in comments in training libraries \parencite{gpt-neox-library}.
    \item It is convenient to borrow architectures from other papers and researchers rarely give substantial thought to whether those choices of model shapes are optimal.
\end{enumerate}

This work attempts to simplify performance tuning for transformer models by carefully considering the architecture of modern GPUs. This paper is also a demonstration of our thesis that \textbf{model dimensions should be chosen with hardware details in mind} to an extent far greater than is typical in deep learning research today.

\begin{figure}[htbp]
\centering
    \includegraphics[width=.95\linewidth,trim=4 2 2 1,clip]{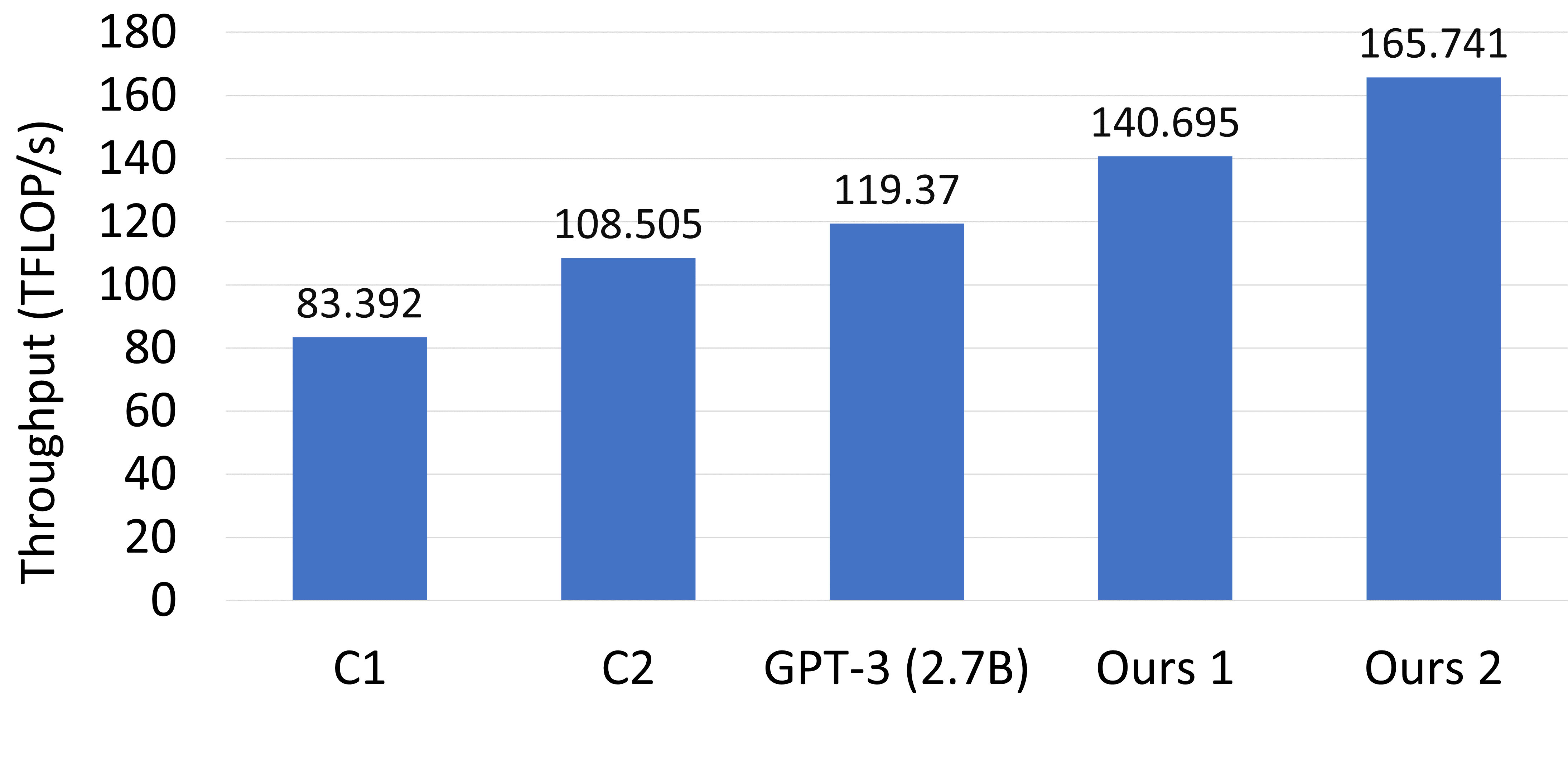}
    \caption{Transformer single-layer throughput of various architectures for a 2.7 billion parameter model (C1 and C2 are defined by this paper as C1: {$h=2560, a=64$}, C2: {$h=2560, a=40$}).}
    \label{fig:gpt_throughput_motivation}
\end{figure}

As shown in Figure \ref{fig:gpt_throughput_motivation}, the runtimes of models with a nearly identical number of parameters but different shapes can vary wildly. 
In this figure, the ``standard architecture'' for a 2.7B transformer model defined by GPT-3 \cite{brown2020language} has been used by OPT\cite{zhang2022opt}, GPT-Neo\cite{black2021gpt}, Cerebras-GPT\cite{dey2023cerebrasgpt}, RedPajama-INCITE\cite{together2023redpajamaincite}, and Pythia\cite{biderman2023pythia}.
Unfortunately the knowledge of how to optimally shape transformer architectures is not widely known, resulting in people often making sub-optimal design decisions. This is exacerbated by the fact that researchers often deliberately copy hyperparameters from other papers for cleaner comparisons, resulting in these sub-optimal choices becoming locked in as the standard. As one example of this, we show that the 2.7 billion parameter model described in \textcite{brown2020language} can be trained almost 20\% faster than the default architecture through minor tweaking of the model shape.

\begin{figure}[htbp]
    \centering
    \includegraphics[width=.95\linewidth,trim=4 3 3 2,clip]{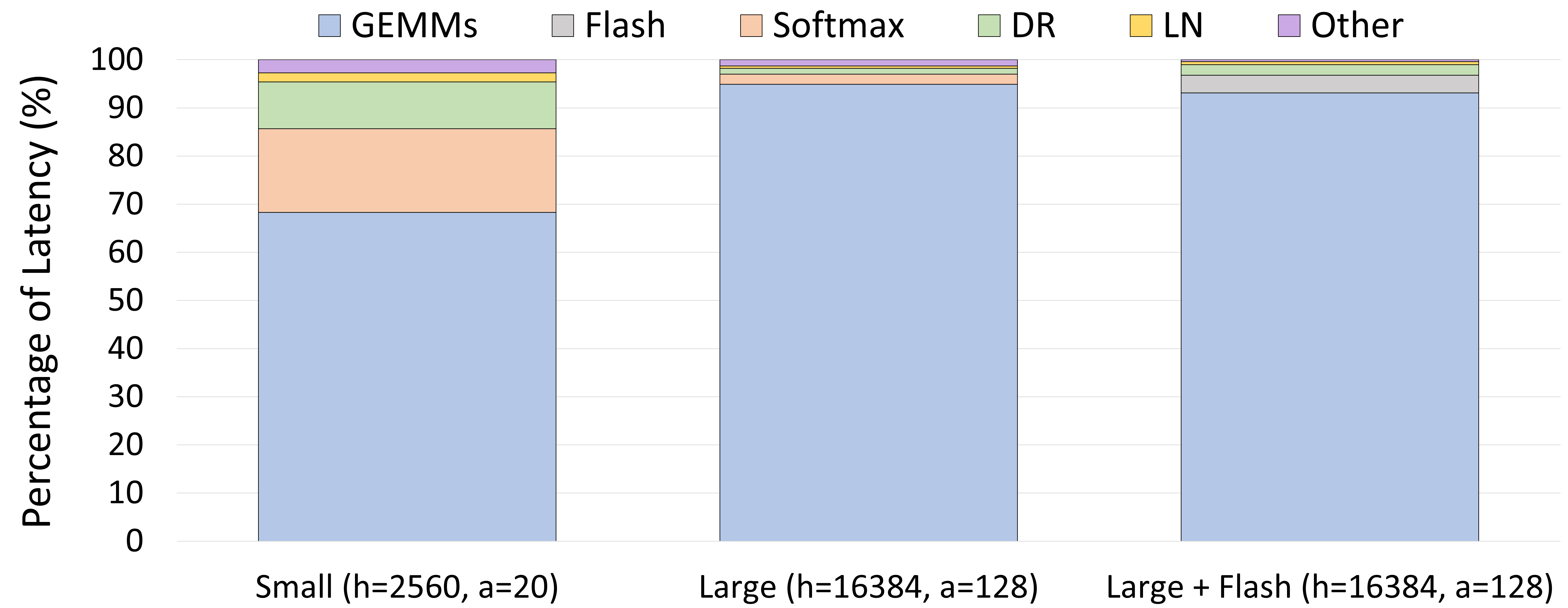}
    \caption{The proportion of latency from each transformer component for one layer of various model sizes}
    \label{fig:proportion-gemms}
\end{figure}

Our analysis makes use of the fact that General Matrix Multiplications (GEMMs) are the lifeblood of modern deep learning. Most widely-used compute-intensive layers in deep learning explicitly use GEMMs (e.g., linear layers or attention layers) or use operators that are eventually lowered into GEMMs (e.g., convolutions). For transformer models, our experiments from Figure \ref{fig:proportion-gemms} show that GEMM kernels regularly account for $68.3\%$ and $94.9\%$ of the total model latency for medium- and large-sized models, respectively. As a result, understanding the performance of GEMMs is crucial to understanding the runtime performance of end-to-end models; this only becomes more important as model size increases. 

On account of their parallel architecture, GPUs are a natural hardware platform for GEMMs. However, the observed throughput for these GEMMs depends on the matrix dimensions due to how the computation is mapped onto the execution units of the GPU (called streaming multiprocessors or SMs for short). As a result, GPU efficiency is sensitive to the model depth and width, which control the arithmetic efficiency of the computation, SM utilization, kernel choice, and the usage of tensor cores versus slower cuda cores. This work tries to determine how best to size models to ensure good performance on GPUs, taking these factors into account. Optimizing model shapes for efficient GEMMs will increase throughput for the entire lifetime of the model, decreasing training time and inference costs\footnote{We expect best results when the inference GPU is the same as the training GPU, but the guidelines we present could also be useful when the two are different.} for production models. 

\subsection{Contributions}
Our contributions are as follows:
\begin{itemize}
    \item We map the transformer model to its underlying matrix multiplications / GEMMs, and show how each component of the transformer model can suffer from using sub-optimal transformer dimensions.
    \item We compile a list of GPU performance factors into one document and explain how to choose optimal GEMM dimensions.
    \item We define rules to ensure transformer models are composed of efficient GEMMs.
\end{itemize}
\section{Related Work}
\label{sec:related-work}

\subsection{GPU Characterization of DNNs}
DL model training involves the heavy use of GPU kernels, and the characterization of such kernel behavior constitutes a large body of prior work that this paper builds upon. GPU kernels, especially GEMM kernels, are key to improving DL training and inference performance. Therefore, characterizing~\parencite{xsp_profiling} and optimizing~\parencite{bytetransformer, aminabadi2022deepspeed, fastertransformer,fang2021turbotransformers} these kernels have received a lot of attention in recent work~\parencite{survey_gpu_optim}.

Beyond GPU kernels, new algorithms and DL training techniques have been developed to optimize I/O~\parencite{dao2022flashattention, dao2023flashattention} and leverage hardware features like Tensor Cores~\parencite{demystify_tc, Raihan2018ModelingDL} as efficiently as possible. In addition to the above studies for DL training, exploiting Tensor Core properties has also shown excellent speedups for scientific applications such as iterative solvers~\parencite{iter_solvers_tc} and sparse linear algebra subroutines~\parencite{tsai2020evaluating}.

\subsection{Comparison Across DL Accelerators}
In recent years, there has emerged a range of acceleration strategies such as wafer-scale (Cerebras), GPUs (AMD and NVIDIA), and tensor processing units (Google). Given this diverse array of new AI accelerators, many pieces of work perform cross-generation and cross-accelerator comparison that have helped elucidate the strengths and weaknesses of each accelerator. Cross-accelerator studies such as 
\textcite{accelerator_evaluations,Wang2019BenchmarkingTG, mlperf} enable HPC and cloud customers to choose an appropriate accelerator for their DL workload.
We seek to extend this particular line of work by evaluating across various datacenter-class NVIDIA (V100, A100, and H100) and AMD GPUs (MI250X). 

\subsection{DL Training Performance Guides}
The most similar effort to our work is a GPU kernel characterization study for RNNs and CNNs performed in~\textcite{yin_cnn}. Since the transformer architecture differs greatly compared to RNNs and CNNs, we believe that our work provides a timely extension. Further, our focus on creating a practical performance guide is similar in nature to the 3D-parallelism optimization for distributed GPU architectures presented in~\textcite{narayanan2021efficient}.

From the above discussion, one can posit that while many papers exist to optimize DL performance on GPUs~\parencite{survey_gpu_optim}, such papers tend to neglect the fundamental effects that GPU properties (e.g. Tensor Cores, tiling, wave quantization, etc.) have on model training. Because of this omission, many disparate DL training groups have rediscovered a similar set of model sizing takeaways~\parencite{karpathy2023tweet, horace2023tweet, shoeybi2019megatron, gpt-neox-library}. We seek to provide explanations for these takeaways from the perspective of fundamental GPU first-principles, and to aggregate these explanations into a concise set of takeaways for efficient transformer training and inference.

\section{Background}
\label{sec:background}

We will now discuss some of the necessary prerequisite material to understand the performance characteristics of the GPU kernels underlying transformer models.

\subsection{GPU Kernels}
\label{sec:background-kernels}

General Matrix Multiplications (GEMMs) serve as a crucial component for many functions in neural networks, including fully-connected layers, recurrent layers like RNNs, LSTMs, GRUs, and convolutional layers. If $A$ is an $m \times k$ matrix and $B$ is a $k \times n$ matrix, then the matrix product $AB$ is a simple GEMM. We can then generalize this to $C = \alpha AB + \beta C$ (in the previous example, $\alpha$ is 1, and $\beta$ is 0). In a fully-connected layer's forward pass, the weight matrix would be argument $A$ and input activations would be argument $B$ ($\alpha$ and $\beta$ would typically be 1 and 0 as before; $\beta$ can be 1 in certain scenarios, such as when adding a skip-connection with a linear operation). 

Matrix-matrix multiplication is a fundamental operation in numerous scientific and engineering applications, particularly in the realm of deep learning. It is a computationally intensive task that requires significant computational resources for large-scale problems. To address this, various algorithms and computational techniques have been developed to optimize matrix-matrix multiplication operations.

Matrix multiplication variants like batched matrix-matrix (BMM) multiplication kernels have also been introduced to improve the throughput of certain common DL operators like attention~\cite{vaswani2017attention}. A general formula for a BMM operation is given by Equation \ref{eq:bmm} below, where $\{A_i\}$ and $\{B_i\}$ are a batch of matrix inputs, $\alpha$ and $\beta$ are scalar inputs, and $\{C_i\}$ is a batch of output matrices.

\begin{equation}
    \label{eq:bmm}
    C_i = \alpha A_i B_i + \beta C_i, \quad i = 1,...N
\end{equation}

\subsection{NVIDIA GEMM Implementation and Performance Factors}

\begin{figure}[htbp]
\vspace{-3ex}
\centering
    \includegraphics[width=.7\linewidth]{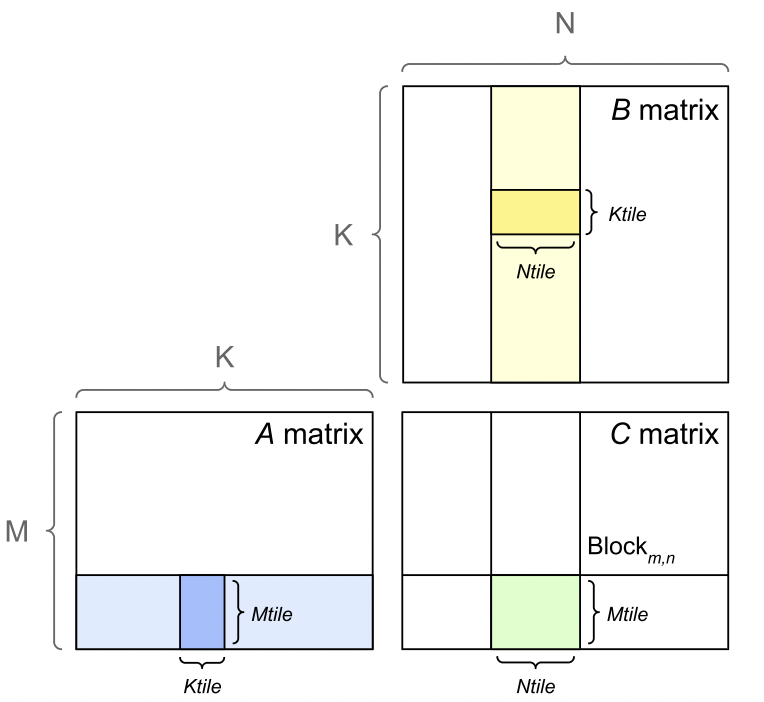}
    \caption{GEMM tiling~\cite{GEMMguide}.}
    \label{fig:tiling}
\end{figure}

There are a number of performance factors to consider when analyzing GEMMs on NVIDIA GPU architectures. NVIDIA GPUs divide the output matrix into regions or tiles as shown in Figure \ref{fig:tiling} and schedule them to one of the available streaming multiprocessors (SM) on the GPU (e.g., A100 GPUs have 108 SMs). Each tile or thread block is processed in a Tensor Core, which NVIDIA introduced for fast tensor operations. NVIDIA Tensor Cores are only available for GEMMs with appropriate dimensions. Tensor Cores can be fully utilized when GEMM dimensions $m$, $k$, and $n$ are multiples of 16 bytes and 128 bytes for V100 and A100 GPUs, respectively. Since a FP16 element is 2 bytes, this corresponds to dimension sizes that are multiples of 8 and 64 elements, respectively. If these dimension sizes are not possible, Tensor Cores perform better with larger multiples of 2 bytes. 

There are multiple tile sizes that the kernel can choose from. If the GEMM size does not divide evenly into the tile size, there will be wasted compute, where the thread block must execute fully on the SM, but only part of the output is necessary. This is called the tile quantization effect, as the output is quantized into discrete tiles. 

Another quantization effect is called wave quantization. As the thread blocks are scheduled to SMs, only 108 thread blocks at a time may be scheduled. If, for example, 109 thread blocks must be scheduled, two rounds, or waves, of thread blocks must be scheduled to GPU. The first wave will have 108 thread blocks, and the second wave will have 1. The second wave will have almost the same latency as the first, but with a small fraction of the useful compute. As the matrix size increases, the last or tail wave grows. The throughput will increase, until a new wave is required. Then, the throughput will drop.

\subsection{Transformer Models}
\label{sec:background-transformer}

In this study, we examine a decoder-only transformer architecture popularized by GPT-2 \parencite{radford2019language}. We focus on this architecture due to its popularity for training very large models \parencite{brown2020language,chowdhery2022palm,smith2022using}
, but most of our conclusions also apply to encoder-only models \parencite{devlin2019bert,liu2019roberta}. Due to the nature of the transition between the encoder and decoder, our analysis will largely not apply to encoder-decoder models \parencite{vaswani2017attention,raffel2020exploring}.
\begin{figure}[htbp]
\vspace{-2ex}
\centering
    \includegraphics[width=.7\linewidth]{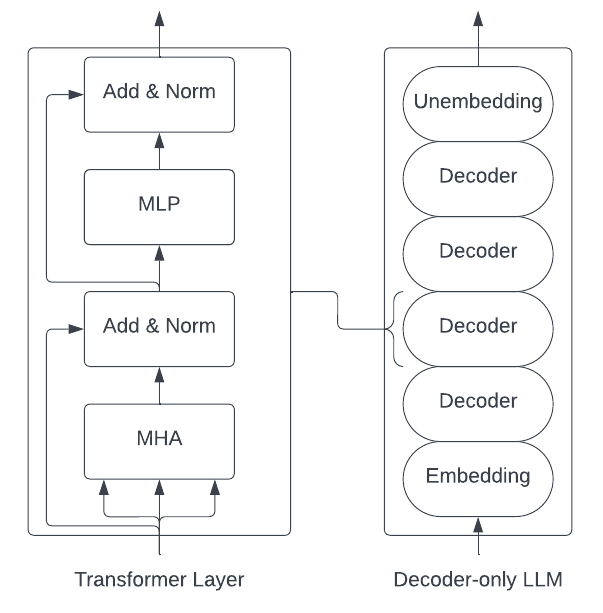}
    \caption{The transformer architecture \parencite{radford2019language}.}
    \label{fig:transformer-arch}
\end{figure}

For a mapping from variables to their definitions, see Table \ref{tab:varnames}. Initially, the network takes in raw input tokens which are then fed into a word embedding table of size $v \times h$. These token embeddings are then merged with learned positional embeddings of size $s \times h$. The output from the embedding layer, which serves as the input for the transformer block, is a 3-D tensor of size $s \times b \times h$. Each layer of the transformer comprises a self-attention block with attention heads, followed by a two-layer multi-layer perceptron (MLP) that expands the hidden size to $4h$ before reducing it back to $h$. The input and output sizes for each transformer layer remain consistent at $s \times b \times h$. The final output from the last transformer layer is projected back into the vocabulary dimension to compute the cross-entropy loss. 

\begin{table}[h]
\resizebox{\columnwidth}{!}{%
\begin{tabular}{l|ll|l}\toprule
\textit{a} & Number of attention heads    & \textit{s} & Sequence length      \\
\textit{b} & Microbatch size              & \textit{t} & Tensor-parallel size \\
\textit{h} & Hidden dimension size        & \textit{v} & Vocabulary size     \\
\textit{L} & Number of transformer layers & 
\end{tabular}%
}
\caption{Variable names.}
\label{tab:varnames}
\end{table}

Each transformer layer consists of the following matrix multiplication operators:
\begin{enumerate}
    \item Attention key, value, query transformations: These can be expressed as a single matrix multiplication of size: $(b \cdot s, h) \times (h, \frac{3h}{t})$. Output is of size $(b \cdot s, \frac{3h}{t})$.
    \item Attention score computation: $b \cdot a/t$ batched matrix multiplications (BMMs), each of size $(s, \frac{h}{a}) \times (\frac{h}{a}, s)$. Output is of size $(\frac{b \cdot a}{t}, s, s)$.
    \item Attention over value computation: $\frac{b \cdot a}{t}$ batched matrix multiplications of size $(s, s) \times (s, \frac{h}{a})$. Output is of size $(\frac{b \cdot a}{t}, s, \frac{h}{a})$.
    \item Post-attention linear projection: a single matrix multiplication of size $(b \cdot s, \frac{h}{t}) \times (\frac{h}{t}, h)$. Output is of size $(b \cdot s, h)$.
    \item Matrix multiplications in the MLP block of size $(b \cdot s, h) \times (h, \frac{4h}{t})$ and $(b \cdot s, \frac{4h}{t}) \times (\frac{4h}{t}, h)$. Outputs are of size $(b \cdot s, \frac{4h}{t})$ and $(b \cdot s, h)$.
\end{enumerate}

The total number of parameters in a transformer can be calculated using the formula $P=12h^2L + 13hL + (v+s)h$. This is commonly approximated as $P = 12h^2L$, omitting the lower-order terms. 

\begin{table}[ht]
\centering
\begin{tabular}{lcc}
 \textbf{Module} & \textbf{GEMM Size} & \textbf{Figure} \\ 
 \midrule
 Input Embedding & --- & --- \\ 
 Layer Norm 1    & --- & --- \\
 $QKV$ Transform & $(b \cdot s, h) \times (h, \frac{3h}{t})$ &  \ref{fig:attn_key_value_query_transform} \\  
 Attention Score & $(\frac{b \cdot a}{t}, s, \frac{h}{a}) \times (\frac{b \cdot a}{t}, \frac{h}{a}, s)$ &  \ref{subfig:attention-kq} \ref{fig:attention-kq-ha64-16384}\\
 Attn over Value & $(\frac{b \cdot a}{t}, s, s) \times (\frac{b \cdot a}{t}, s, \frac{h}{a})$ &  \ref{subfig:attention-val} \ref{fig:attention-val-ha64-16384}\\
 Linear Projection & $(b \cdot s, \frac{h}{t}) \times (\frac{h}{t}, h)$ &  \ref{fig:attn_linproj} \\
 Layer Norm 2 & --- & --- \\
 MLP $h$ to $4h$ & $(b \cdot s, h) \times (h, \frac{4h}{t})$ &  \ref{fig:mlp_h_to_4h} \\
 MLP $4h$ to $h$ & $(b \cdot s, \frac{4h}{t}) \times (\frac{4h}{t}, h)$ &  \ref{fig:mlp_4h_to_h} \\
 Linear Output & $(b \cdot s,v) \times (v,h)$ &  \ref{fig:vocab_sweep} \\ 
\bottomrule
\end{tabular}
\caption{Summary of operators in the transformer layer considered in this paper, along with the size of the GEMMs used to execute these operators.}
\label{tab:operators}
\end{table}

Here, we make the assumption that the projection weight dimension in the multi-headed attention block is $h / a$, which is the default in existing implementations like Megatron~\parencite{shoeybi2019megatron} and GPT-NeoX \parencite{gpt-neox-library}.

The total number of compute operations needed to perform a forward pass for training is
then $24bsh^2 + 4bs^2h = 24bsh^2 \left(1 + \frac{s}{6h}\right)$.

\textit{Parallelization Across GPUs.} Due to the extreme size of modern transformer models, and the additional buffers and activations needed for training, it is common to split transformers across multiple GPUs using tensor and pipeline parallelism~\parencite{shoeybi2019megatron, narayanan2021efficient}. Since this paper focuses on the computations being done on a single GPU, we will largely ignore parallelism. When we speak of the hidden size of a model, that should be understood to mean the hidden size \textit{per GPU}. For example, with $t$-way tensor parallelism, the hidden size per GPU is typically $h/t$. We leave an analysis of the implications of pipeline and sequence parallelism on optimal model shapes to future work.
\begin{table*}[htb]
\centering
\resizebox{\textwidth}{!}{%
\begin{tabular}{lccccc}
\toprule
& \textbf{GPU Vendor} & \textbf{GPU} & \textbf{CPU}              & \textbf{Inter-node Interconnect} & \textbf{Intra-node Interconnect} \\ \midrule
AWS p4d                  & NVIDIA              & 8x(A100 40GB)    & Intel Cascade Lake 8275CL & Amazon EFA [400 Gbps]            & NVLINK [600 GBps]                \\ \midrule
ORNL Summit    & NVIDIA              & 6x(V100 16GB)    & IBM POWER9                & InfiniBand EDR [200 Gbps]        & NVLINK (2x3) [100 GBps]          \\ \midrule
SDSC Expanse  & NVIDIA              & 4x(V100 32GB)    & AMD EPYC 7742             & InfiniBand HDR [200 Gbps]        & NVLINK [100 GBps]                \\ \bottomrule
\end{tabular}
}
\caption{Hardware systems used in this paper.}
\label{tab:hardware}
\end{table*}

\section{Experimental  Setup}
\label{sec:systems}

\subsection{Hardware Setup}
\label{sec:hw}

All experimental results were measured on one of the systems described in Table ~\ref{tab:hardware}. We used compute from a wide variety of sources such as Oak Ridge National Laboratory (ORNL), the San Diego Supercomputing Center (SDSC), and cloud providers such as AWS and Cirrascale. In order to increase the coverage of our takeaways as much as possible, we have included a diverse range of systems in this study.

\subsection{Software Setup}
\label{sec:sw}

Each hardware setup has used slightly different software. For the V100 experiments, we used PyTorch 1.12.1 and CUDA 11.3. For the A100 experiments, we used PyTorch 1.13.1, CUDA 11.7. For H100 experiments, we used PyTorch 2.1.0 and CUDA 12.2.2. For MI250X experiments, we used PyTorch 2.1.1 and ROCM 5.6.0. All transformer implementations are ported from GPT-NeoX~\parencite{gpt-neox-library}. 
\section{GEMM Results}
\label{sec:results-gemm}
\begin{figure}[h]
    \centering
    \subfloat[$(m, 4096) \times (4096, m)$]{
        \includegraphics[width=0.8\linewidth]{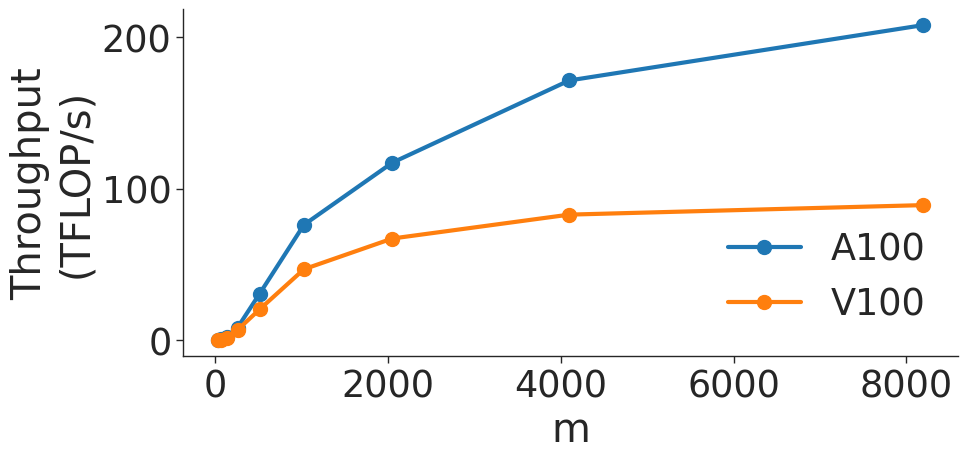}
        \label{fig:first}
    }\\
    \subfloat[$(27648, 4096) \times (4096, k)$]{
        \includegraphics[width=0.8\linewidth]{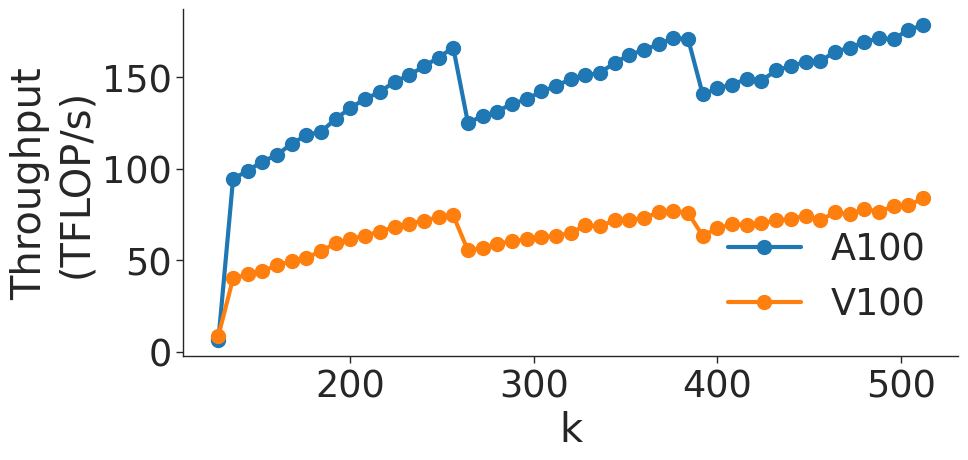}
    }\\
    \subfloat[$(2304, 4096) \times (4096, k)$.]{
        \includegraphics[width=0.8\linewidth]{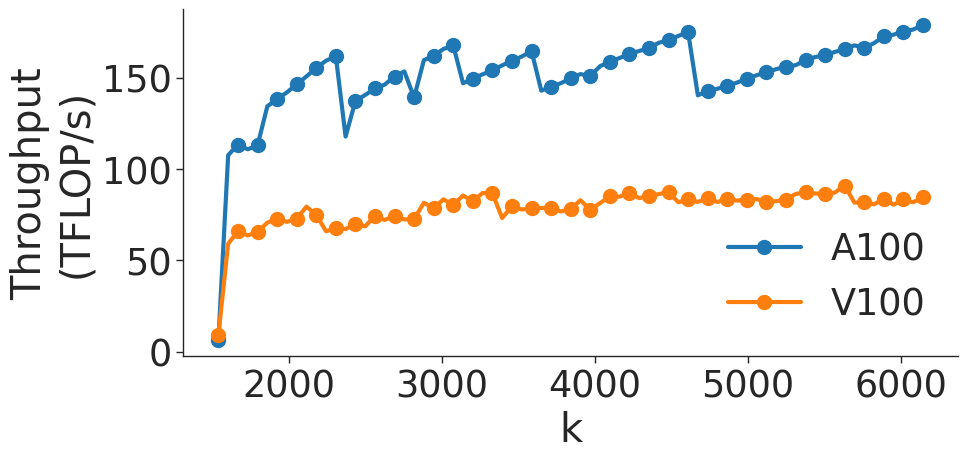}
    }
    \caption{
       Throughput (in teraFLOP/s) for matrix multiplication computations
       of various sizes.
    }
    \label{fig:matrix_multiplication_performance}
\end{figure}

Figure~\ref{fig:matrix_multiplication_performance} shows the throughput
(in teraFLOP/s) of matrix multiplication computations of various sizes
on two types of NVIDIA GPUs. As the GEMM size increases, the operation becomes more computationally intensive and uses memory more efficiently (GEMMs are memory-bound for small matrices). As shown in Figure~\ref{fig:matrix_multiplication_performance}a, throughput of the GEMM kernel increases with matrix size as the kernel becomes compute-bound. However, wave quantization inefficiencies reduce the throughput when the GEMM size crosses certain thresholds. The effects of wave quantization can be seen clearly in Figure \ref{fig:matrix_multiplication_performance}b. Additionally, when the size of the GEMM is sufficiently large, PyTorch may automatically choose a tile size that decreases quantization effects. In Figure~\ref{fig:matrix_multiplication_performance}c, the effects of wave quantization are lessened, as PyTorch is able to better balance the improvements from GEMM parallelization and inefficiencies from wave quantization to improve throughput. 

\begin{figure*}[htbp]
    \centering
    \subfloat[$(b, m, m) \times (b, m, m)$ BMM on V100 GPU.]{
        \includegraphics[width=0.45\linewidth]{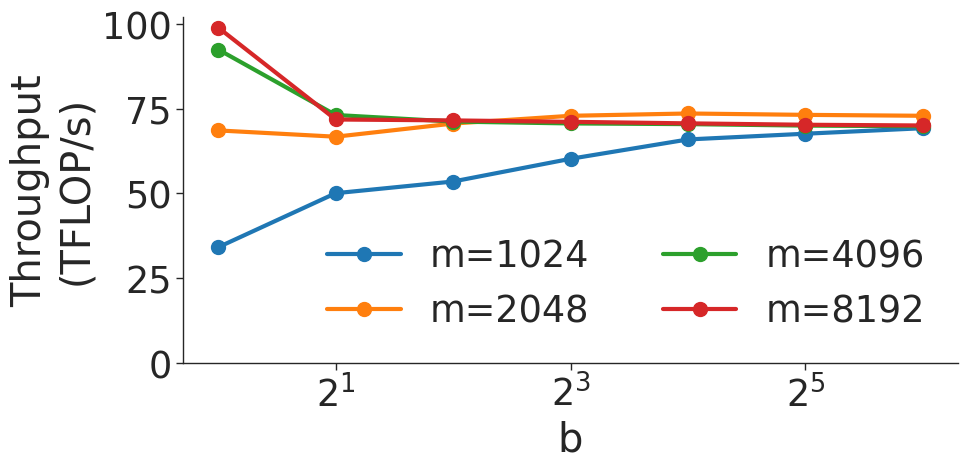}
    }
    \subfloat[$(b, m, m) \times (b, m, m)$ BMM on A100 GPU.]{
        \includegraphics[width=0.45\linewidth]{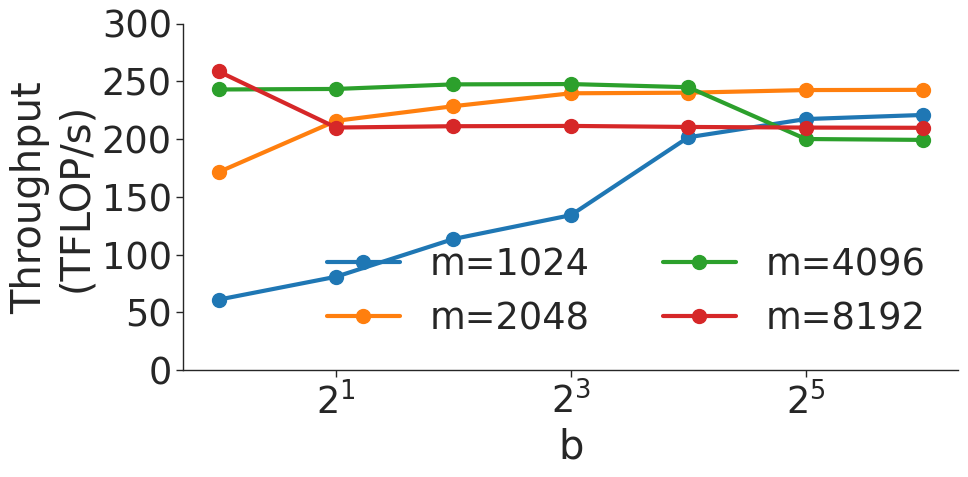}
    }\\
    \subfloat[$(b, m, 4096) \times (b, 4096, m)$ BMM on V100 GPU.]{
        \includegraphics[width=0.45\linewidth]{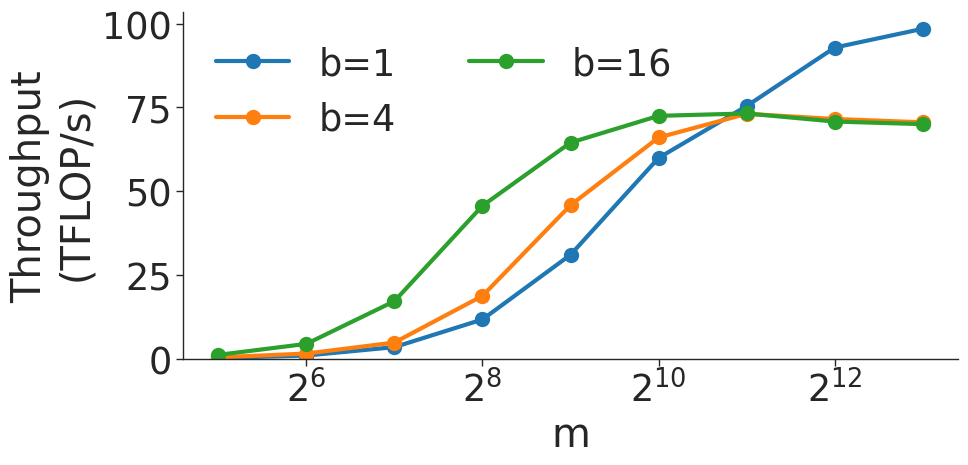}
    }
    \subfloat[$(b, m, 4096) \times (b, 4096, m)$ BMM on A100 GPU.]{
        \includegraphics[width=0.45\linewidth]{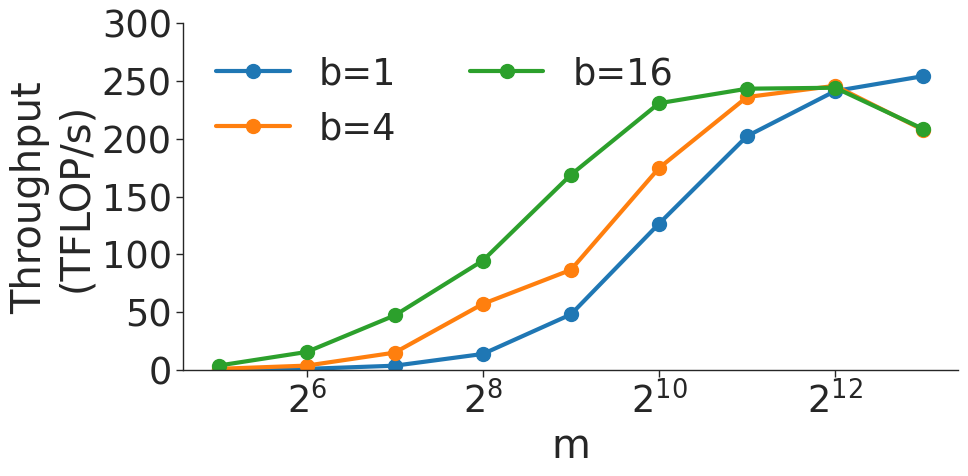}
    }
    \caption{
       Throughput (in teraFLOP/s) for batched matrix multiplication (BMM)
       computations with various dimensions.
    }
    \label{fig:batched_matrix_multiplication_performance}
\end{figure*}

Figure~\ref{fig:batched_matrix_multiplication_performance} shows the
throughput (in teraFLOP/s) of batched matrix multiplication (BMMs) computations
of various sizes. Since BMMs are composed of GEMMs, the same wave quantization effects would apply (though they do not for these BMM sizes and on these GPU architectures). BMM throughput also increases as the size of the BMM and arithmetic intensity increases.
\section{Transformer Results}
\label{sec:results-dl}

\subsection{Transformer as a Series of GEMMs}

The settings of the various hyperparameters in the transformer layer controlling its shape all have
an impact on its observed end-to-end throughput.
Some of these hyperparameters can affect performance in subtle ways. The purpose of this section is to map GEMM performance to transformer throughput, use these mappings to explain the performance effects of relevant hyperparameters, and finally to boil down these effects into a series of practical takeaways.

\begin{figure*}[htbp]
    \centering
    \captionsetup{width=.9\linewidth}
    \subfloat[Attention key-query score GEMM throughput for 32 attention heads.]{
        \includegraphics[width=0.45\linewidth]{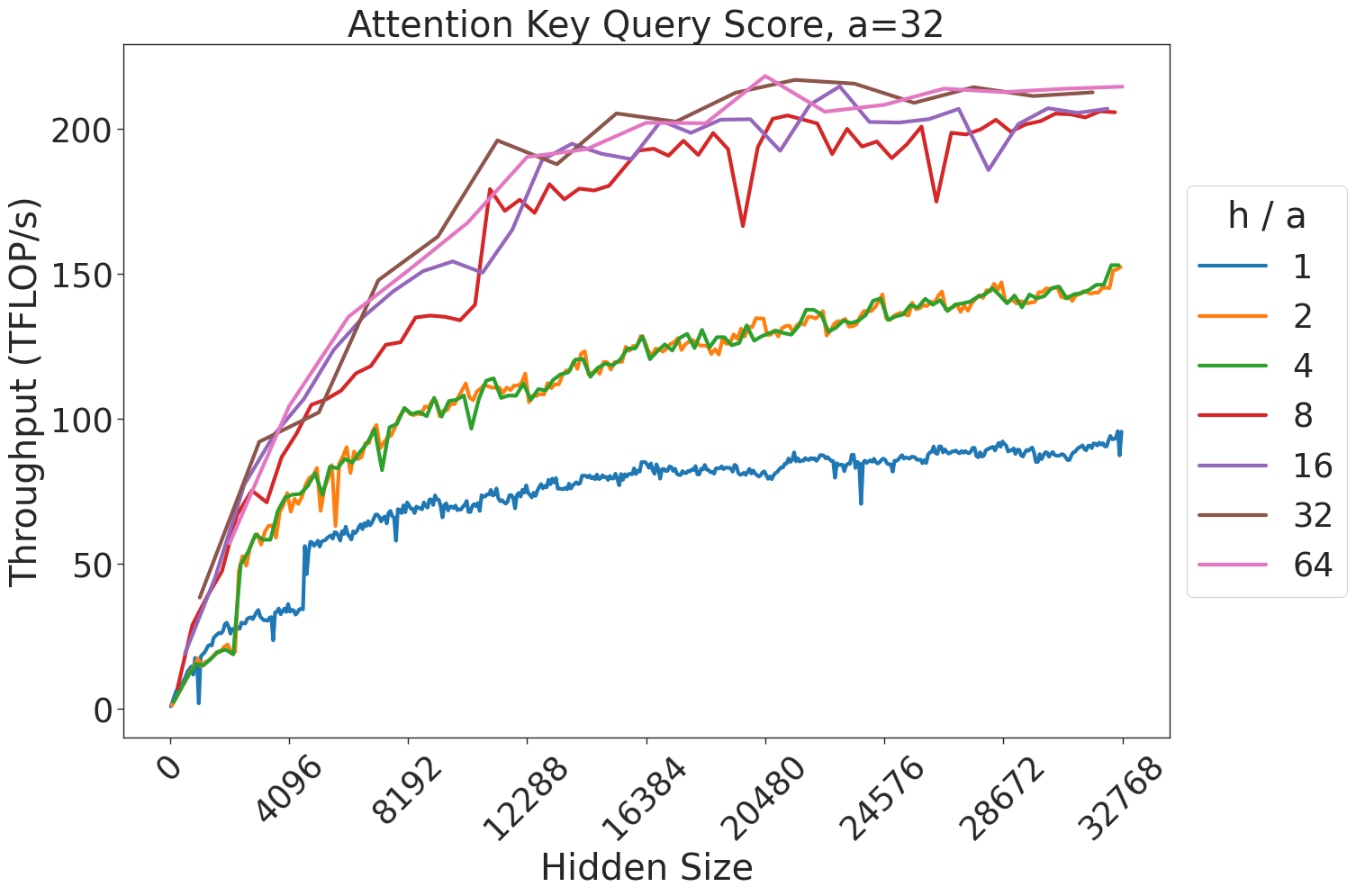}
    \label{subfig:attention-kq}
    }
    \subfloat[Attention over value GEMM throughput for 32 attention heads.]{
        \includegraphics[width=0.45\linewidth]{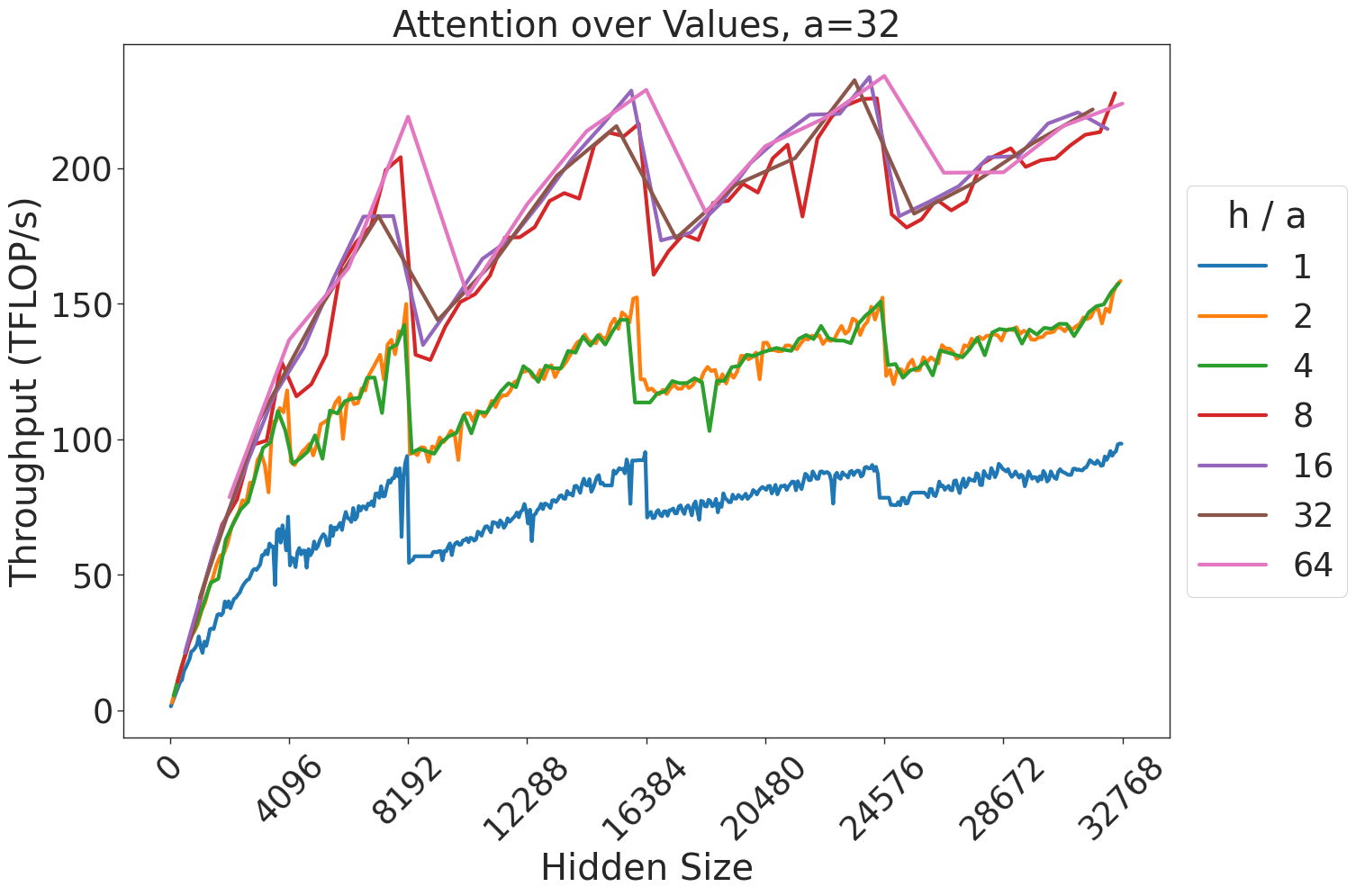}
    \label{subfig:attention-val}
    }
    \label{fig:attention-kq-val-results}
    \caption{
       Attention GEMM performance on A100 GPUs. Each plot is a single series (i.e. if we didn't split, there would be three regions with spikes), but split by the largest power of two that divides $h/a$ to demonstrate that more powers of two leads to better performance up to $h/a=64$.
    }
\end{figure*}

\begin{figure}[htbp]
\centering
    \includegraphics[width=.8\linewidth]{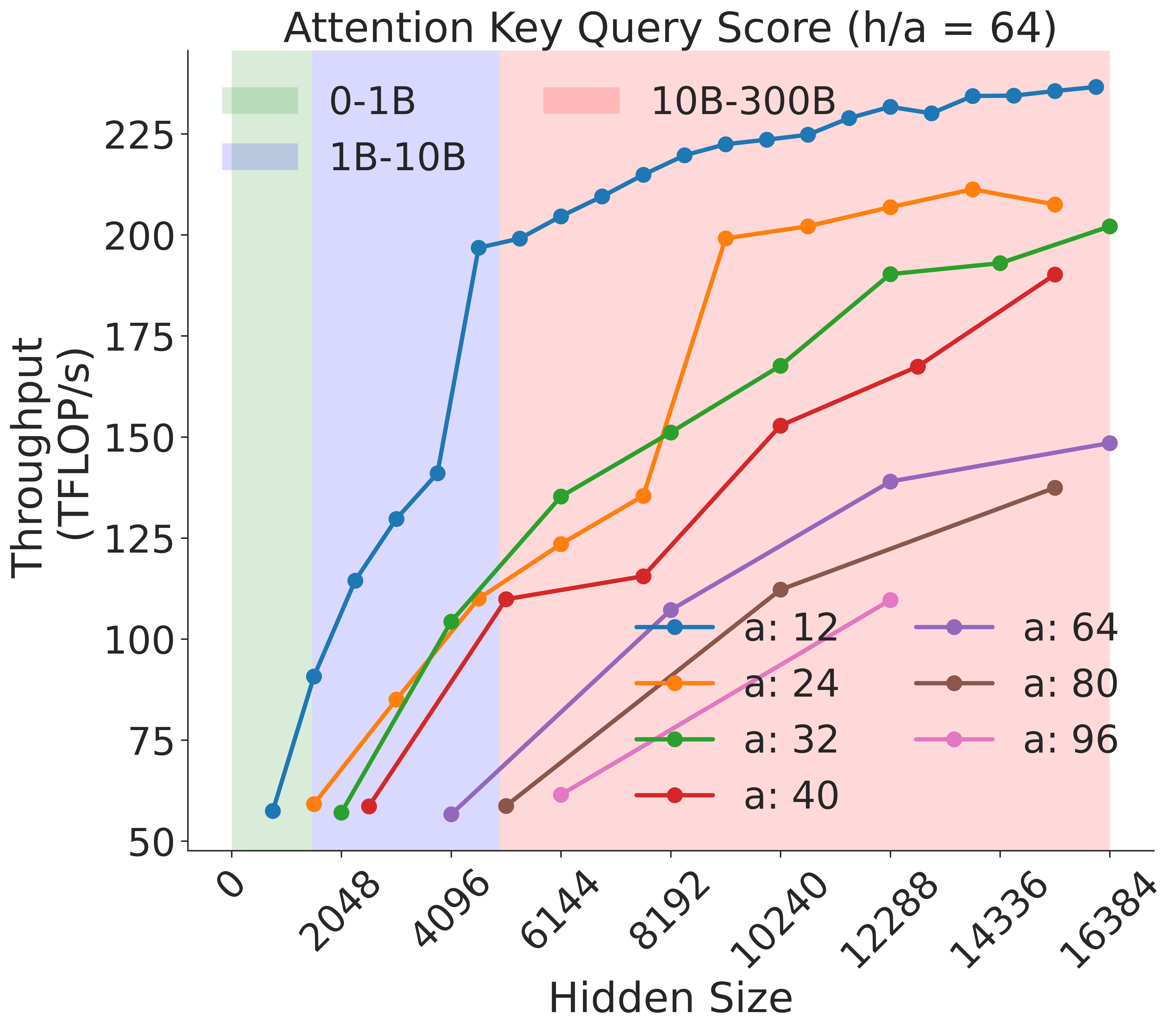}
    \caption{Attention key-query score GEMM throughput assuming fixed ratio of $\frac{h}{a}=64$ on A100 GPU}
    \label{fig:attention-kq-ha64-16384}
\end{figure}

\begin{figure}[htbp]
\centering
    \includegraphics[width=.8\linewidth]{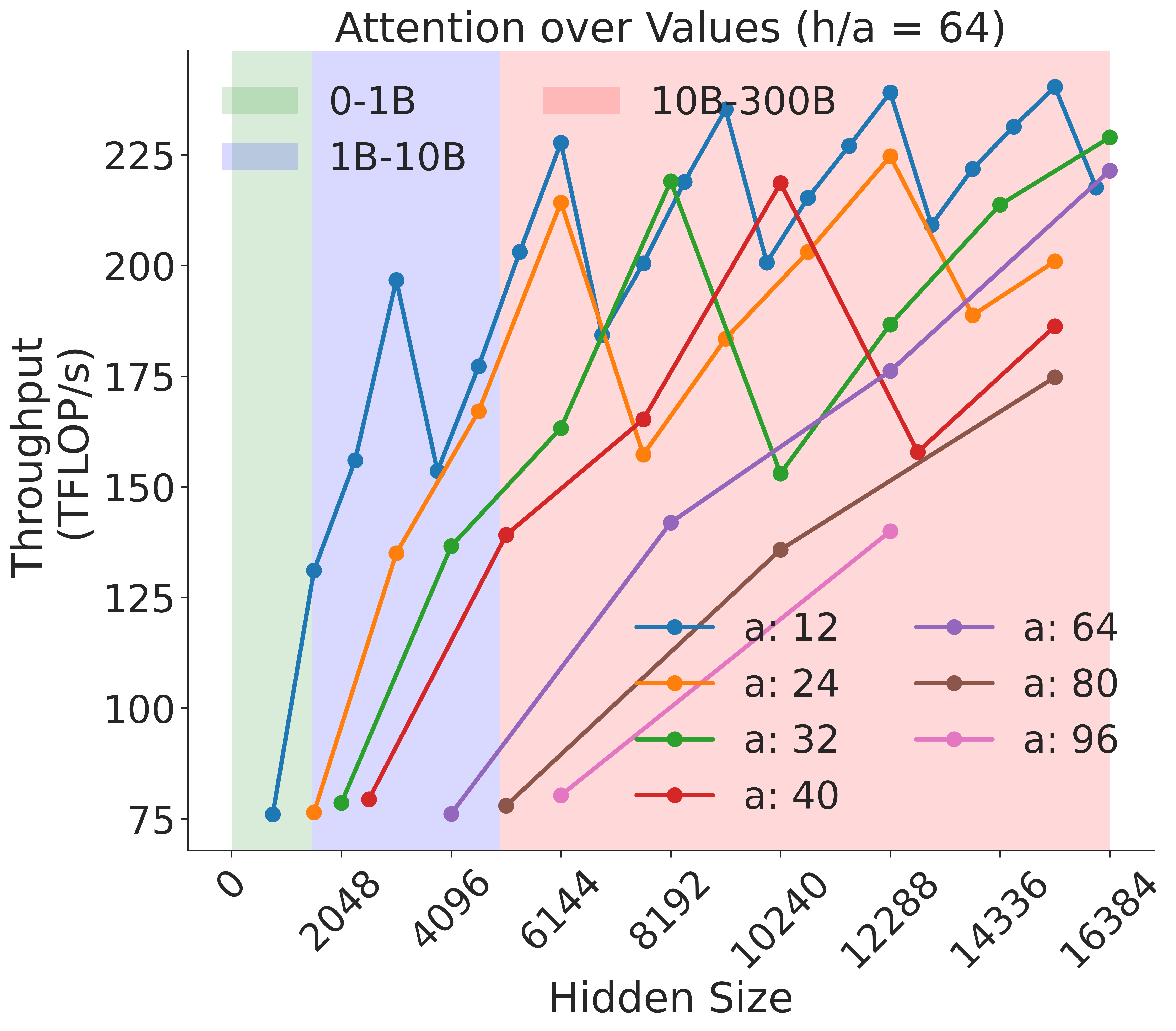}
    \caption{Attention over value GEMM throughput assuming fixed ratio of $\frac{h}{a}=64$ on A100 GPU.}
    \label{fig:attention-val-ha64-16384}
\end{figure}

\begin{figure}[htbp]
    \centering
    \subfloat[MLP h to 4h Block]{
    \label{fig:mlp_h_to_4h}
        \includegraphics[width=.8\linewidth]{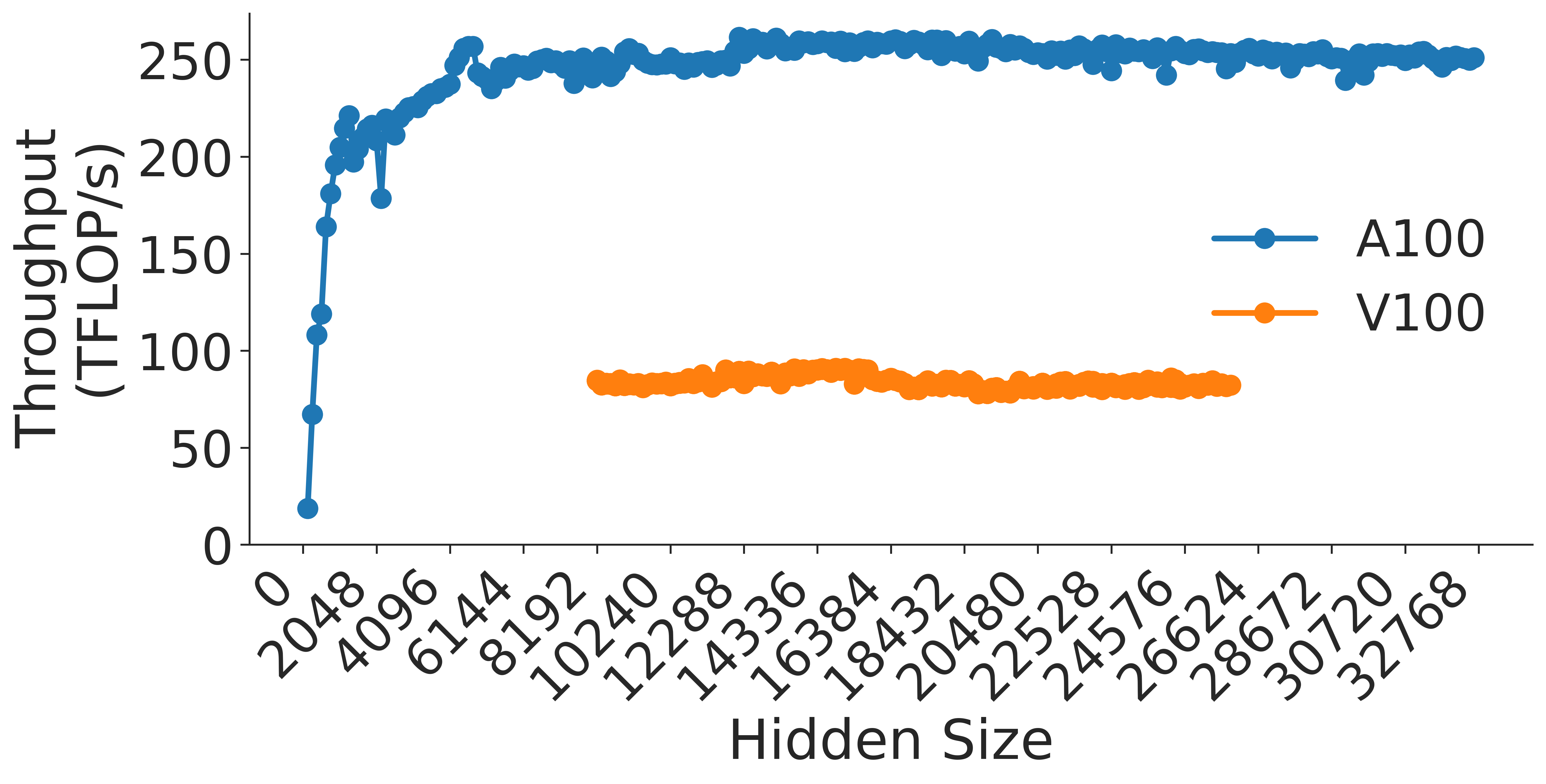}
    }\\
    \subfloat[MLP 4h to h Block]{
    \label{fig:mlp_4h_to_h}
    \includegraphics[width=.8\linewidth]{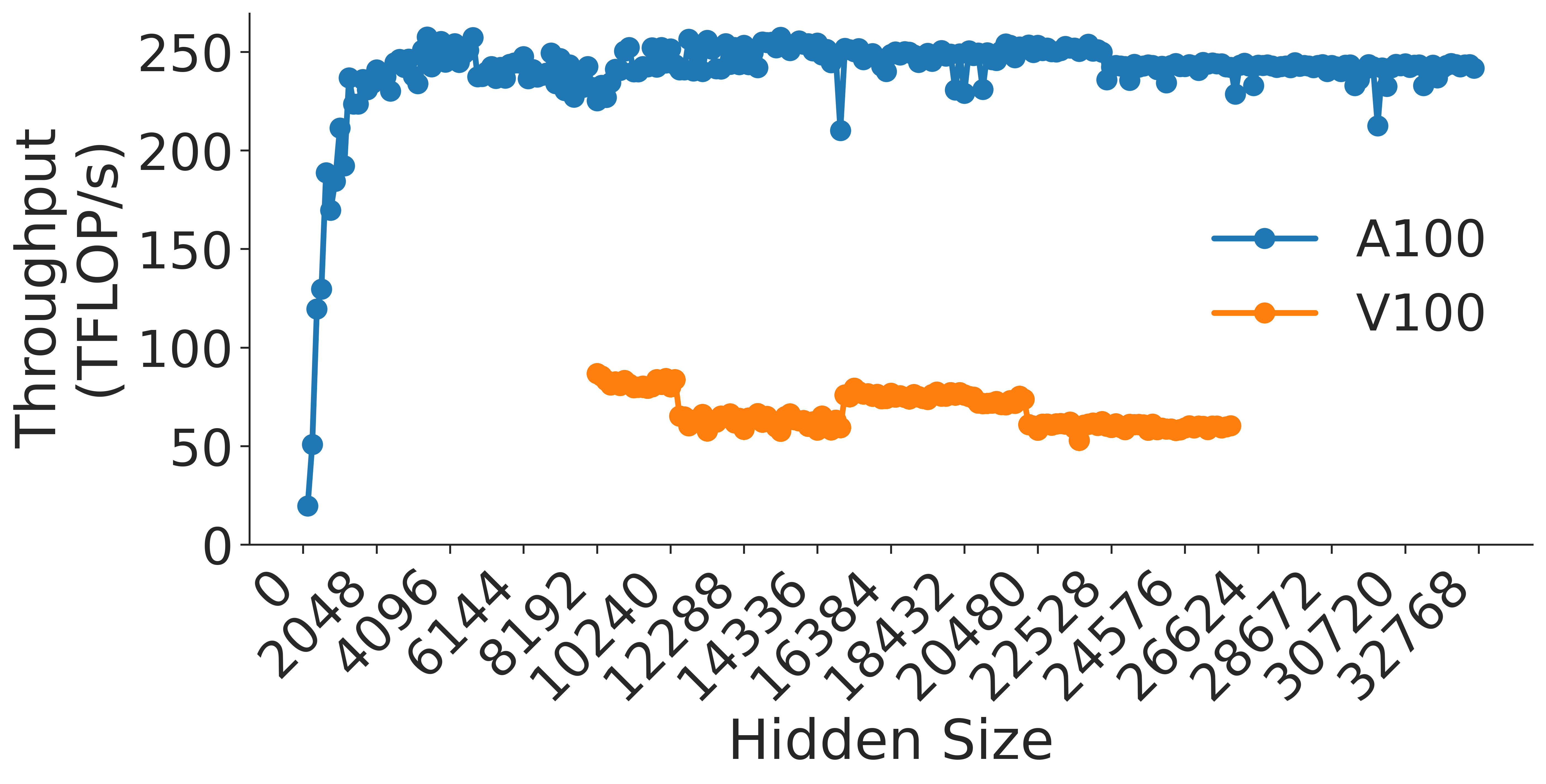}
    }\\
    \caption{
       Throughput (in teraFLOP/s) for multilayer perceptrons (MLP) for each transformer layer as a function of hidden dimension for $a=128$.
    }
    \label{fig:mlp-perf}
\end{figure}

For example, let us consider the attention block on an A100 GPU. The number of attention
heads affects the number of independent matrix multiplications in the
BMM, as well as the size of each matrix multiplication computation.
Figure~\ref{fig:batched_matrix_multiplication_performance} shows the effect of the
number of attention heads and the hidden size on the throughput of the BMM used in
attention key-query score computation and attention over value computation.
NVIDIA Tensor Cores are more efficient when the dimensions of the matrices $m$, $n$, and $k$  are multiples of 128 bytes for A100 GPUs. Therefore, efficiency is maximized when matrix sizes are multiples of 64 FP16 elements. If this cannot be achieved, sizes that are multiples of larger powers of 2 perform better, as shown in Figures ~\ref{subfig:attention-kq} and ~\ref{subfig:attention-val}, where the matrix dimension of interest is of size $h/a$. Figures \ref{fig:attention-kq-ha64-16384} and \ref{fig:attention-val-ha64-16384} show how decreasing the number of attention heads for any given hidden size results in more efficient GEMMs.  Because a decrease in $a$ is an increase in $h/a$ and these two GEMMs are memory bound, an increase in component matrices size creates much more efficient GEMMs.
Figure \ref{fig:attention-val-ha64-16384} also clearly shows the effects of wave quantization in the peaks and valleys within any given line. Since each line moves in steps of $64h/a$, the BMMs corresponding to each line grow at different rates. This causes the period of the wave quantization effect to appear different for each $a$ value.

\begin{figure}[htbp]
\centering
    \includegraphics[width=.95\linewidth,trim=4 6 4 6,clip]{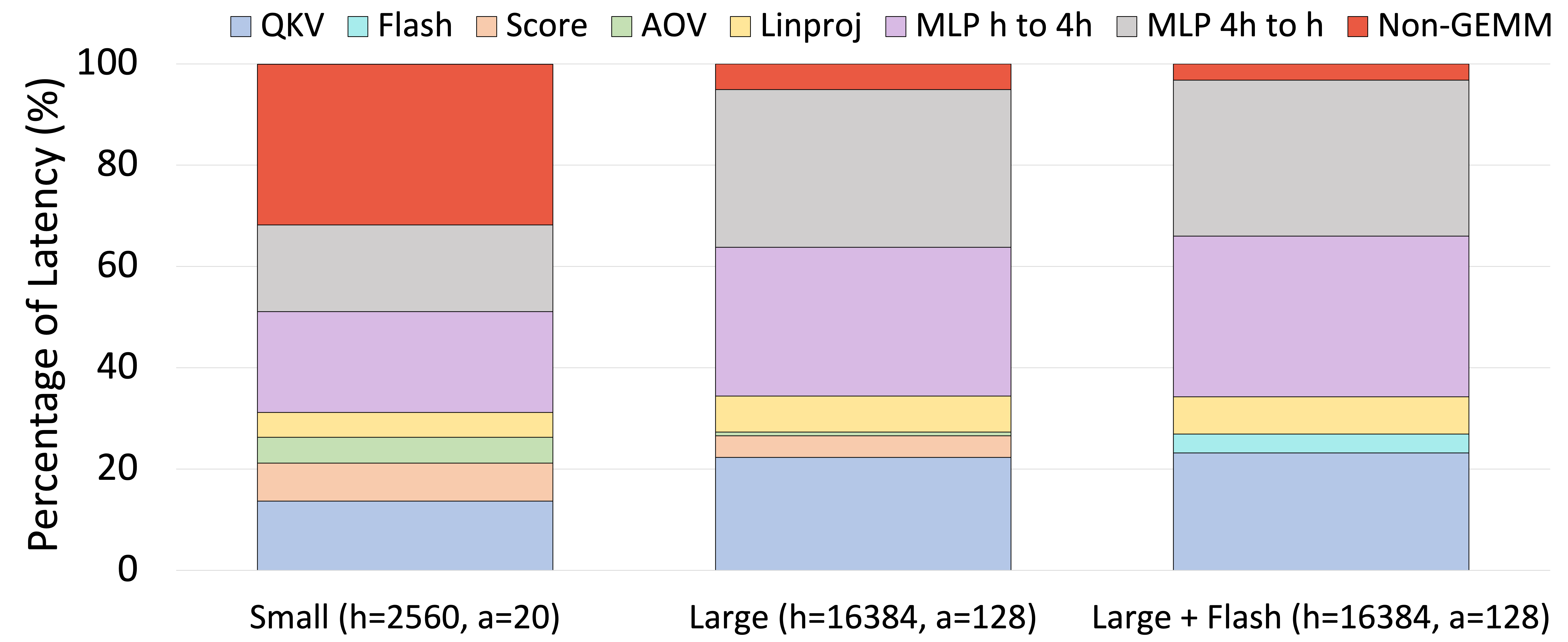}
    \caption{The proportion of latency of each GEMM module for one layer of various model sizes.}
    \label{fig:prop_individual}
\end{figure}

Figure \ref{fig:prop_individual} shows the proportion of latency spent in each transformer GEMM; consequently, it also shows the most relevant GEMMs to optimize in the transformer module. As the size of the model grows, it is even more important to optimize GEMM operations. For the largest models, the $QKV$ transformation in the attention block along with the MLP block are the most prevalent GEMMs. Therefore, the overall latency of the model would benefit most from optimizing these kernels. Attention over value (AOV) computation is the smallest GEMM computation in large transformer models; however, optimizing attention key-query score computation will have similar benefits to attention over value computation, so both can be optimized at the same time. 



\subsection{Analysis}

To recap, we have the following requirements to efficiently run GEMMs on NVIDIA GPUs:
\begin{itemize}
    \item \textbf{Tensor Core Requirement:} Ensure the inner and outer dimension of the GEMM is divisible by 128 bytes (64 FP16 elements).
    \item \textbf{Tile Quantization:} To use the most efficient tile size ensure that the output matrix is divisible into $128 \times 256$ blocks.
    \item \textbf{Wave Quantization:} Ensure that the number of blocks that the output matrix is divided into is divisible by the number of streaming multiprocessors (80 for V100s, 108 for A100s, and 144 for H100s).
\end{itemize}

While tile quantization is relevant to GEMM performance, tile quantization is hard to observe by the user. If the GEMM does not divide evenly into the tile size, a tile without a full compute load will execute. However, this tile will execute concurrently with other tiles in the same wave. In effect, the kernel will run with the same latency as a kernel with a larger problem size. 

Wave quantization is more easily observable. There will be no wave quantization inefficiency when a matrix of size $(X,Y)$ satisfies the following constraints on its size (assuming a tile size of $t_1 \times t_2$):

$$
\left\lceil{\frac{X}{t_1}}\right\rceil \cdot \left\lceil{\frac{Y}{t_2}}\right\rceil \equiv 0 \text{ or }
\left\lceil{\frac{X}{t_2}}\right\rceil \cdot \left\lceil{\frac{Y}{t_1}}\right\rceil \equiv 0 \pmod{\#SMs}
$$

Assuming a tile size of $128 \times 256$ which is the most efficient, there is not a transformer configuration with GEMMs that fill tensor core requirements without wave quantization inefficiency. Further, PyTorch's linear algebra backend can use different tile sizes for each GEMM. Therefore, PyTorch is unable to efficiently overcome the effects of wave quantization.

Therefore to ensure the best performance from transformer models, ensure:
\begin{itemize}
    \item The vocabulary size should be divisible by $64$.
    \item The microbatch size $b$ should be as large as possible~\parencite{nado2021large}.
    \item $b \cdot s$, $\frac{h}{a}$, and $\frac{h}{t}$ should be divisible by a power of two, though there is no further benefit to going beyond $64$.
    \item $(b \cdot a)/t$ should be an integer.
    \item $t$ should be as small as possible~\parencite{narayanan2021efficient}.
\end{itemize}

Importantly, the microbatch size $b$ does not itself need to be divisible by a large power of 2 since the sequence length $s$ is a large power of two.

Whether it is optimal to train using pipeline parallelism depends on additional details of the computing set-up, most notable the speed and bandwidth of internode connections. We note that this is further evidence for our thesis that model dimensions should be chosen with hardware details in mind, but leave an analysis of this phenomenon to future work. In all cases it is optimal for the number of layers to be divisible by the number of pipeline parallel stages.

Using these recommendations we can achieve a 1.18$\times$ speed-up on a widely used model architecture introduced by \textcite{brown2020language}. GPT-3 2.7B's architecture was copied for many other models including GPT-Neo 2.7B \parencite{black2021gpt}, OPT 2.7B \parencite{zhang2022opt}, RedPajama 3B \parencite{together2023redpajama}, and Pythia 2.8B \parencite{biderman2023pythia}, but possesses an inefficiency. It features $32$ attention heads and a hidden dimension of $2560$, resulting in a head dimension of $h/a = 2560/32 = 80$ which is not a multiple of $64$. This can be addressed either by increasing the side of the hidden dimension to $4096$ or by decreasing the number of heads to $20$. Increasing the hidden dimension to $4096$ doubles the number of parameters to $6.7$ billion, so instead we decrease the number of heads. These results are shown in Figure~\ref{fig:gpt_throughput_motivation}.


To raise $h/a$, the easiest solution is to decrease $a$, but decreasing $a$ may lead to a drop in model accuracy. Fortunately, as shown in Figure \ref{fig:prop_individual}, only a small portion of the latency of large models is the attention score computation and attention over value computation GEMMs, so an increase in the latency of these components will have only a small effect on the end-to-end model performance. Therefore, we recommend either using FlashAttention v2 (see Section~\ref{sec:flash}) for small models to mitigate these effects, or increasing $h$ as much as possible to reach the saturation point shown in Figures ~\ref{fig:mlp_h_to_4h} and ~\ref{fig:mlp_4h_to_h}.

\subsection{Architectural Modifications}

While decoder-only architectures are largely standardized and follow the GPT-2 architecture \parencite{radford2019language} described in the previous section, there are some architectural modifications that are popular in recent work. Here we briefly describe them and how they affect our overall discussion.

\subsubsection{Parallel Layers} Parallel attention and MLP layers were introduced by \textcite{gpt-j}. Instead of computing attention and MLPs sequentially  ($y = x + \MLP(\Ln(x + \Attn(\Ln(x))))$), the transformer block is formulated as: $$y = x + \MLP(\Ln(x)) + \Attn(\Ln(x)).$$ While this computation is represented as being in parallel, in practice the two branches are not computed simultaneously. Instead, a speed-up is achieved by fusing the MLP and Attention blocks into a single kernel. We recommend using parallel attention as the default best practice, though it does not impact our analysis at all.

\subsubsection{Alternative Positional Embeddings} While the original positional embeddings used in transformers are pointwise operations \parencite{vaswani2017attention}, today other approaches such as Rotary \parencite{su2021roformer} and ALiBi \parencite{press2021train} embeddings are more popular. While point-wise operations are slightly faster than the GEMM necessary for Rotary and ALiBi embeddings, the improved model accuracy that Rotary or ALiBi embeddings bring are generally considered well worth it. Recently, custom kernels for rotary embeddings have been introduced, further reducing their costs. We recommend using rotary or ALiBi embeddings as best practice. Using these embeddings again does not impact our analysis.

\subsubsection{FlashAttention}
\label{sec:flash}
\begin{figure}[htbp]
\centering
    \includegraphics[width=.9\linewidth]{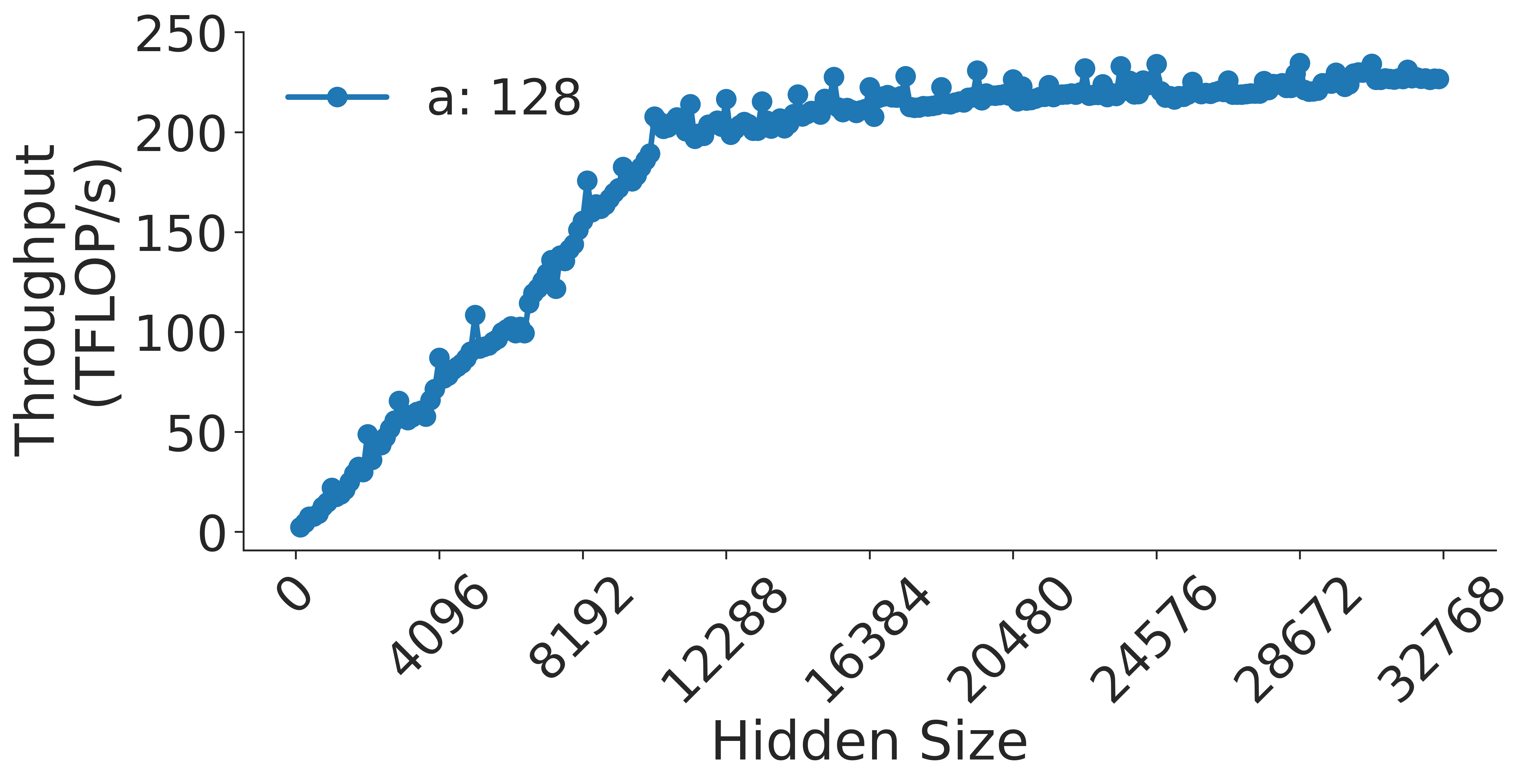}
        \caption{Sweep over hidden dimension for FlashAttention (v2)~\cite{dao2023flashattention} on NVIDIA A100 GPU.}
    \label{fig:flash-hdim-sweep}
\end{figure}

FlashAttention \cite{dao2022flashattention} and FlashAttention 2 \cite{dao2023flashattention} are novel attention kernels that are widely popular for training large language models. In order to see its impact on the attention calculation sizing, we set $a=128$ and sweep over the hidden dimension in Figure~\ref{fig:flash-hdim-sweep}. We find that FlashAttention follows a roofline model, which simplifies our attention takeaways to only require that $h$ be as large as possible; the takeaways for MLPs remain unchanged.

\subsubsection{SwiGLU and $8h/3$ MLPs}

Models such as PaLM, LLaMA, and Mistral use the SwiGLU \textcite{shazeer2020glu} activation function in place of the more common GLU activation function. While the choice of activation function is generally irrelevant to our analysis, this activation function has an extra parameter compared to other commonly used options. Consequently, its common to adjust the projection factor for the MLP block from $dim_{MLP} = 4\cdot dim_{Attn}$ to $dim_{MLP} = \frac{8}{3}\cdot dim_{Attn}$ to preserve the ratio of the total number of parameters in the attention and MLP blocks. This change has substantial implications for our analysis, which we discuss it detail in Section \ref{sec:swiglu}.
\section{Case Studies}
\label{sec:case-studies}

Finally, we present a series of case studies illustrating how we use the principles described in this paper in practice. These demonstrate real-world challenges we have encountered in training large language models with tens of billions of parameters.

\subsection{6-GPU Nodes}

While the most common data-center scale computing set-up is to have 8 GPUs per node, some machines such as Oak Ridge National Lab's Summit supercomputer feature six. This presents a multi-layer challenge to training language models when the tensor parallel degree is equal to the number of GPUs on a single node, which is commonly the most efficient 3D-parallelism scheme~\cite{narayanan2021efficient}. This often causes $h/t$ to no longer have a factor of some power of two, which greatly improves performance as we demonstrated above. Therefore:
\begin{enumerate}
    \item Model architectures common on 8-GPU nodes may not be possible on 6-GPU nodes.
    \item Even when they are possible, model architectures common on 8-GPU nodes may not be efficient on 6-GPU nodes.
    \item If concessions are made to ameliorate \#1 and \#2, they may cause problems in deployment if downstream users wish to use the model designed for a 6-GPU node on a 2-GPU, 4-GPU, or 8-GPU node.
\end{enumerate}

Several large transformers on Summit have been trained, such as the INCITE RedPajama 3B and 7B~\cite{together2023redpajama}, and such model designers must make a choice. Does one choose the most efficient hyperparameters for pretraining only (which would involve a tensor-parallel degree of 6 and therefore a hidden dimension divisible by 6 and 64), or should the pretraining team choose a set of hyperparameters that are more amenable to the node architectures commonly used for finetuning or inference?

\subsection{SwiGLU Activation Functions}\label{sec:swiglu}

Recently the SwiGLU activation function has become popular for training language models. The SwiGLU function contains an additional learned matrix in its activation function, so that now the MLP block contains 3 matrices instead of the original 2. To preserve the total number of parameters in the MLP block the paper that introduces SwiGLU proposes to use $d_{ff}=\frac{8}{3}h$ instead of the typical $d_{ff}=4h$. 

If you followed the recommendations in this paper for finding the value of $h$ that would lead to the best $matmul$ performance, you will realize that $\frac{8}{3}h$ is likely to result in a much slower MLP block, because $\frac{1}{3}$ will break all the alignments.

In order to overcome this problem one only needs to realize that the $\frac{8}{3}$ coefficient is only a suggestion and thus it's possible to find other coefficients that would lead to better-shaped MLP matrices. In fact if you look at the publicly available LLama-2 models, its 7B variant uses $\frac{11008}{4096}=2.6875$ as a coefficient, which is quite close to $\frac{8}{3}=2.667$, and its 70B variant uses a much larger $\frac{28672}{8192}=3.5$ coefficient. Here the 70B variant ended up with an MLP block that contains significantly more parameters than a typical transformer block that doesn't use SwiGLU.

Now that we know the recommended coefficient isn't exact and since a good $h$ has already been chosen, one can now search for a good nearby number that still leads to high-performance GEMMs in the MLP. 
Running a brute-force search reveals that Llama-2-7B's intermediate size is indeed one of the best performing sizes in its range.

\subsection{Inference}

\begin{figure}[htbp]
\centering
    \includegraphics[width=.95\linewidth]{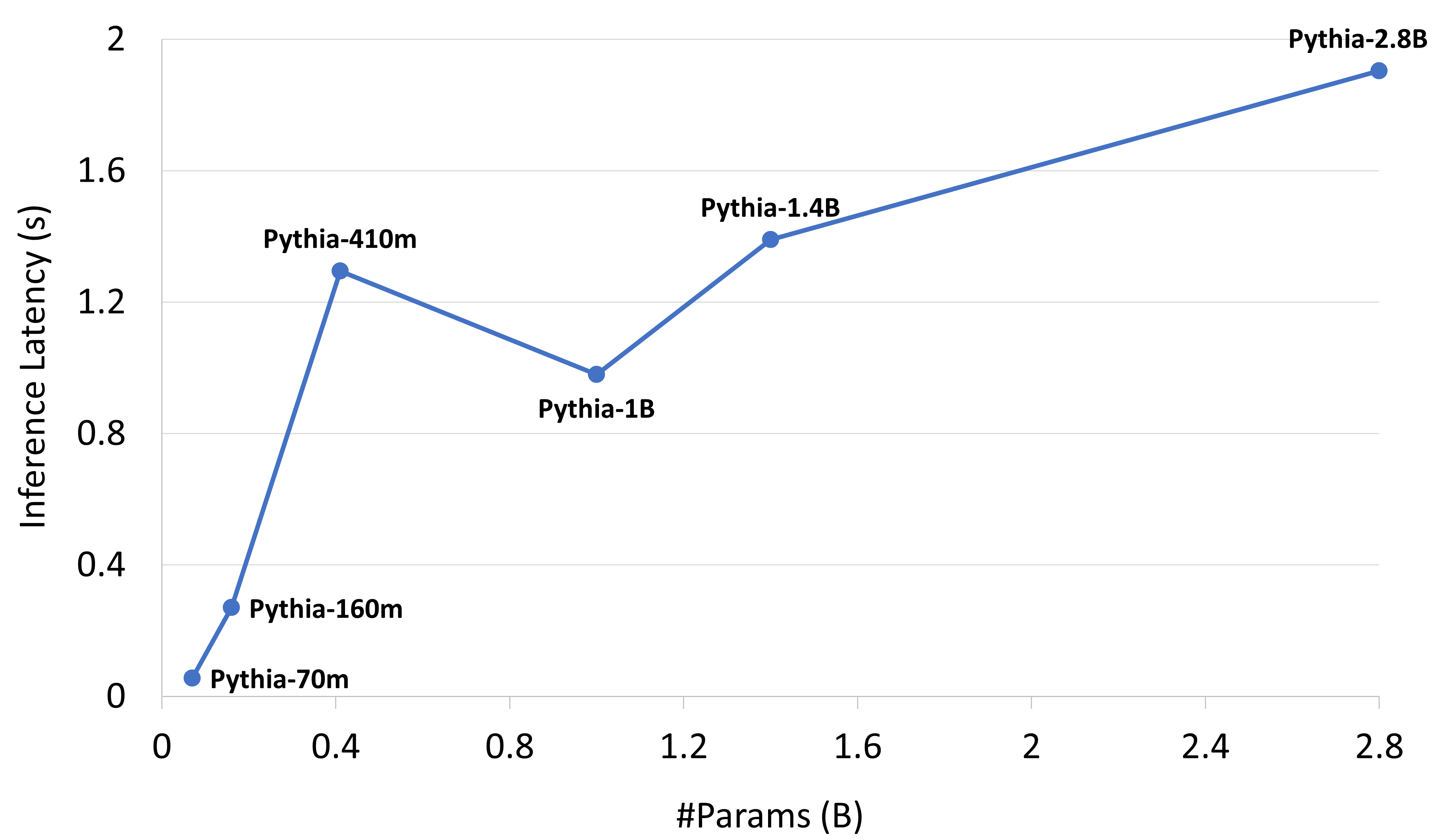}
    \caption{Inference latency of Pythia suite using DeepSpeed-MII \cite{deepspeed-mii}. Pythia-410M / Pythia-1B are off-trend due to their sizing.}
    \label{fig:inference}
\end{figure}

In order to demonstrate that 1) models trained efficiently on a given GPU will also infer efficiently on the same GPU, since the underlying forward-pass GEMMs are the same, and 2) our sizing recommendations are kernel-invariant, we have run inference benchmarks using DeepSpeed-MII~\cite{deepspeed-mii} and the Pythia~\cite{biderman2023pythia} suite. We show in Figure~\ref{fig:inference} that Pythia-1B is significantly more efficient at inference time than Pythia-410M due to its fewer attention heads and layers than Pythia-410M, and a larger hidden dimension. Despite these architectural changes, the test loss of Pythia-1B is on-trend with the rest of the suite while having significantly higher training and inference throughput.

\section{Discussion}
\label{sec:discussion}
In the current landscape of AI hardware, Transformer workloads stand out as a pivotal target. They constitute a significant component (e.g., BERT, GPT-3) of the MLCommons benchmarks, capturing the attention of major hardware vendors and data centers. Notably, these benchmarks have been integrated as a crucial metric for procurement \cite{olcf6-rfp} in the upcoming Exascale supercomputer at Oak Ridge Leadership Computing Facility. Our analysis strongly suggests that leveraging representative GEMM kernels holds promise as a reliable performance indicator for Transformer-based workloads. Consequently, these kernels should be embraced as a benchmarking tool for hardware co-design. The advantages stem from several key points:
\begin{enumerate}
    \item The optimizations made at the GEMM level exhibit a demonstrable transferability to various applications, as evidenced in Sec.\ref{sec:results-dl}.
    \item Benchmarking at the kernel level proves to be more cost-effective and time-efficient.
    \item This approach remains model-agnostic, accommodating diverse architectures like GPT-NeoX, Pythia, and OPT, as long as they are based on the Transformer architecture.
\end{enumerate}
This assertion finds partial validation in the observed correlation between MLCommons benchmarks and our findings. To illustrate, consider the performance of BERT benchmarks, which demonstrates a consistent 3:1 ratio between H100- and A100-based systems. Notably, this aligns harmoniously with our observed kernel throughput for the respective hardware configurations (see Sec.\ref{sec:results-gemm}). 
\section{Conclusion}
\label{sec:conclusion}

State-of-the-art deep learning (DL) models are driving breakthroughs in existing fields and paving the way towards new areas of study. However, while the transformer model is at the forefront of this DL explosion, few transformer architectures consider their underlying hardware. We believe that instead of creating new designs to improve efficiency, many practitioners would be better served by slightly modifying their existing architectures to maximally utilize the underlying hardware. Well informed hyperparameter choices improve training and inference throughput throughout a model's lifetime. We demonstrate that minor modifications to the model architecture improve GPU throughput by up to 38.9\% while maintaining accuracy. Since we have explained how to motivate model hyperparameters from a GPU architecture standpoint, this paper can be used to guide future model design while clarifying the relevant first principles necessary to extend such hyperparameter choices to future architectures. 

\section{Acknowledgements}
\noindent We are grateful to Stability AI for providing the compute required for A100 evaluations, Oak Ridge National Lab (ORNL) for providing the compute required for 6-V100 special-case evaluations, and the San Diego Supercomputing Center (SDSC) for providing the compute required for general V100 evalutions.

We thank Horace He and various members of the EleutherAI Discord Server for their feedback.

\printbibliography
\clearpage
\appendix
\section{Misc}
\begin{figure}[htbp]
\centering
    \includegraphics[width=.8\linewidth]{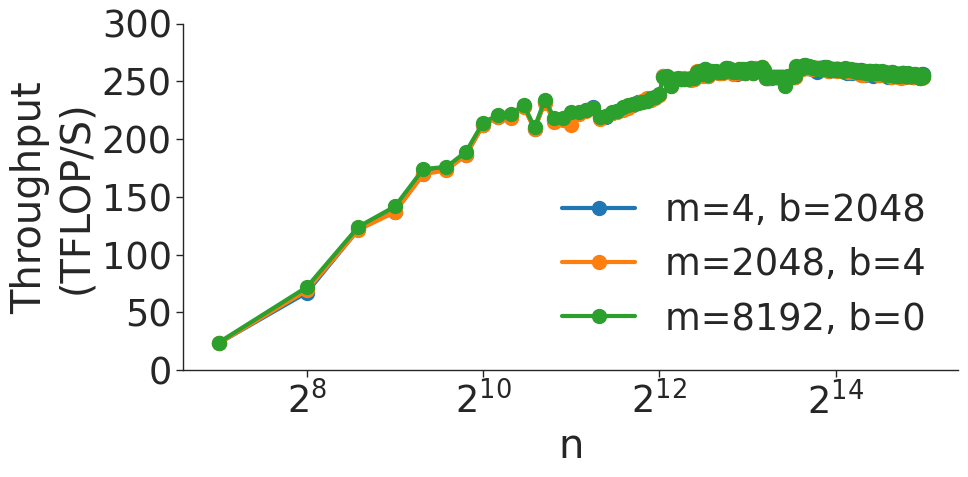}
    \caption{GEMMs with different ordering of dimensions.}
    \vspace{-2ex}
    \label{fig:bmm=gemm}
\end{figure}

When using PyTorch to invoke GEMMs, we use \texttt{torch.nn.functional.linear}. This function accepts 2 tensors as parameters, where one tensor can be 3 dimensional. Figure \ref{fig:bmm=gemm} shows how the ordering of a tensor's dimensions impacts performance. We benchmark GEMMs of size $(2048,4,n)\times(n,3n)$, $(4,2048,n)\times(n,3n)$, and $(8192,n)\times(n,3n)$. This shows that the ordering of the batched dimension does not affect performance. The batched implementation is also the same speed as a 2-dimensional GEMM, so these implementation details do not affect performance. Therefore we can represent GEMMs between 3 and 2 dimensional tensors as GEMMs between two 2-dimensional tensors. 

\begin{figure}[h]
    \centering
    \subfloat[Sweep over vocabulary size]{
    \includegraphics[width=0.8\linewidth]{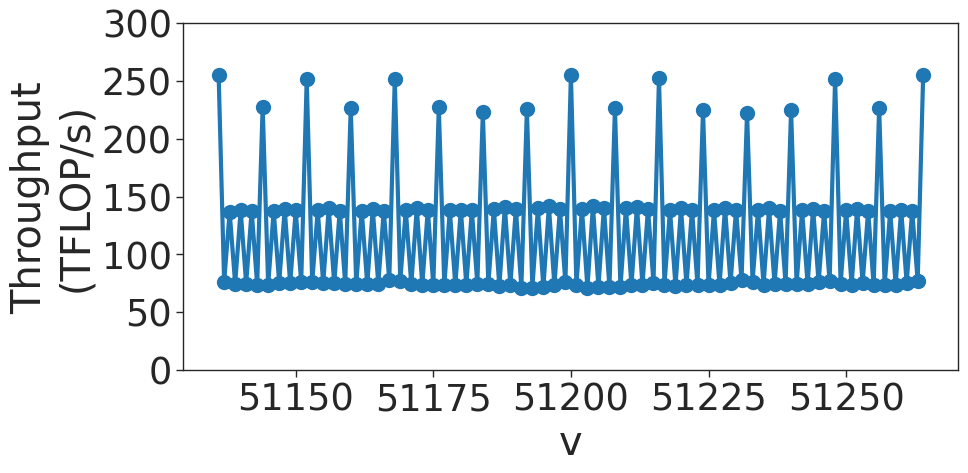}
    \label{fig:vocab_v_sweep}
    }\\
    \subfloat[Zoomed-in sweep over vocabulary size]{
    \includegraphics[width=0.8\linewidth]{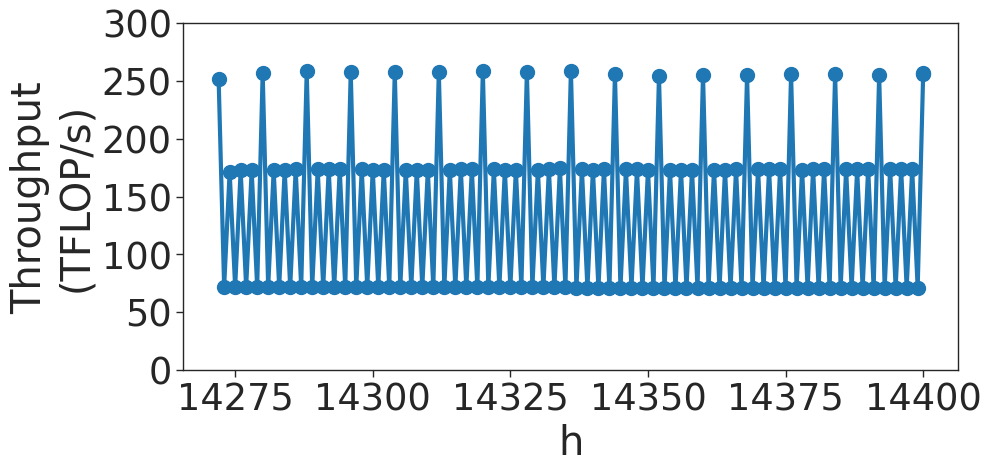}
    \label{fig:vocab_h_sweep}
    }
    \caption{Vocabulary embedding transformation.}
    \label{fig:vocab_sweep}
\end{figure}

A series of benchmarks are shown in Figure \ref{fig:attn_key_value_query_transform} through Figure \ref{fig:vocab_sweep}. These figures show the performance of transformer GEMMs listed in Table \ref{tab:operators}. In each of the figures, throughput for a transformer with 128 attention heads is plotted against hidden size. Performance generally increases with hidden size, as the size of each GEMM is growing. However, in GEMMs where one dimension is of size $h/a$, Attention Score Computation and Attention Over Value, throughput depends on the highest power of 2 that divides $h/a$, as described in secion VI.A.

\begin{figure}[htbp]
\centering
    \includegraphics[width=.8\linewidth]{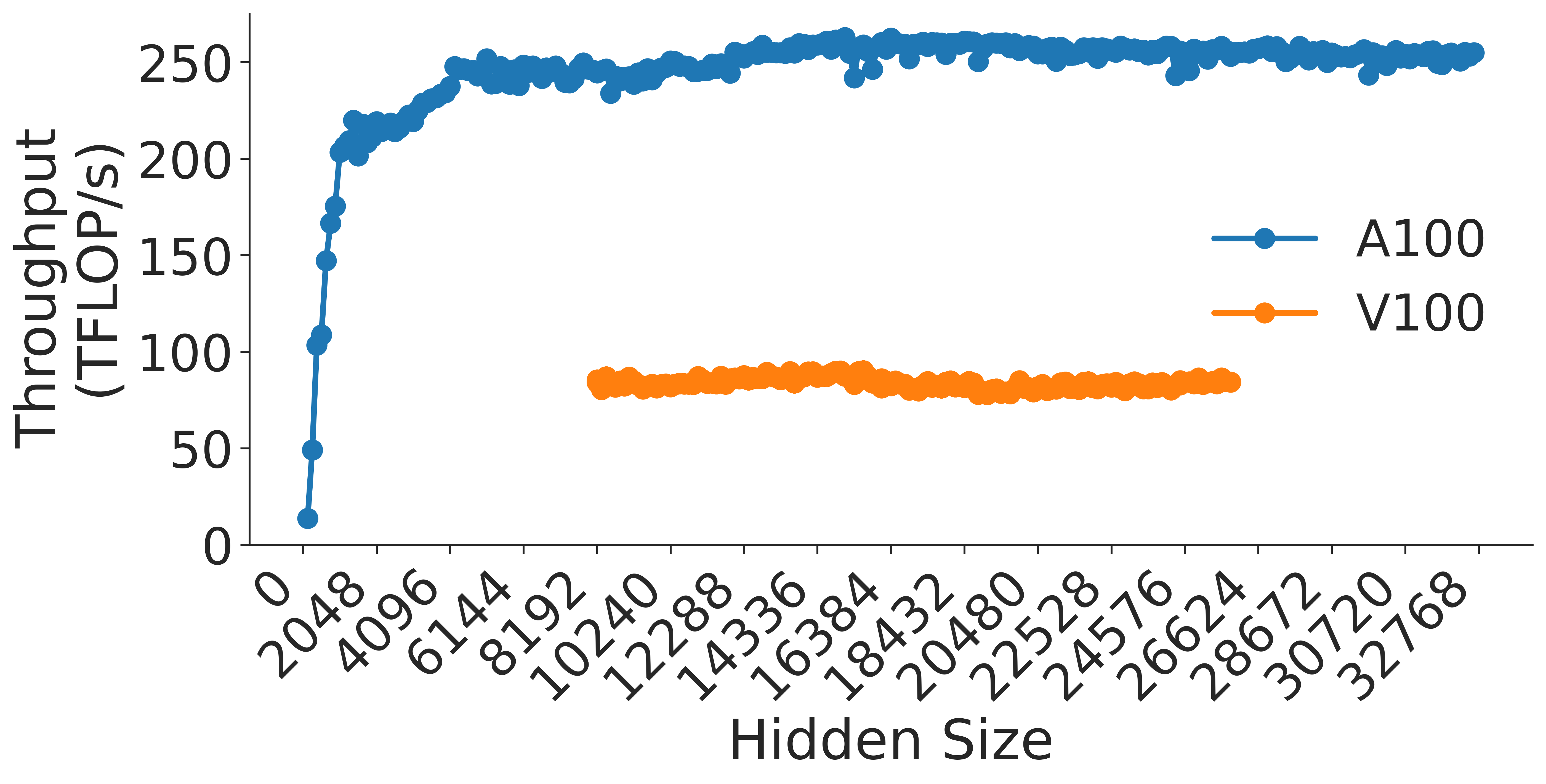}
    \caption{Attention $QKV$ transform.}
    \vspace{-2ex}
    \label{fig:attn_key_value_query_transform}
\end{figure}

\begin{figure}[htbp]
\centering
    \includegraphics[width=.8\linewidth]{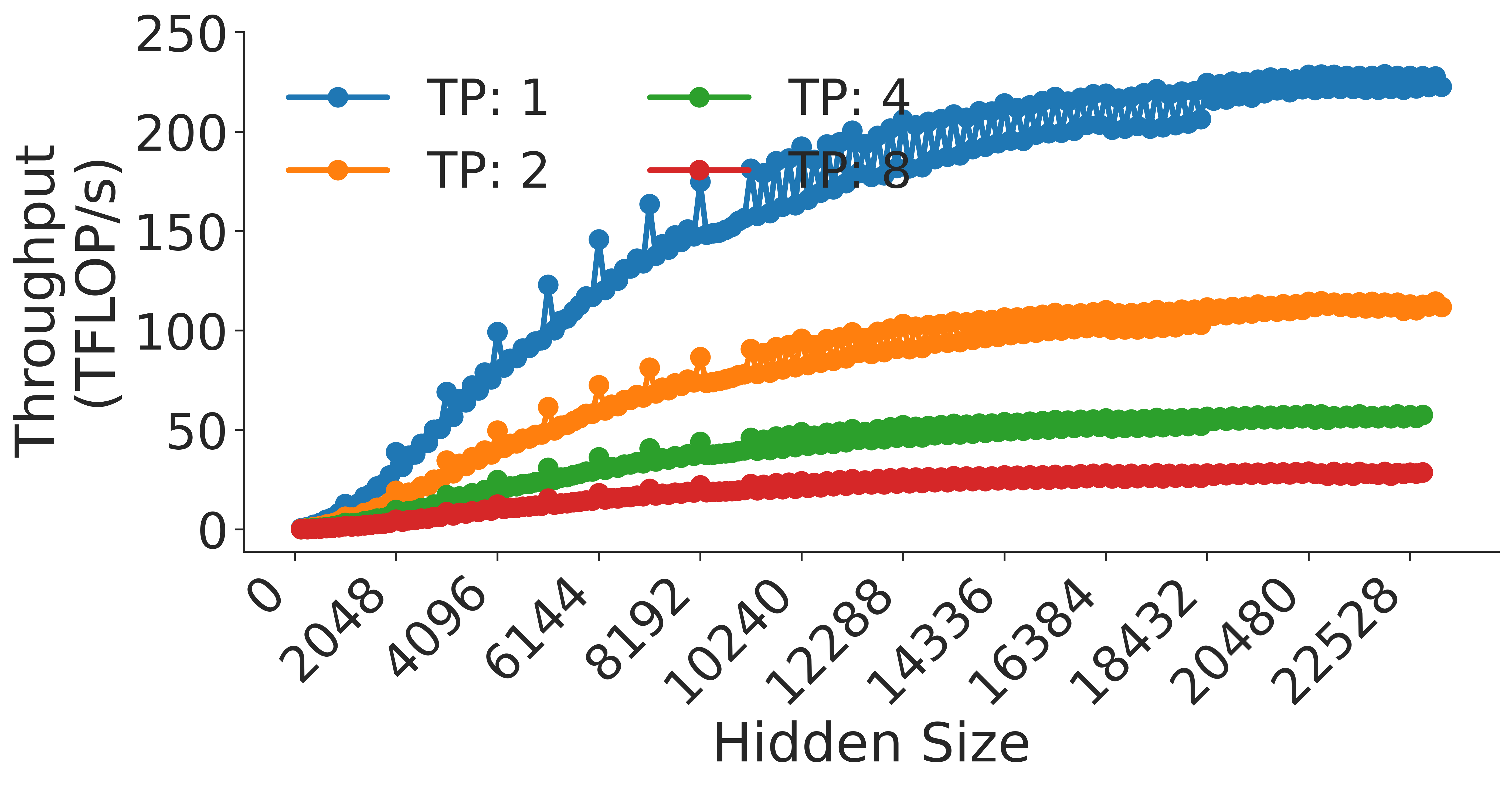}
    \caption{Attention $QKV$ transform with different TP sizes.}
    \vspace{-2ex}
    \label{fig:attn_key_value_query_transform}
\end{figure}

\begin{figure}[htbp]
\centering
    \includegraphics[width=.8\linewidth]{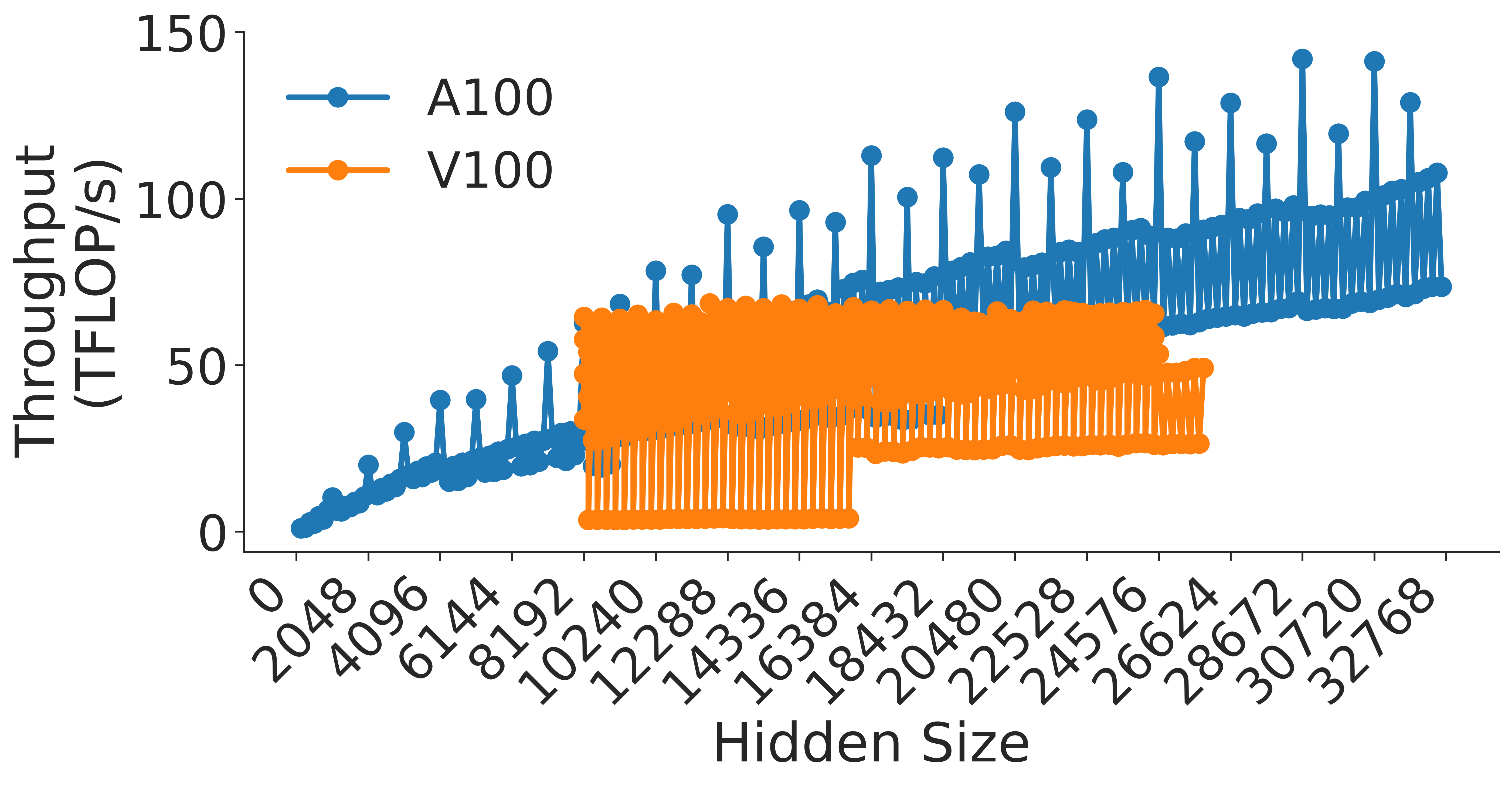}
    \caption{Attention key-query score computation ($KQ^T$).}
    \vspace{-2ex}
    \label{fig:attn_key_query_prob}
\end{figure}

\begin{figure}[htbp]
\centering
    \includegraphics[width=.8\linewidth]{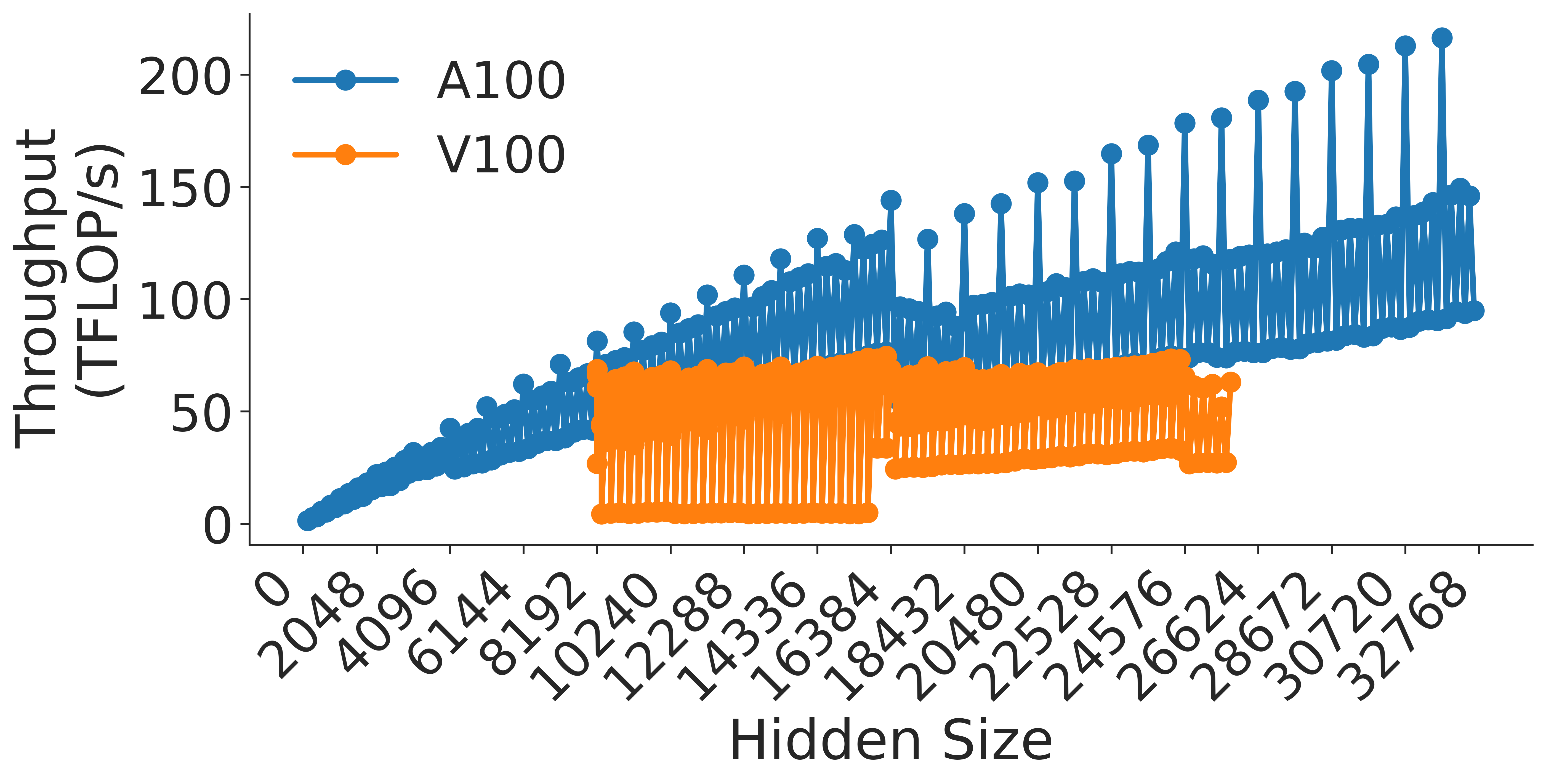}
    \caption{Attention score times values.}
    \vspace{-2ex}
    \label{fig:attn_prob_times_values}
\end{figure}

\begin{figure}[htbp]
\centering
    \includegraphics[width=.8\linewidth]{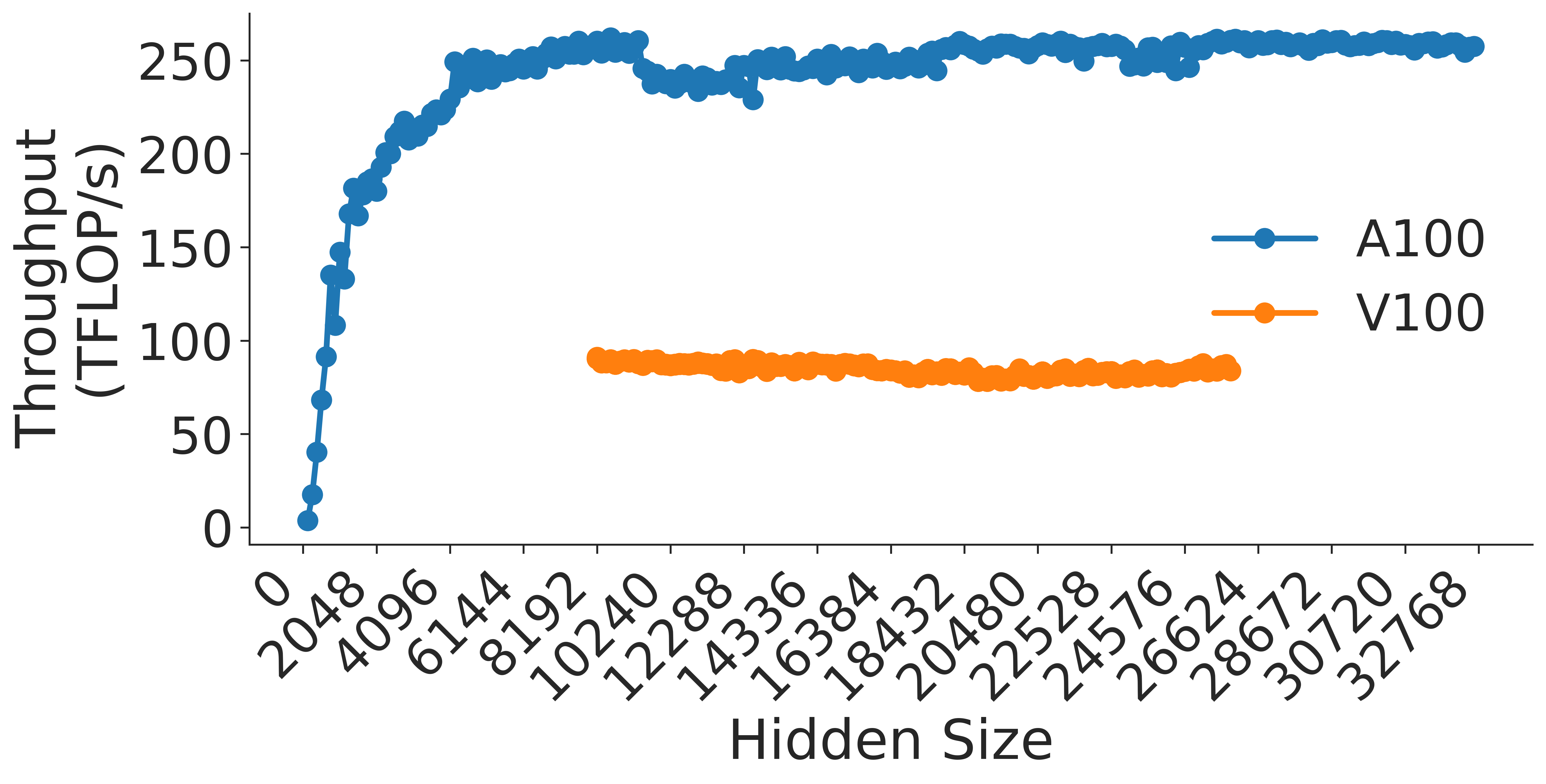}
    \caption{Post-attention linear projection.}
    \vspace{-2ex}
    \label{fig:attn_linproj}
\end{figure}

Figure \ref{fig:vocab_sweep} shows how the size of the vocab and the hidden dimension affects the logit layer, which is a linear layer at the end of the transformer model. The performance of the logit layer is maximized when $v$ is a multiple of 64, therefore it is best to pad the vocab size to the nearest multiple of 64. Likewise, the layer also performs best with a hidden size that is a multiple of 64. 

\section{A100 Results}

\begin{figure}[htbp]
\centering
    \includegraphics[width=.8\linewidth]{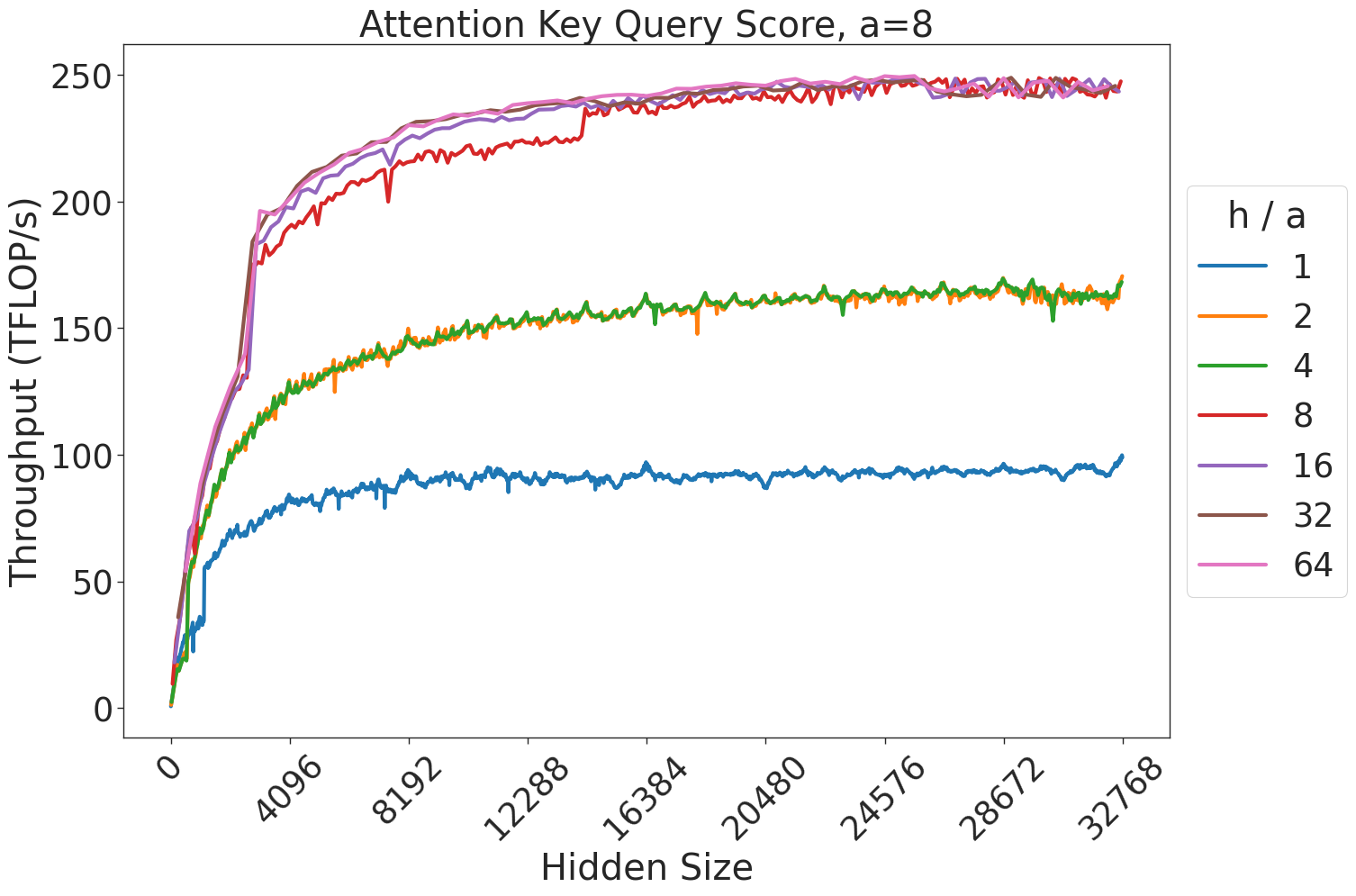}
    \caption{Attention key-query score GEMM throughput for 8 attention heads.}
    \vspace{-2ex}
    \label{fig:attention-kq-a8}
\end{figure}

\begin{figure}[htbp]
\centering
    \includegraphics[width=.8\linewidth]{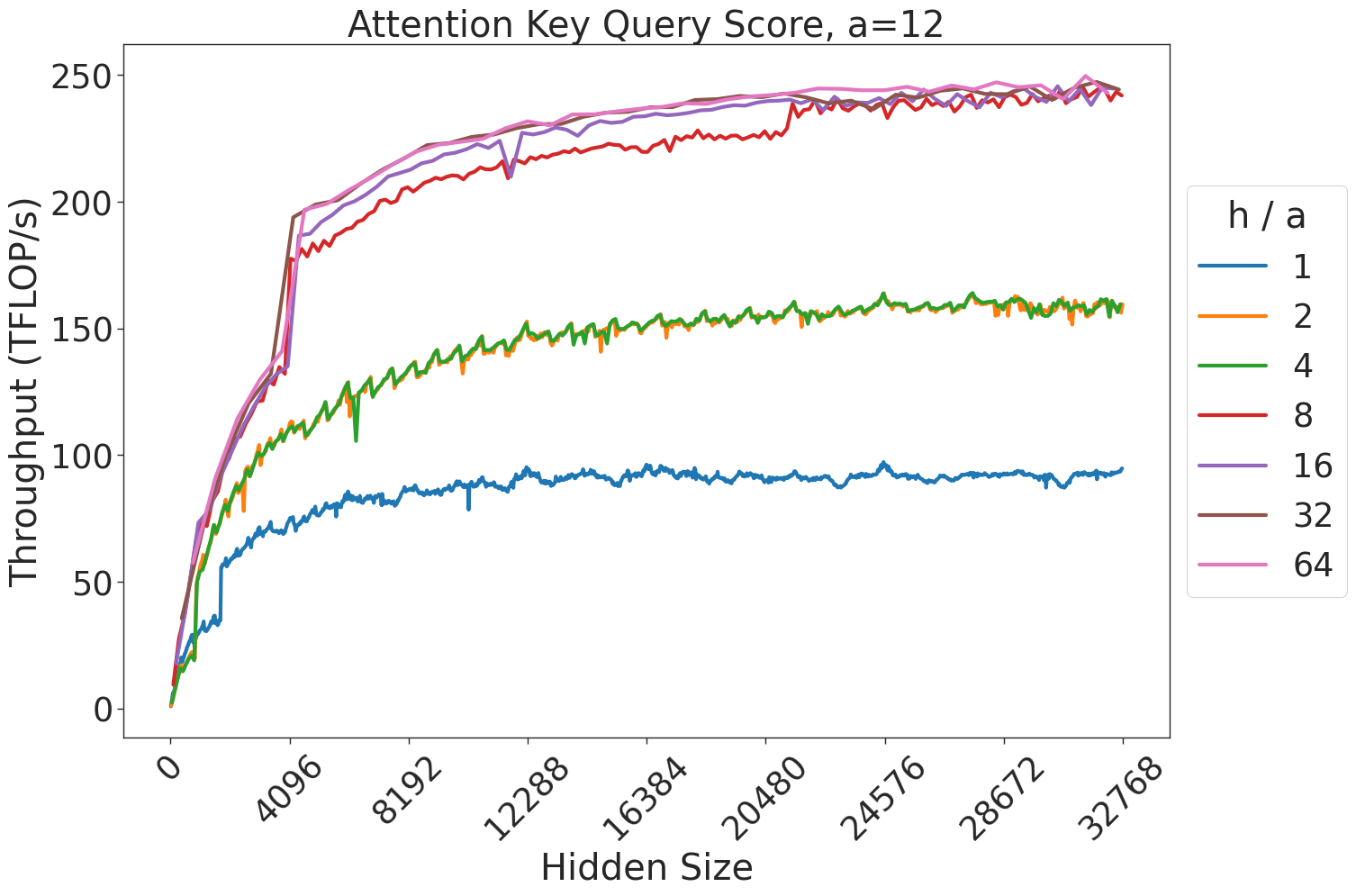}
    \caption{Attention key-query score GEMM throughput for 12 attention heads.}
    \vspace{-2ex}
    \label{fig:attention-kq-a12}
\end{figure}

\begin{figure}[htbp]
\centering
    \includegraphics[width=.8\linewidth]{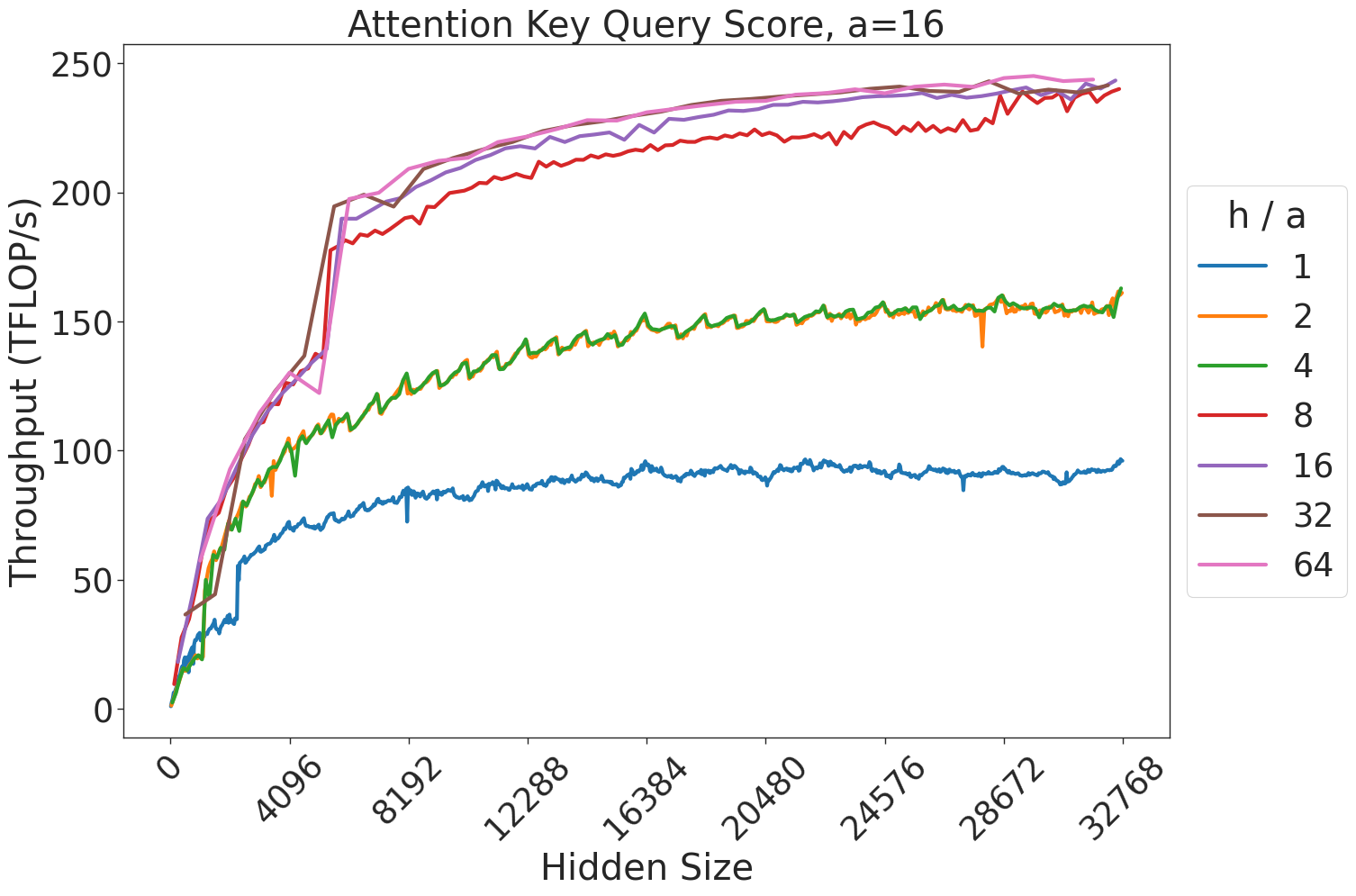}
    \caption{Attention key-query score GEMM throughput for 16 attention heads.}
    \vspace{-2ex}
    \label{fig:attention-kq-a16}
\end{figure}

\begin{figure}[htbp]
\centering
    \includegraphics[width=.8\linewidth]{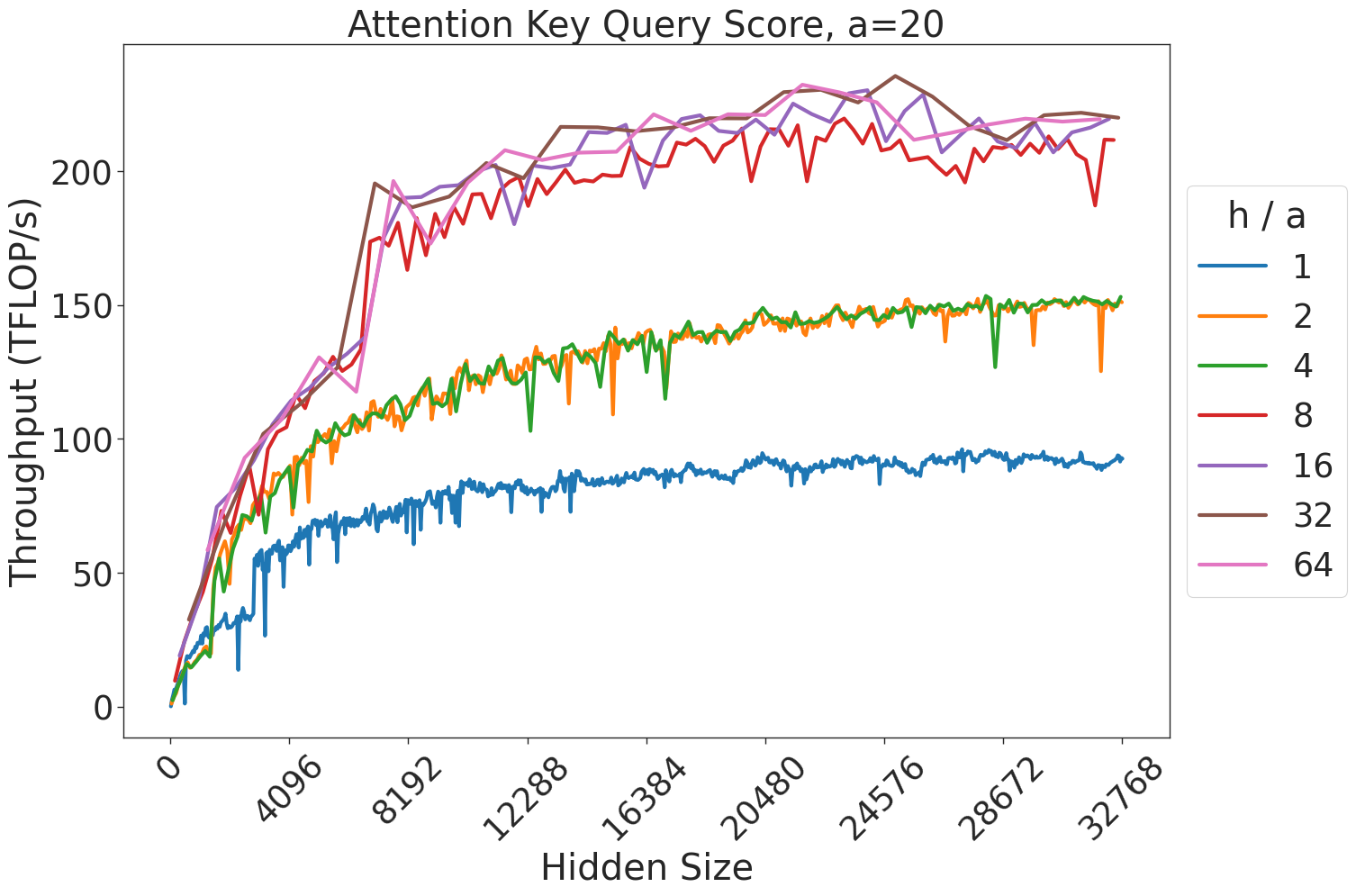}
    \caption{Attention key-query score GEMM throughput for 20 attention heads.}
    \vspace{-2ex}
    \label{fig:attention-kq-a20}
\end{figure}

\begin{figure}[htbp]
\centering
    \includegraphics[width=.8\linewidth]{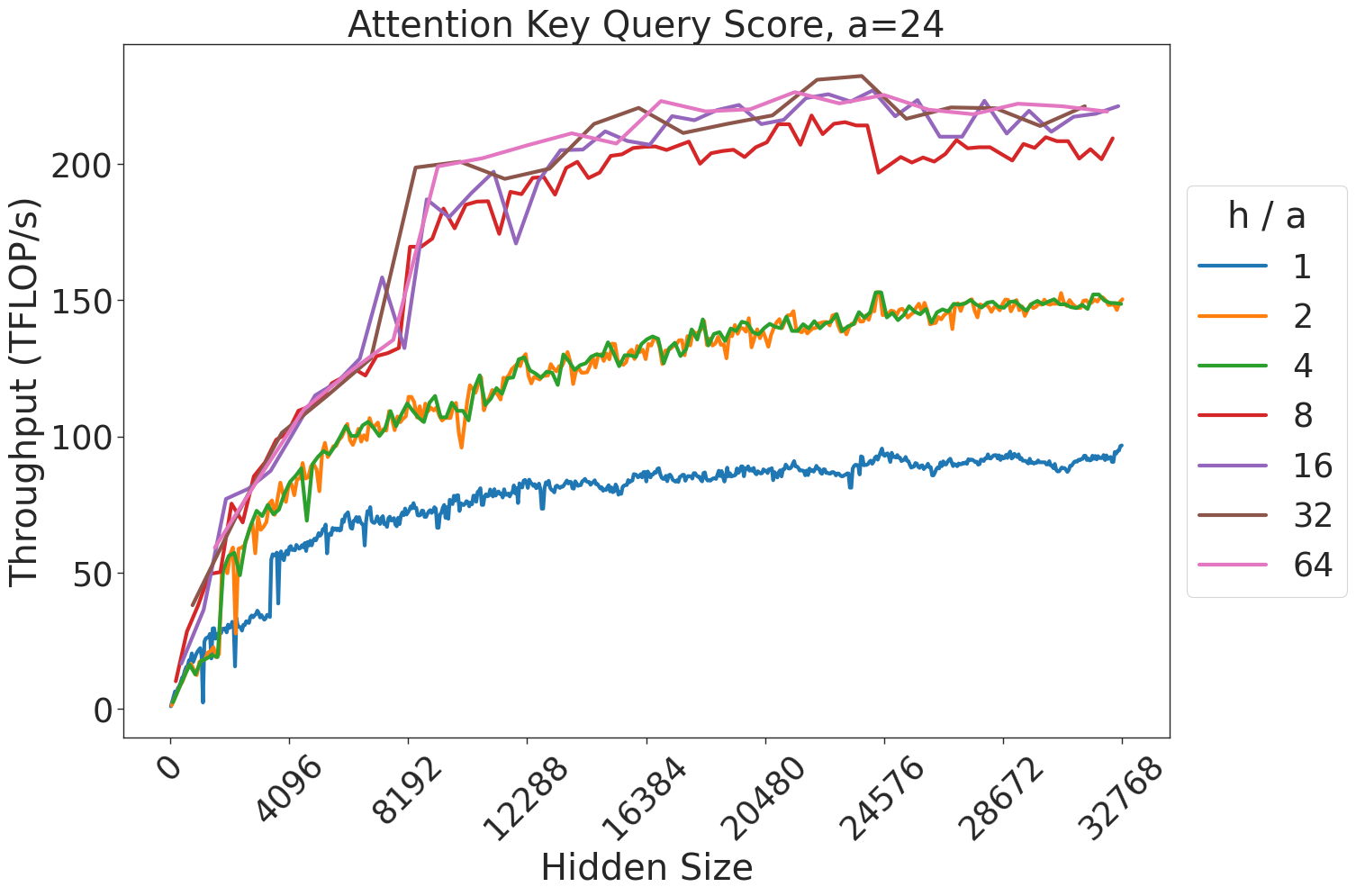}
    \caption{Attention key-query score GEMM throughput for 24 attention heads.}
    \vspace{-2ex}
    \label{fig:attention-kq-a24}
\end{figure}

\begin{figure}[htbp]
\centering
    \includegraphics[width=.8\linewidth]{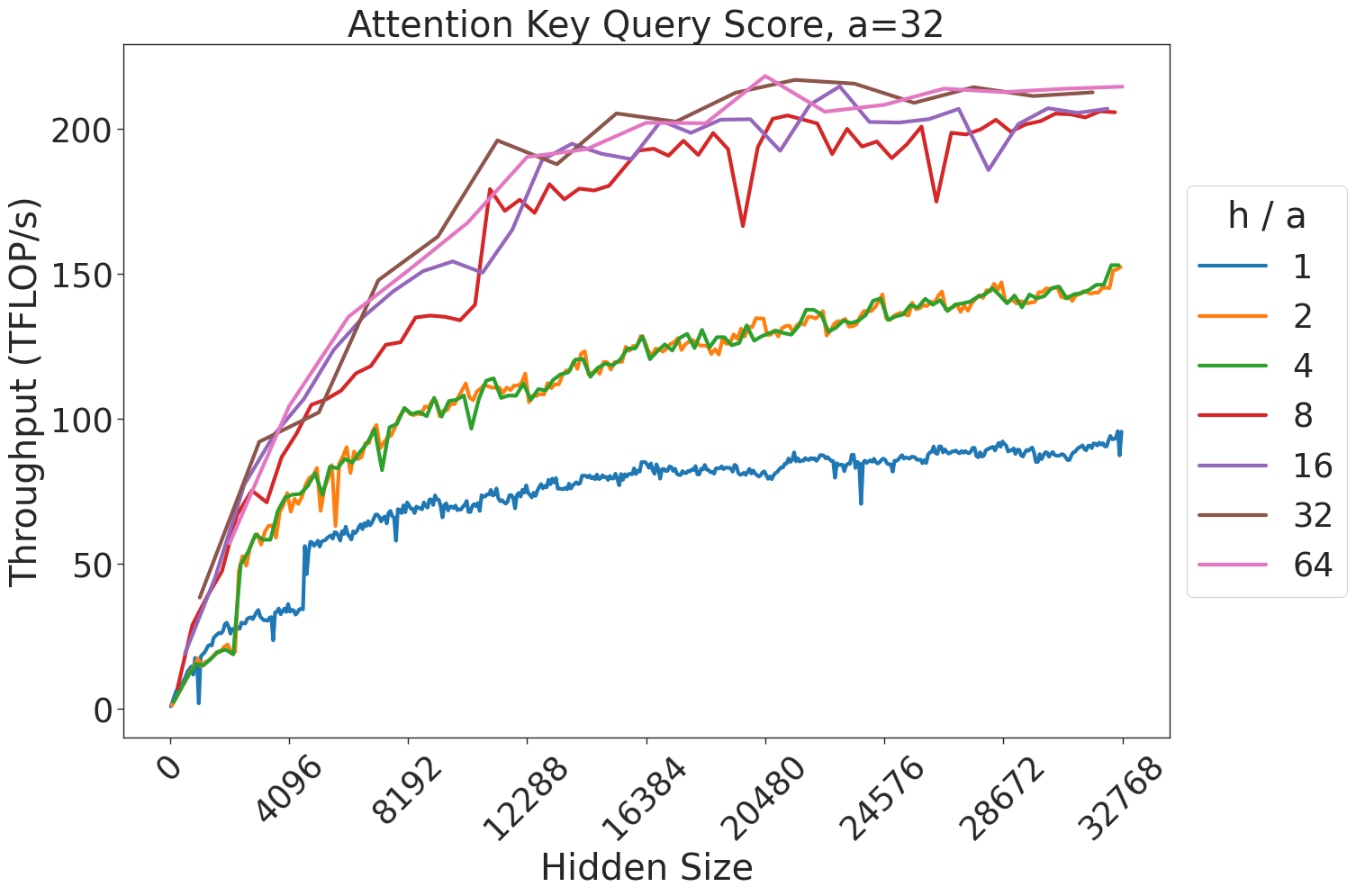}
    \caption{Attention key-query score GEMM throughput for 32 attention heads.}
    \vspace{-2ex}
    \label{fig:attention-kq-a32}
\end{figure}

\begin{figure}[htbp]
\centering
    \includegraphics[width=.8\linewidth]{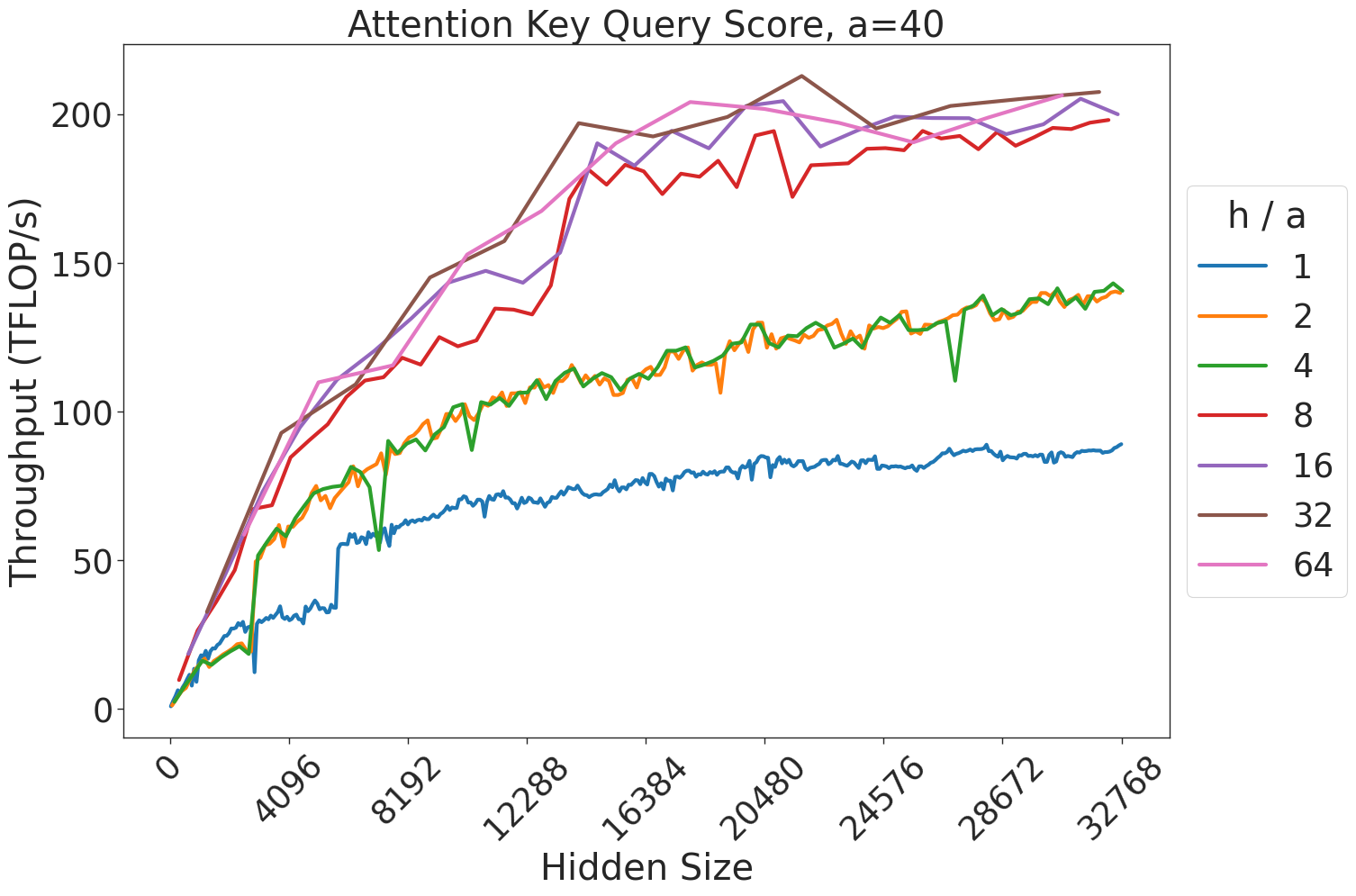}
    \caption{Attention key-query score GEMM throughput for 40 attention heads.}
    \vspace{-2ex}
    \label{fig:attention-kq-a40}
\end{figure}

\begin{figure}[htbp]
\centering
    \includegraphics[width=.8\linewidth]{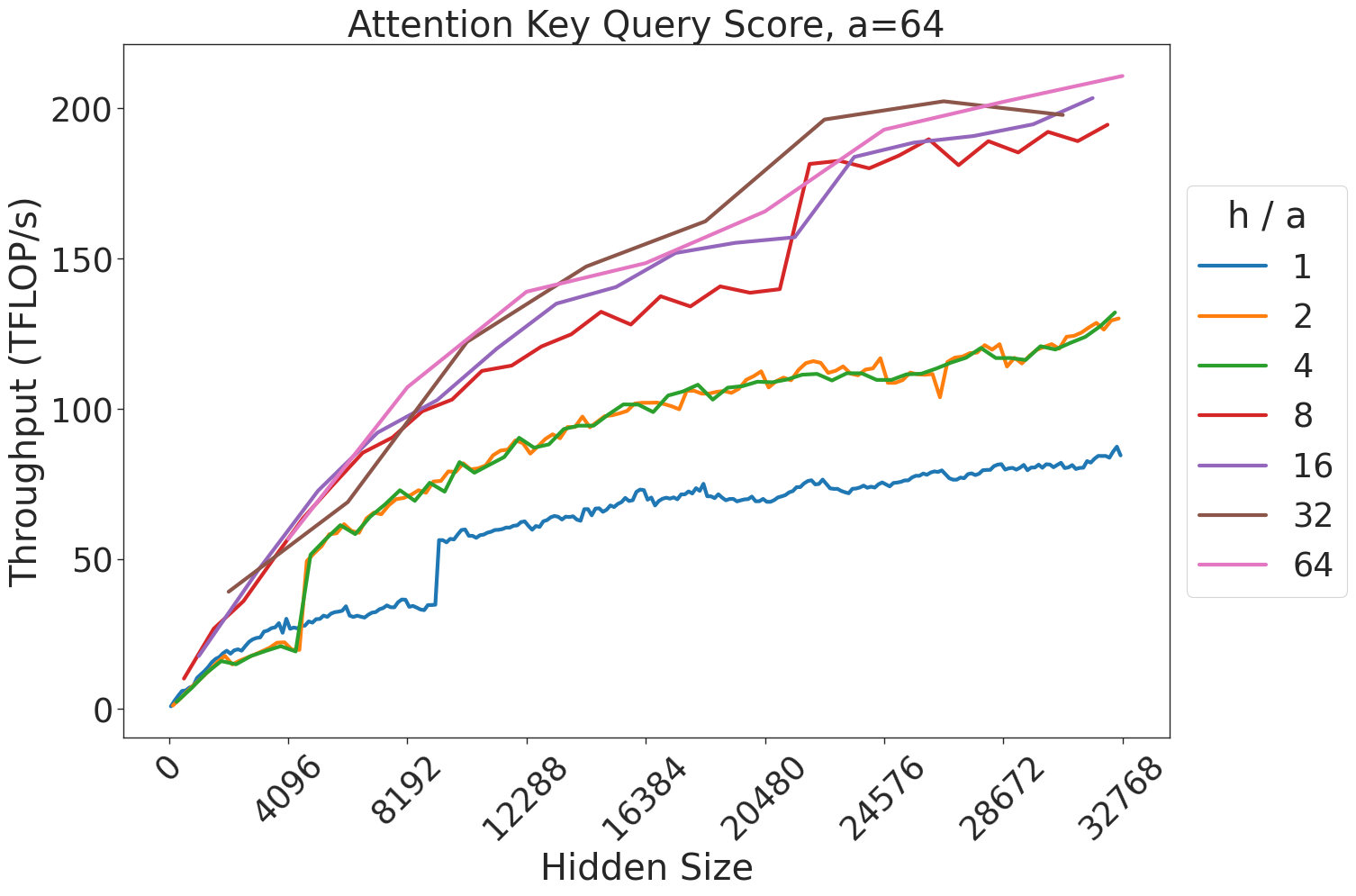}
    \caption{Attention key-query score GEMM throughput for 64 attention heads.}
    \vspace{-2ex}
    \label{fig:attention-kq-a64}
\end{figure}

\begin{figure}[htbp]
\centering
    \includegraphics[width=.8\linewidth]{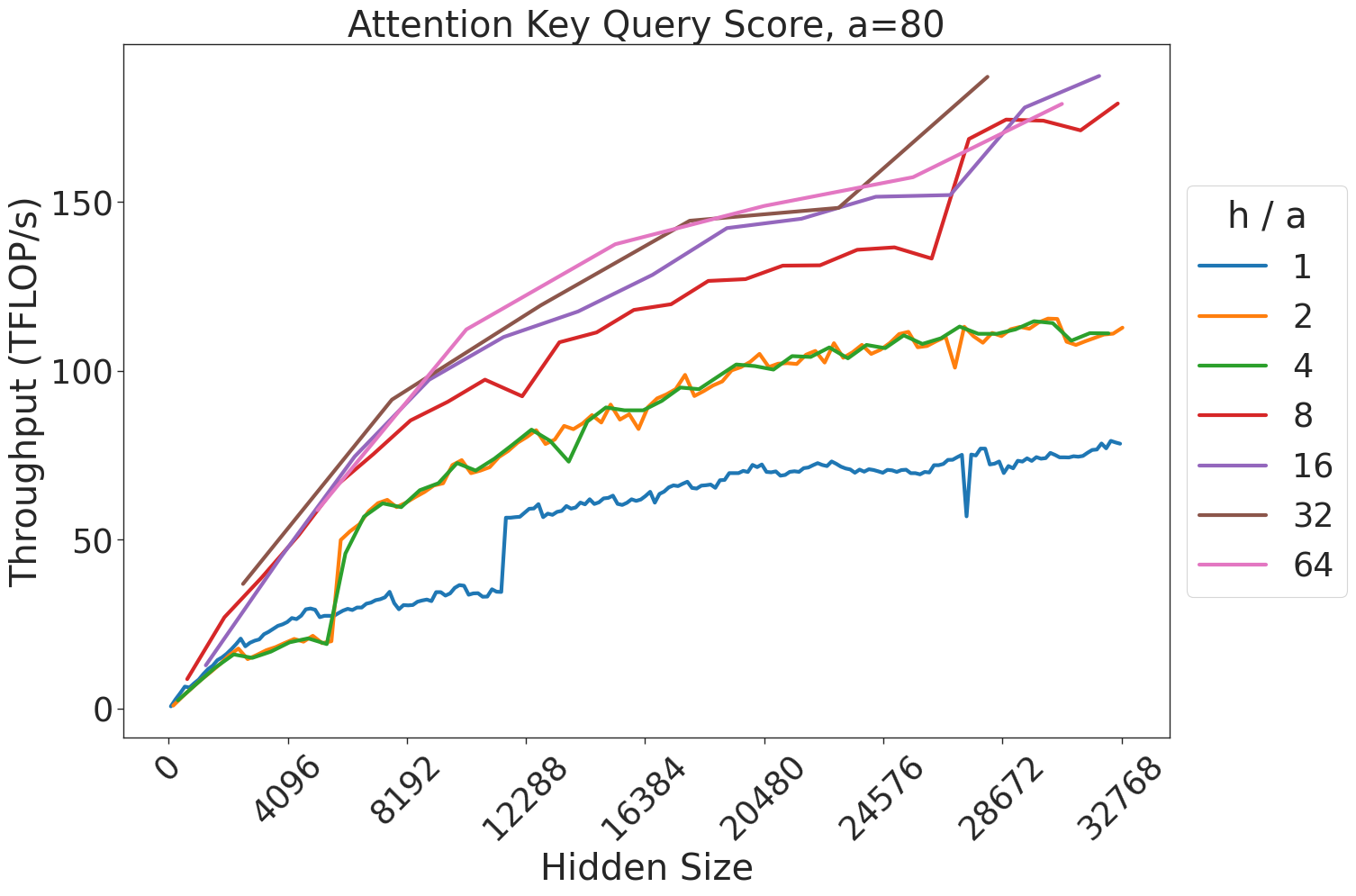}
    \caption{Attention key-query score GEMM throughput for 80 attention heads.}
    \vspace{-2ex}
    \label{fig:attention-kq-a80}
\end{figure}

\begin{figure}[htbp]
\centering
    \includegraphics[width=.8\linewidth]{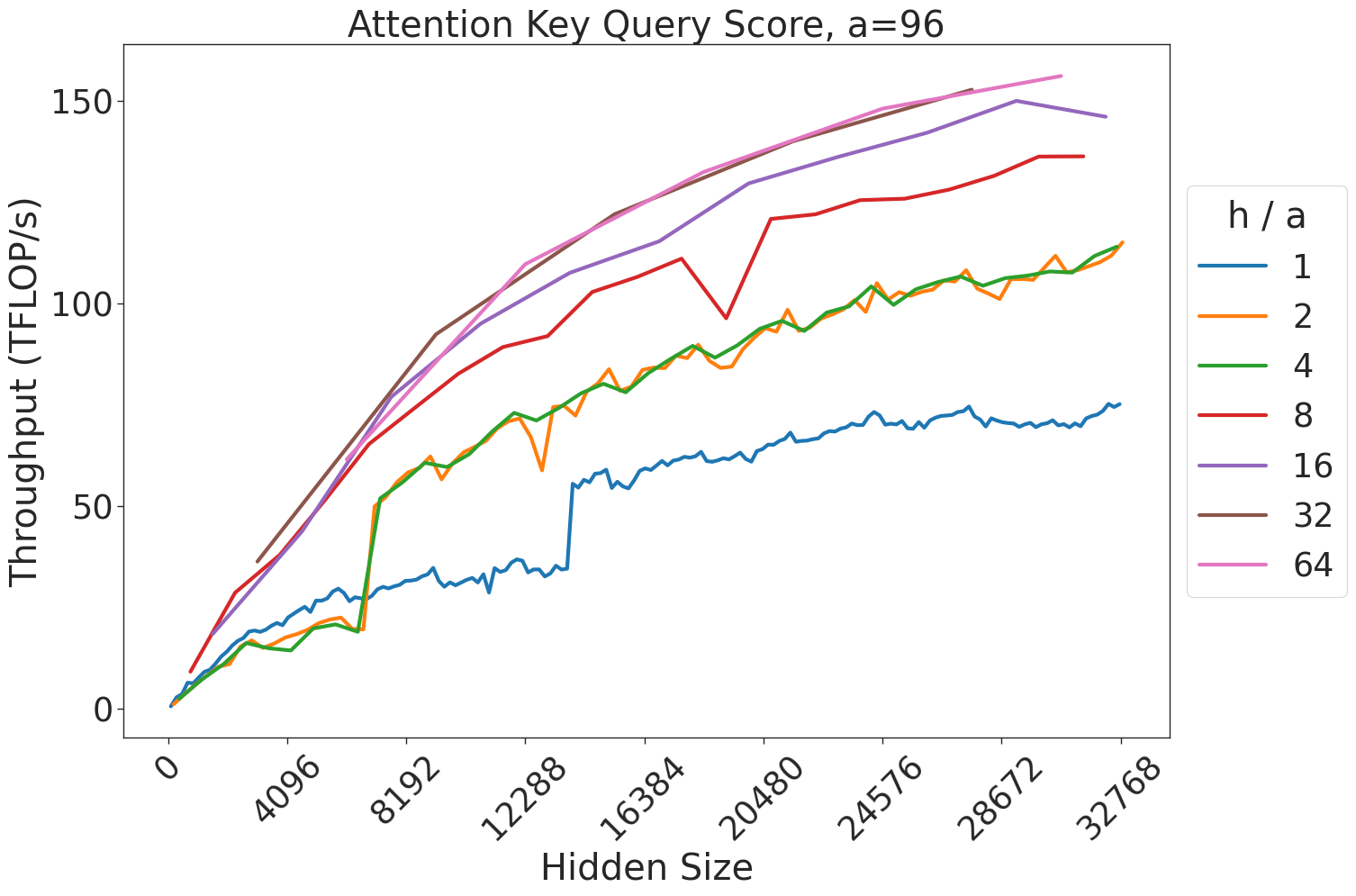}
    \caption{Attention key-query score GEMM throughput for 96 attention heads.}
    \vspace{-2ex}
    \label{fig:attention-kq-a96}
\end{figure}

\begin{figure}[htbp]
\centering
    \includegraphics[width=.8\linewidth]{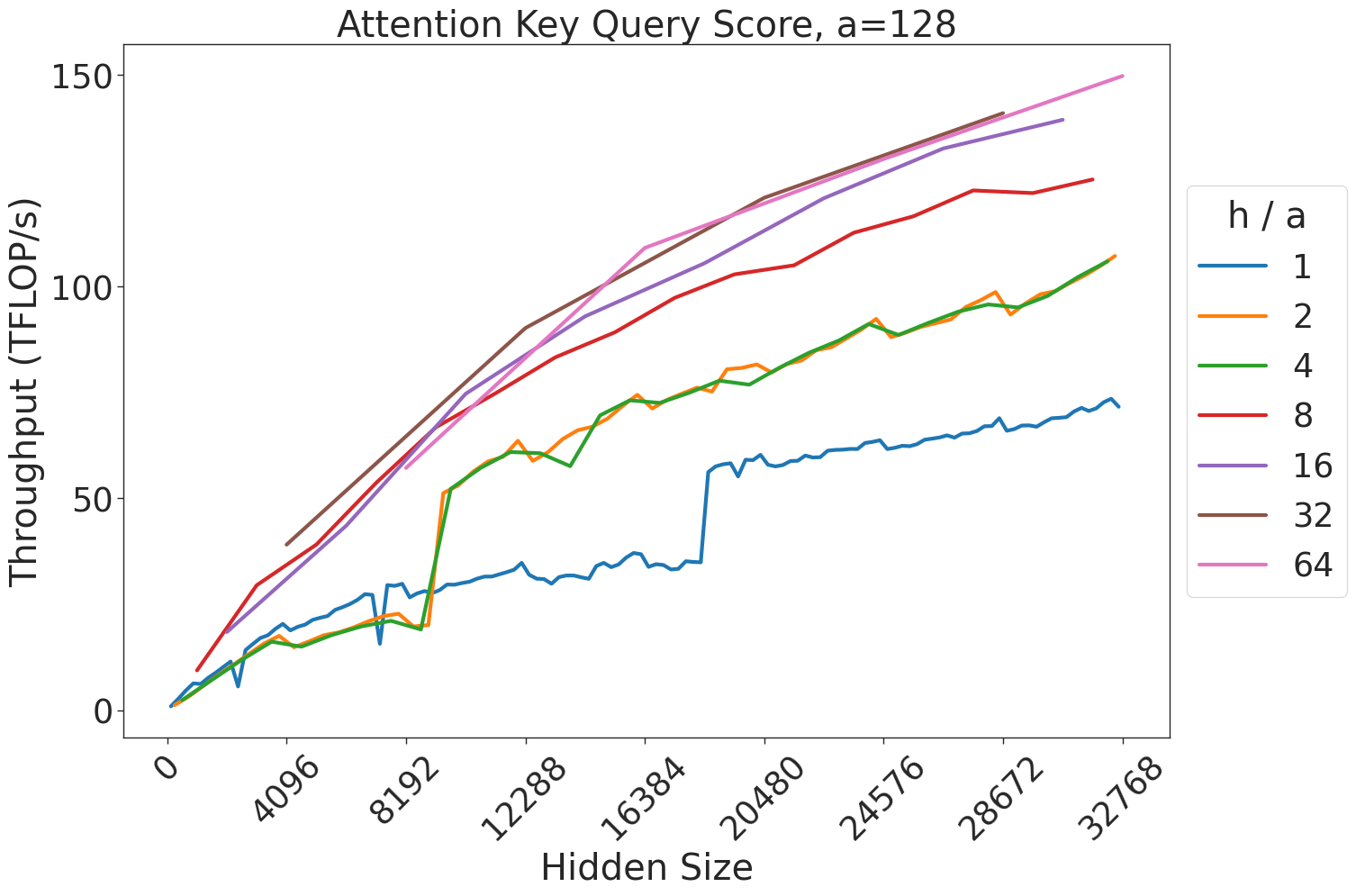}
    \caption{Attention key-query score GEMM throughput for 128 attention heads.}
    \vspace{-2ex}
    \label{fig:attention-kq-a128}
\end{figure}

\begin{figure}[htbp]
\centering
    \includegraphics[width=.8\linewidth]{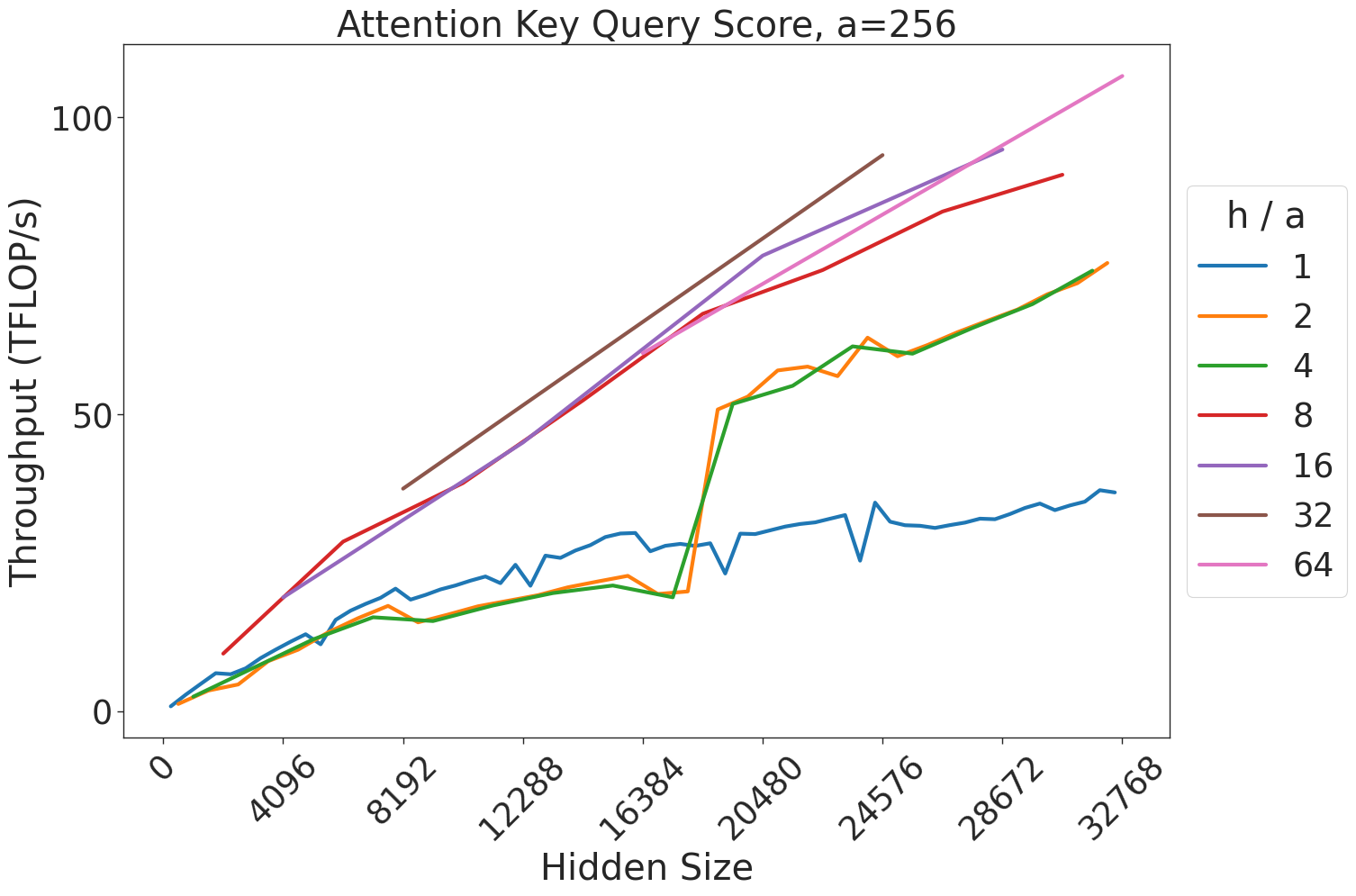}
    \caption{Attention key-query score GEMM throughput for 256 attention heads.}
    \vspace{-2ex}
    \label{fig:attention-kq-a256}
\end{figure}

\begin{figure}[htbp]
\centering
    \includegraphics[width=.8\linewidth]{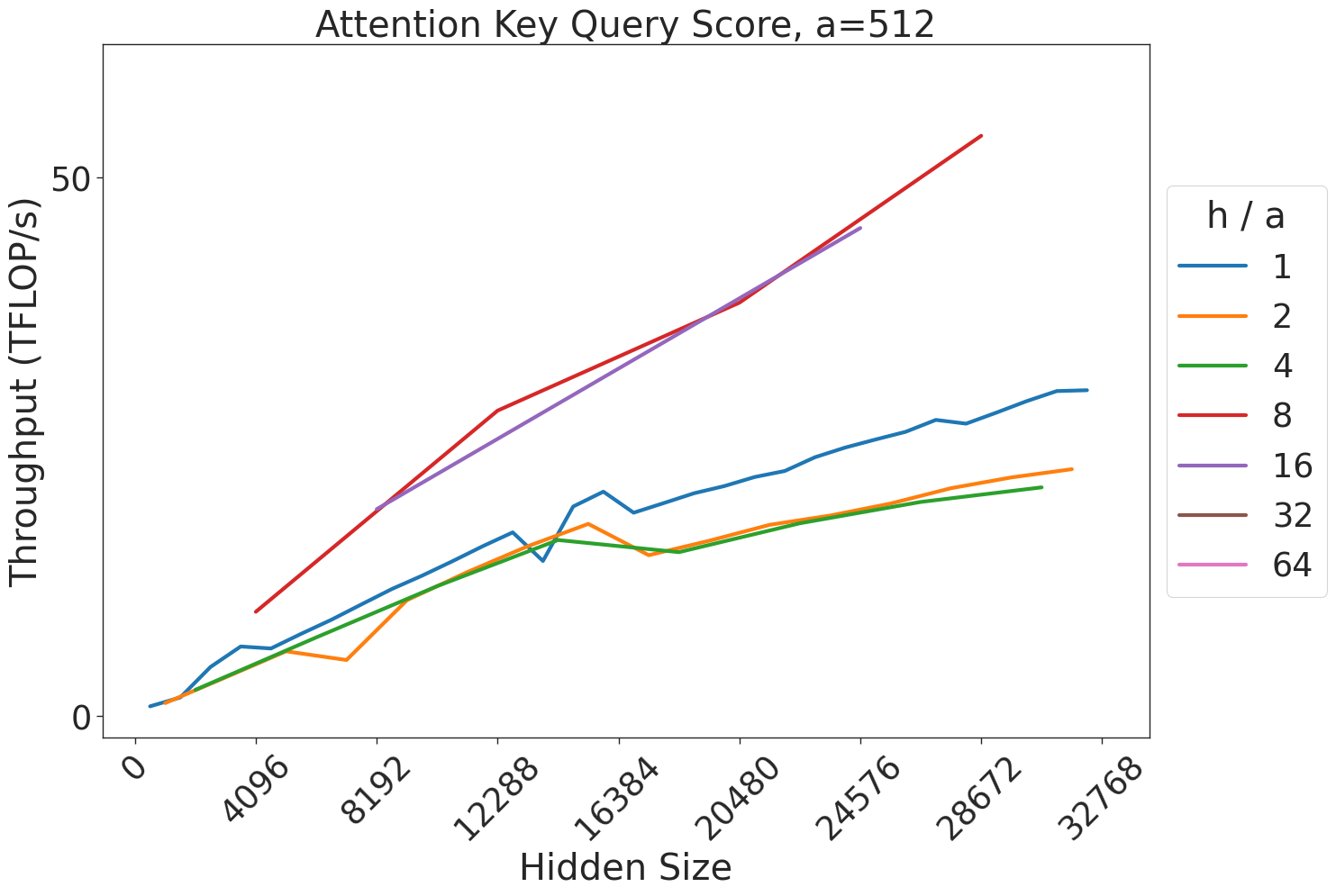}
    \caption{Attention key-query score GEMM throughput for 512 attention heads.}
    \vspace{-2ex}
    \label{fig:attention-kq-a512}
\end{figure}

\begin{figure}[htbp]
\centering
    \includegraphics[width=.8\linewidth]{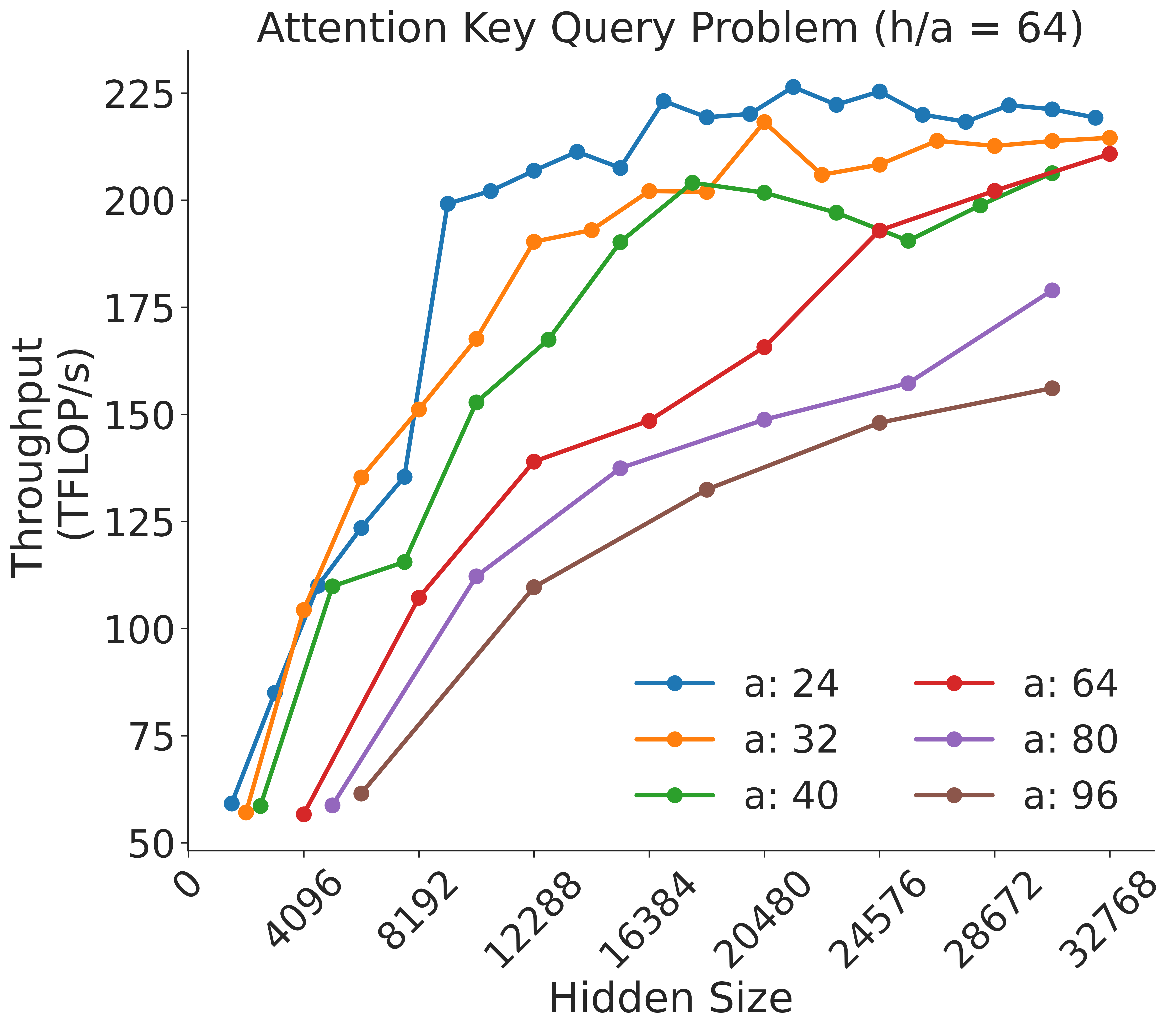}
    \caption{Attention key-query score GEMM throughput assuming fixed ratio of $\frac{h}{a}=64$.}
    \vspace{-2ex}
    \label{fig:attention-kq-ha64}
\end{figure}

\begin{figure}[htbp]
\centering
    \includegraphics[width=.8\linewidth]{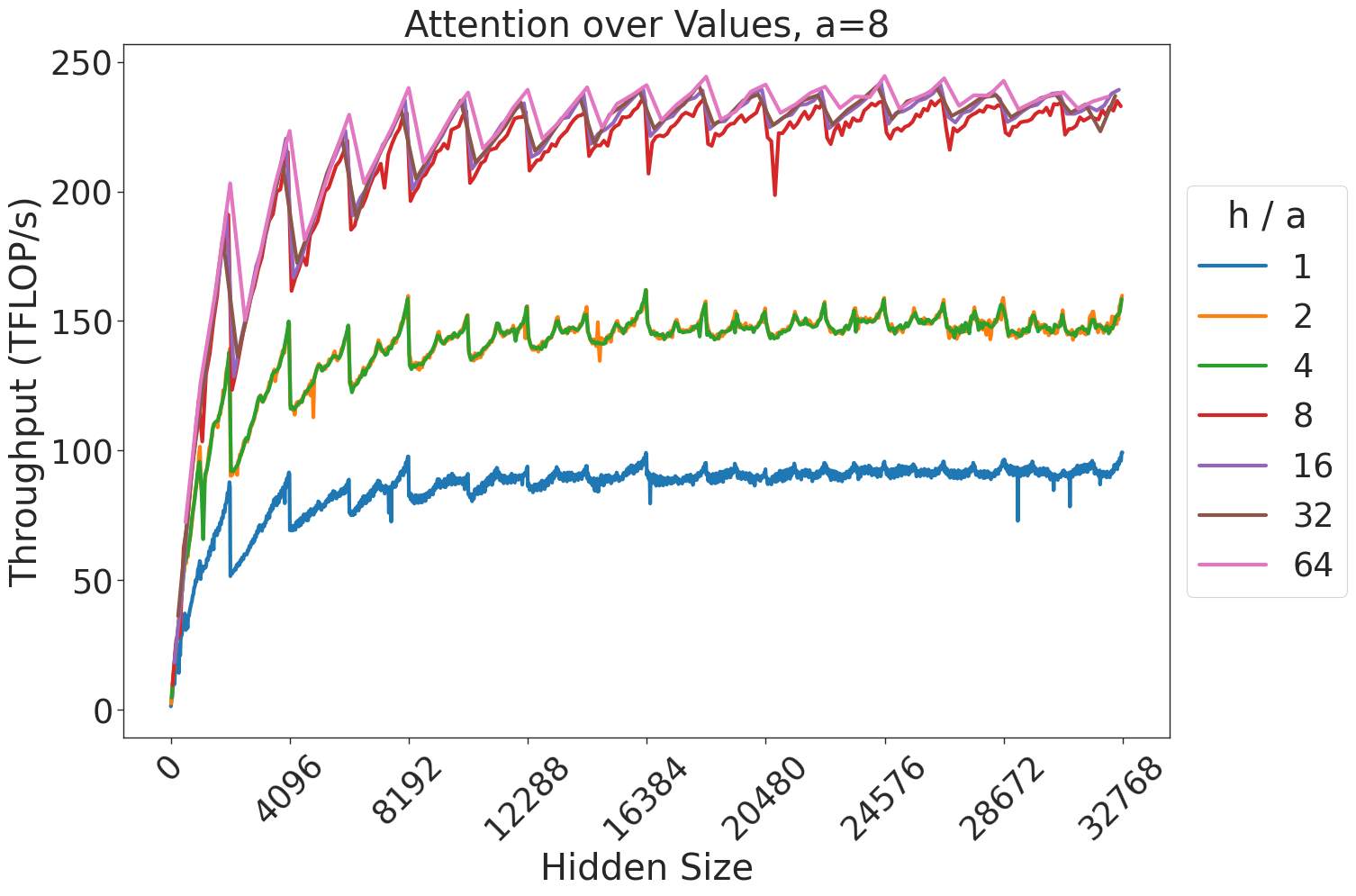}
    \caption{Attention over value GEMM throughput for 8 attention heads.}
    \vspace{-2ex}
    \label{fig:attention-val-8}
\end{figure}

\begin{figure}[htbp]
\centering
    \includegraphics[width=.8\linewidth]{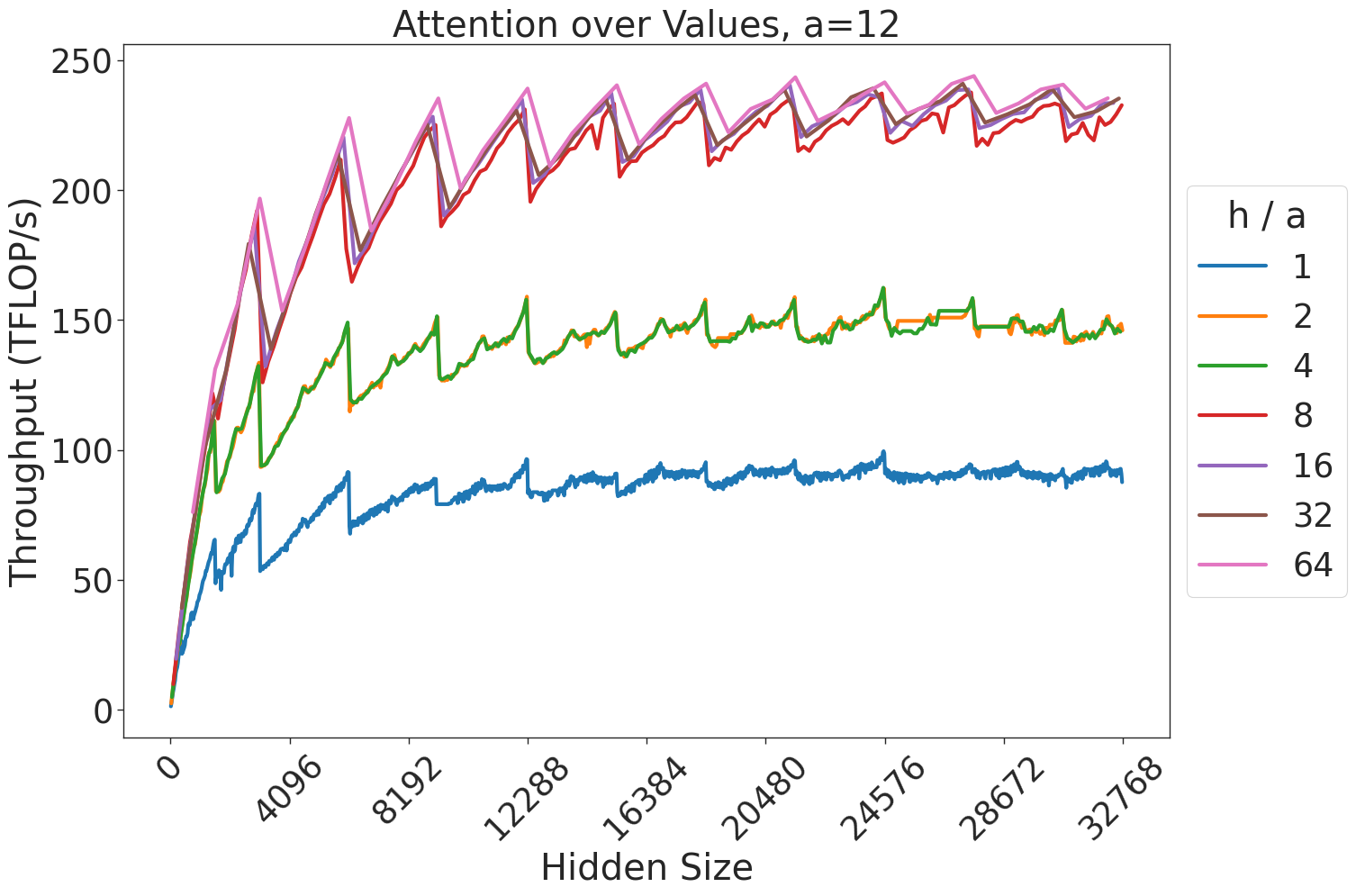}
    \caption{Attention over value GEMM throughput for 12 attention heads.}
    \vspace{-2ex}
    \label{fig:attention-val-12}
\end{figure}

\begin{figure}[htbp]
\centering
    \includegraphics[width=.8\linewidth]{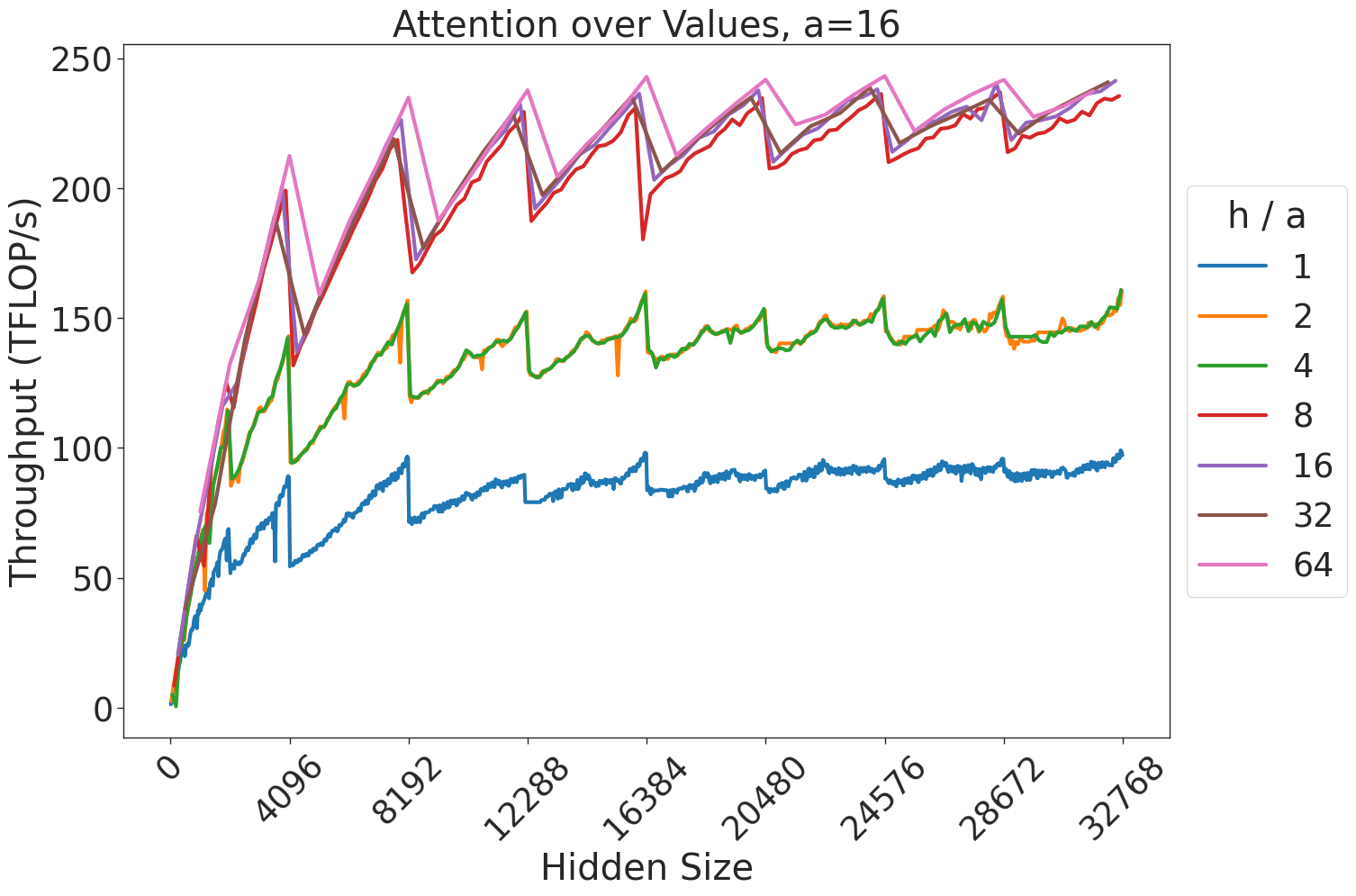}
    \caption{Attention over value GEMM throughput for 16 attention heads.}
    \vspace{-2ex}
    \label{fig:attention-val-16}
\end{figure}

\begin{figure}[htbp]
\centering
    \includegraphics[width=.8\linewidth]{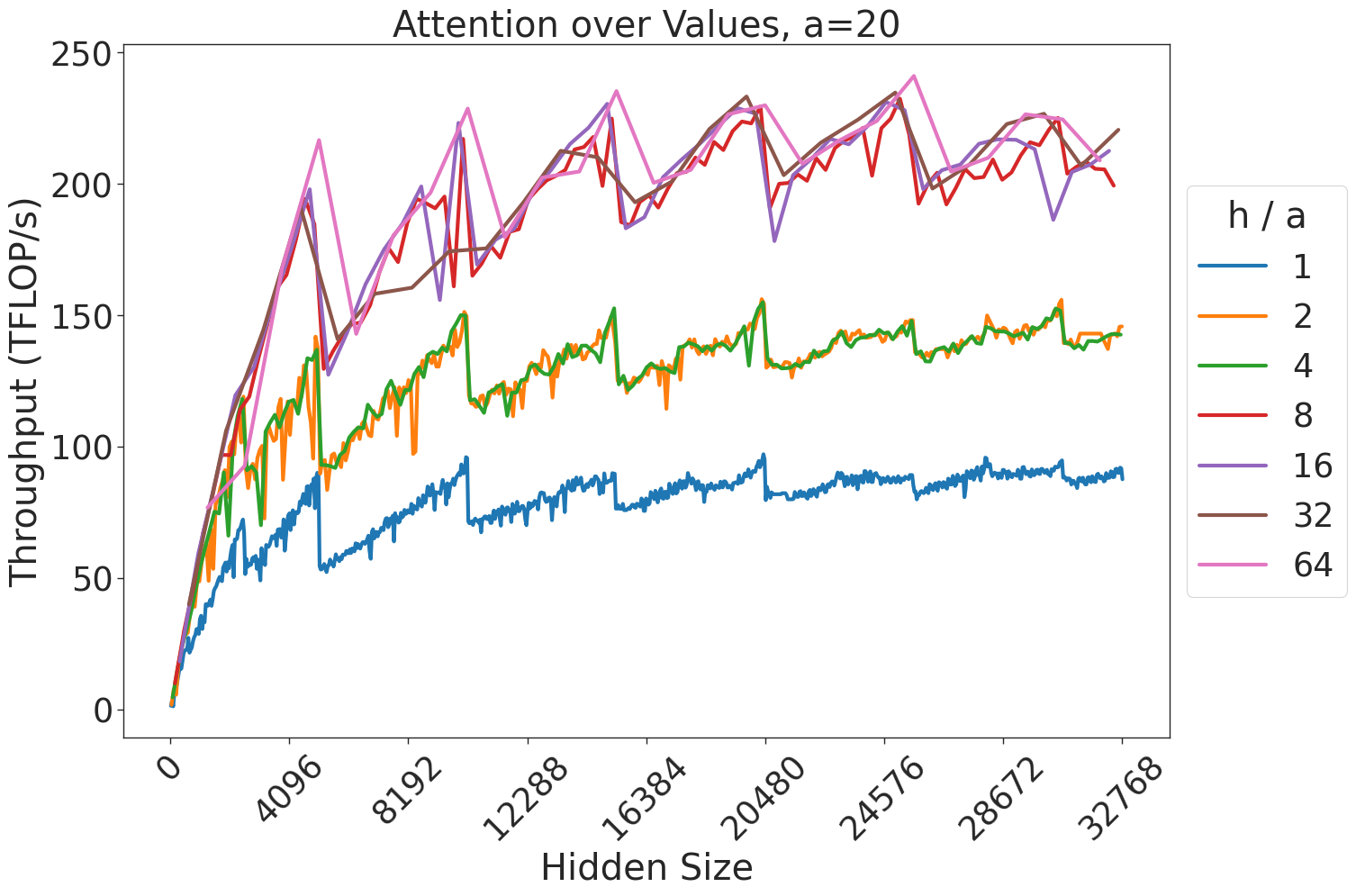}
    \caption{Attention over value GEMM throughput for 20 attention heads.}
    \vspace{-2ex}
    \label{fig:attention-val-20}
\end{figure}

\begin{figure}[htbp]
\centering
    \includegraphics[width=.8\linewidth]{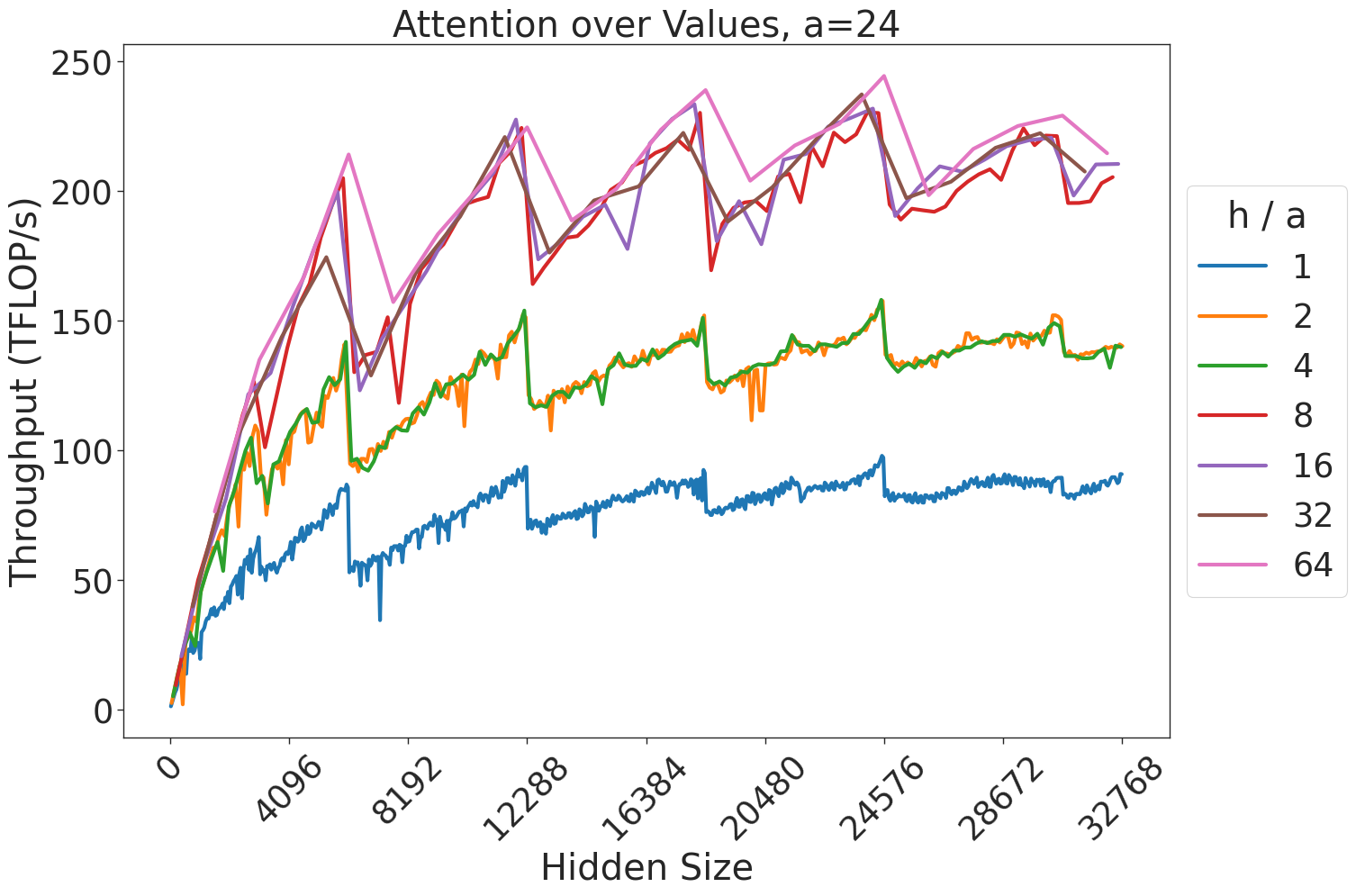}
    \caption{Attention over value GEMM throughput for 24 attention heads.}
    \vspace{-2ex}
    \label{fig:attention-val-24}
\end{figure}

\begin{figure}[htbp]
\centering
    \includegraphics[width=.8\linewidth]{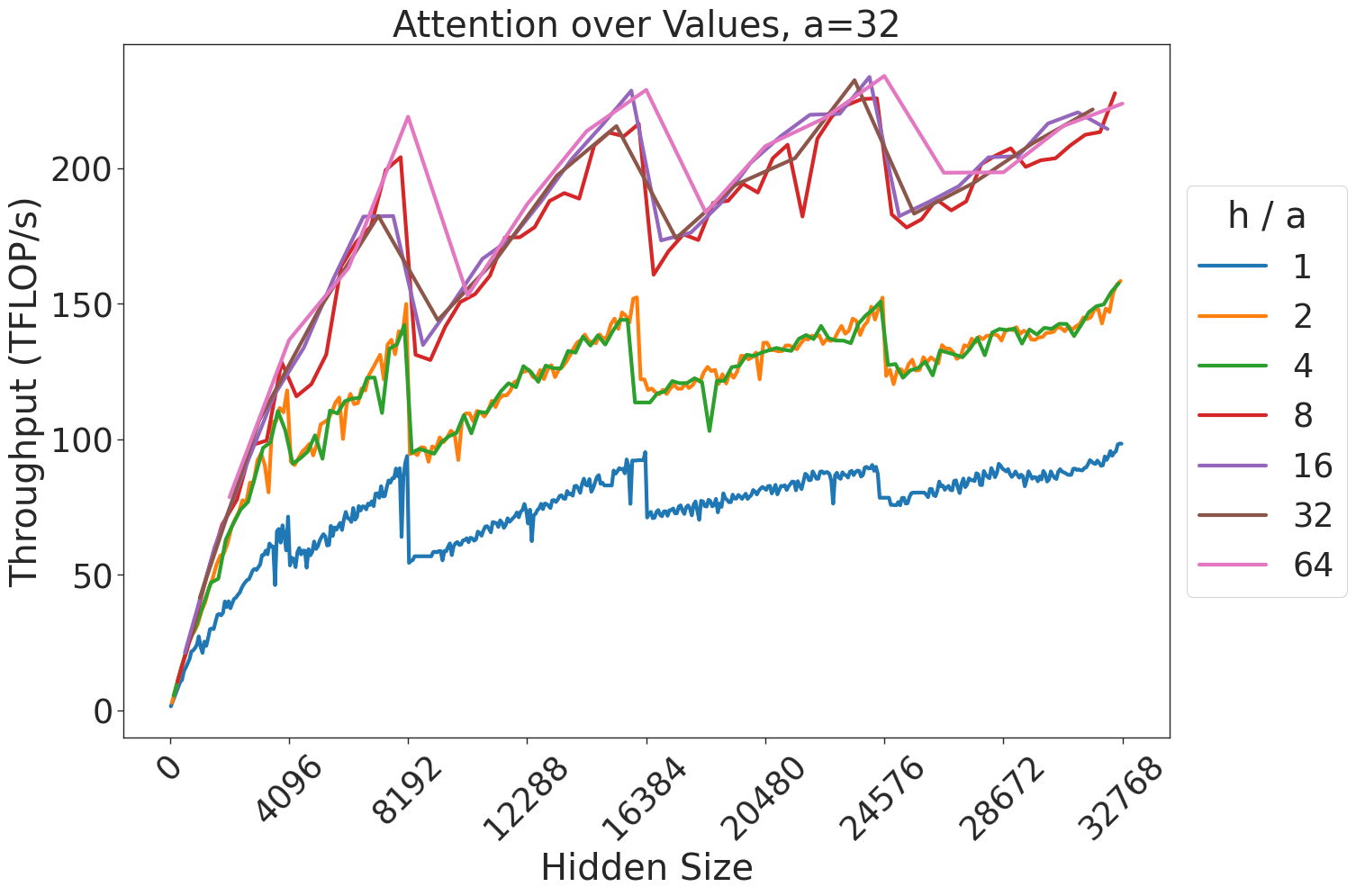}
    \caption{Attention over value GEMM throughput for 32 attention heads.}
    \vspace{-2ex}
    \label{fig:attention-val-32}
\end{figure}

\begin{figure}[htbp]
\centering
    \includegraphics[width=.8\linewidth]{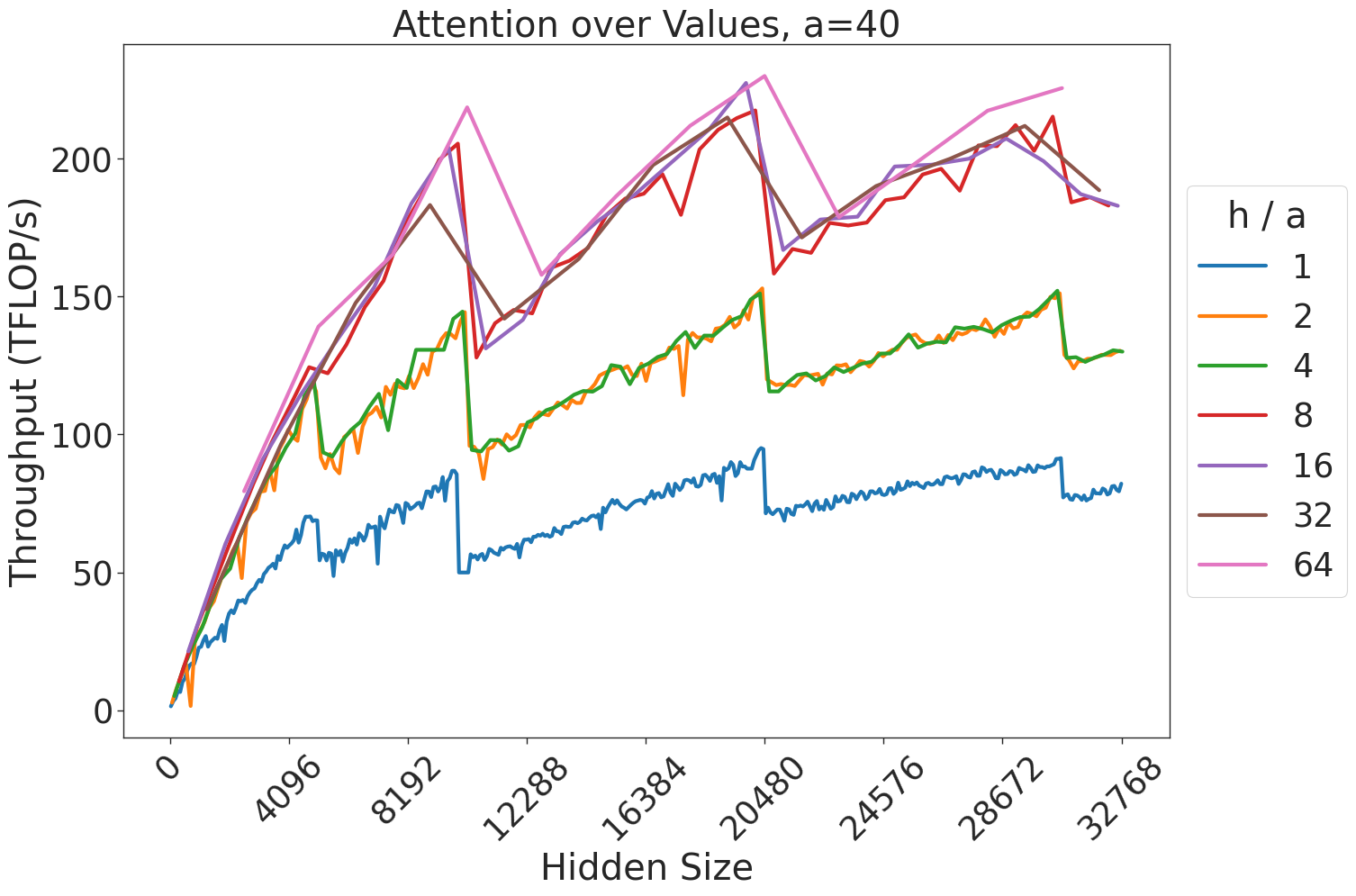}
    \caption{Attention over value GEMM throughput for 40 attention heads.}
    \vspace{-2ex}
    \label{fig:attention-val-40}
\end{figure}

\begin{figure}[htbp]
\centering
    \includegraphics[width=.8\linewidth]{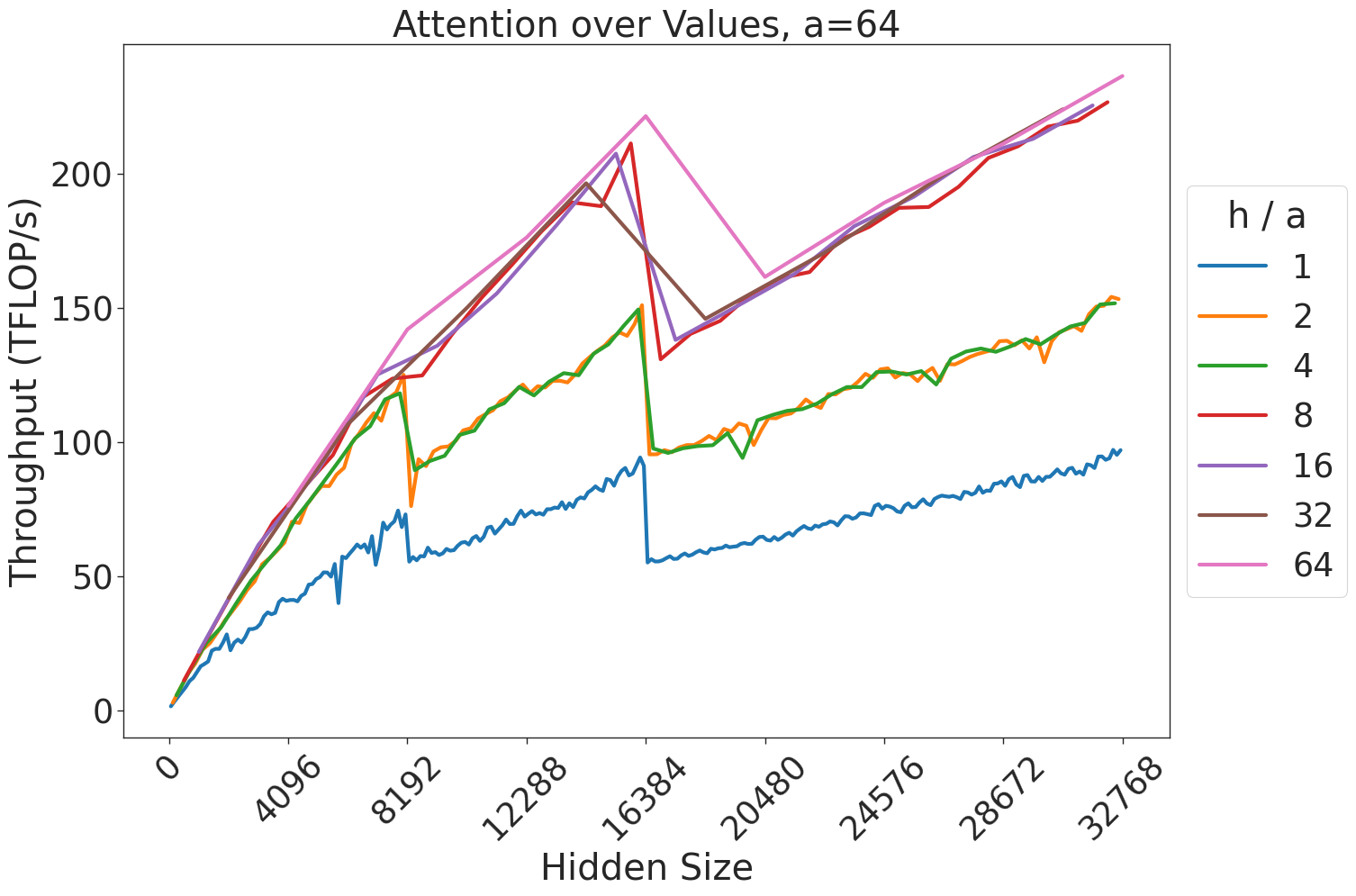}
    \caption{Attention over value GEMM throughput for 64 attention heads.}
    \vspace{-2ex}
    \label{fig:attention-val-64}
\end{figure}

\begin{figure}[htbp]
\centering
    \includegraphics[width=.8\linewidth]{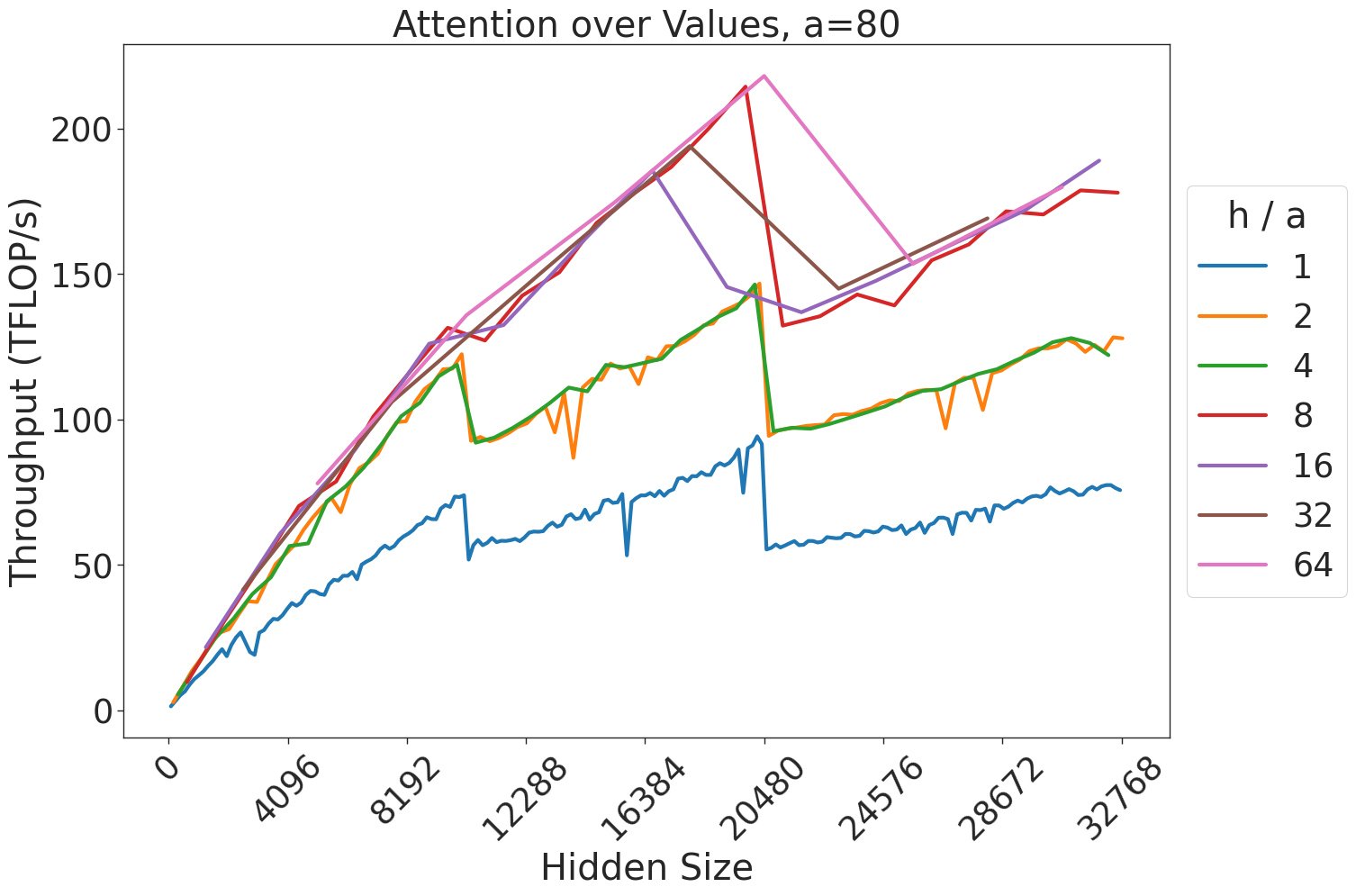}
    \caption{Attention over value GEMM throughput for 80 attention heads.}
    \vspace{-2ex}
    \label{fig:attention-val-80}
\end{figure}

\begin{figure}[htbp]
\centering
    \includegraphics[width=.8\linewidth]{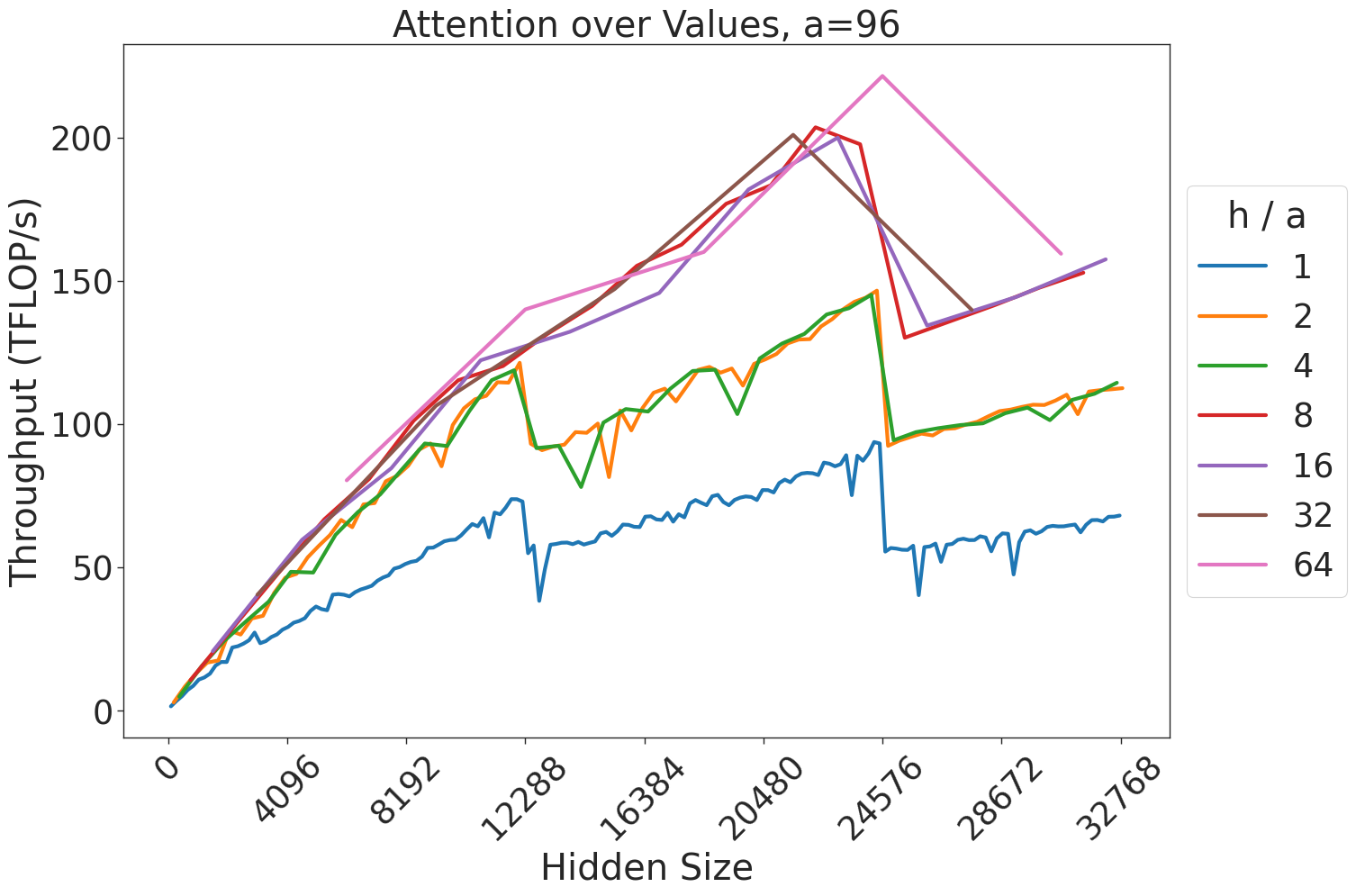}
    \caption{Attention over value GEMM throughput for 96 attention heads.}
    \vspace{-2ex}
    \label{fig:attention-val-96}
\end{figure}

\begin{figure}[htbp]
\centering
    \includegraphics[width=.8\linewidth]{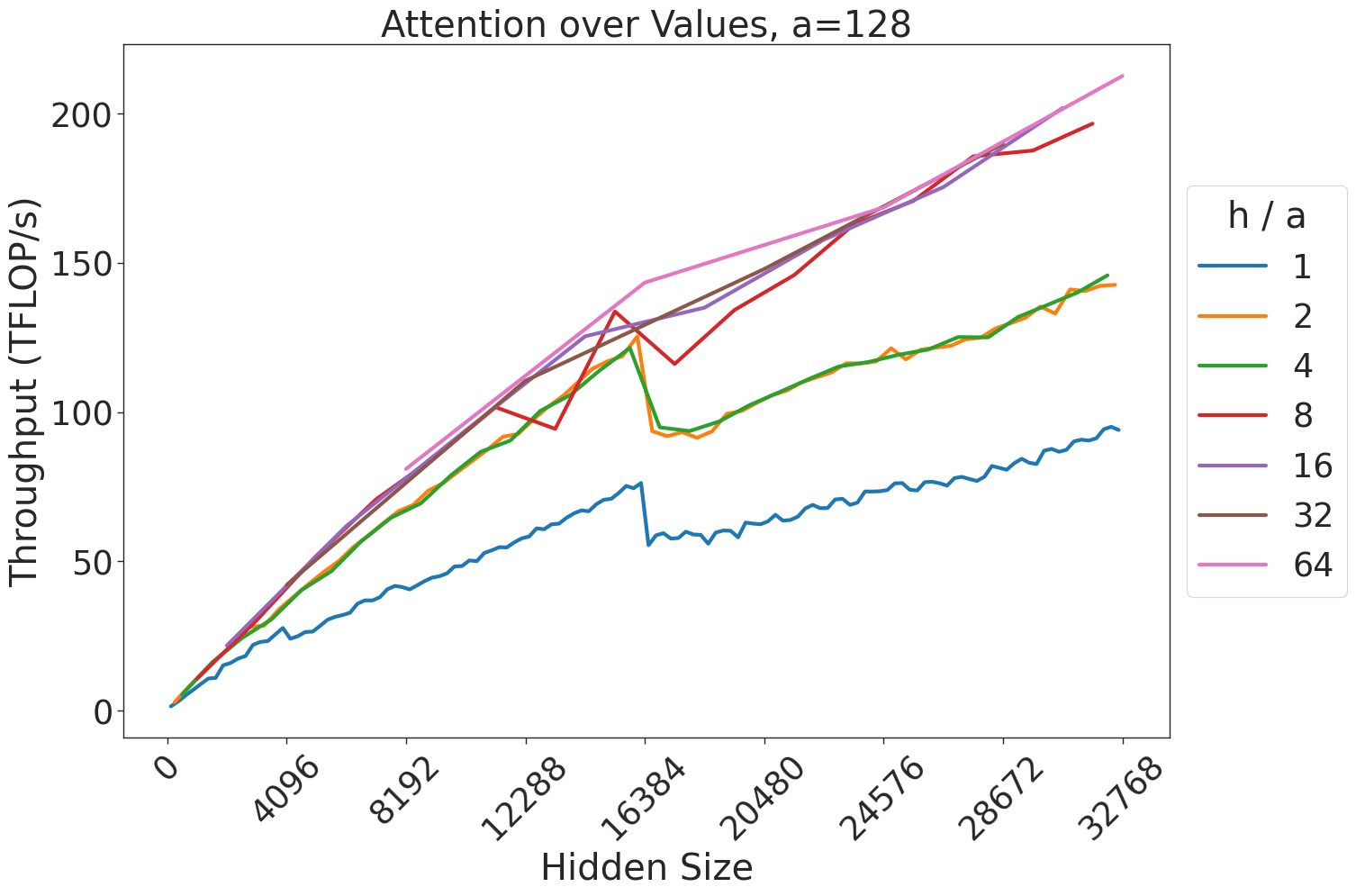}
    \caption{Attention over value GEMM throughput for 128 attention heads.}
    \vspace{-2ex}
    \label{fig:attention-val-128}
\end{figure}

\begin{figure}[htbp]
\centering
    \includegraphics[width=.8\linewidth]{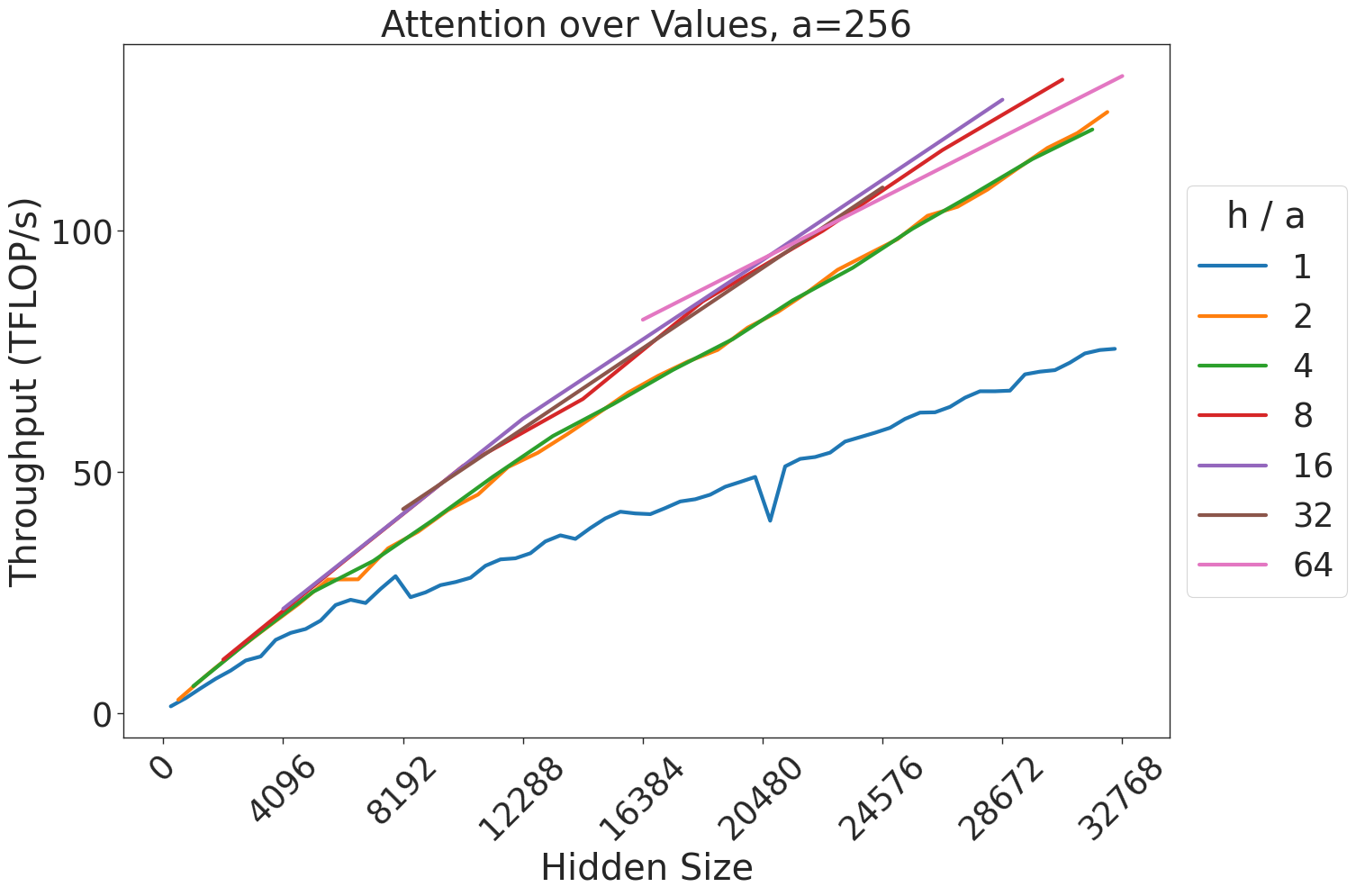}
    \caption{Attention over value GEMM throughput for 256 attention heads.}
    \vspace{-2ex}
    \label{fig:attention-val-256}
\end{figure}

\begin{figure}[htbp]
\centering
    \includegraphics[width=.8\linewidth]{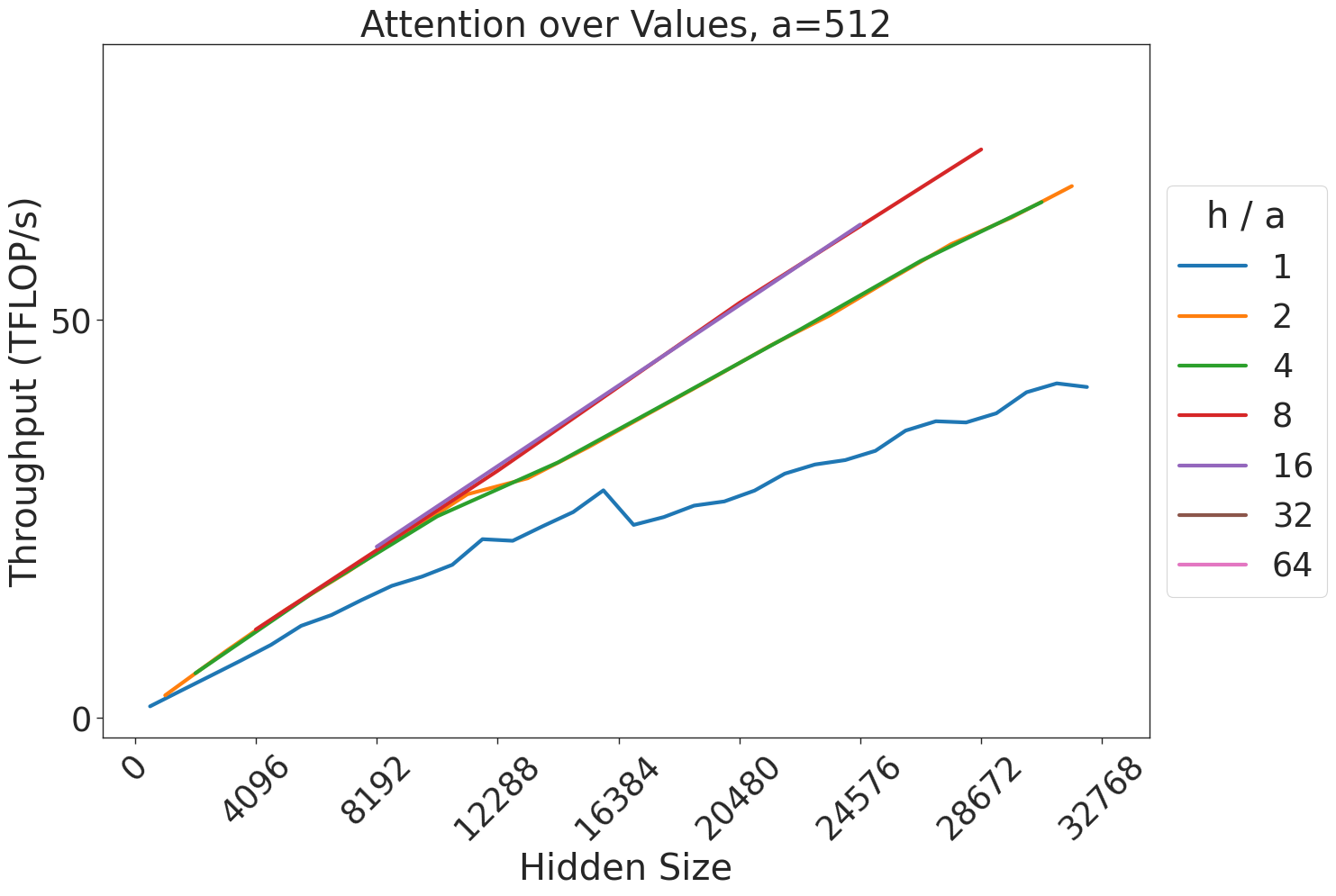}
    \caption{Attention over value GEMM throughput for 512 attention heads.}
    \vspace{-2ex}
    \label{fig:attention-val-512}
\end{figure}

Figures \ref{fig:attention-kq-a8} through \ref{fig:attention-val-512} show the performance of the Attention Key-Query Score and Attention Over Value computations for various numbers of attention heads. In each of these figures, we highlight the trend observed when using tensor cores. Each color is represented in the legend as a power of 2, which designates the highest power of 2 that divides $h/a$. This shows how using a value of $h/a$ where the highest power of 2 multiple is 3 or less can impact performance greatly. Figures \ref{fig:attention-kq-ha64} and \ref{fig:attention-val-ha64-16384} show that in general, throughput increases with hidden size and decreases with the number of attention heads. Some of these figures also show the effects of wave quantization. 

\clearpage



\end{document}